\documentclass{jfm}
\usepackage{graphicx}
\usepackage{epstopdf,epsfig}
\usepackage{amsmath}
\usepackage{amssymb} % Use symbol
\usepackage{subcaption}
\usepackage{xcolor}
\usepackage{tikz}
\usepackage{newtxtext}
\usepackage{newtxmath}
\usepackage{natbib}
\usepackage{hyperref}
\hypersetup{
    colorlinks = true,
    urlcolor   = blue,
    citecolor  = black,
}

\graphicspath{ {Images/} }

\shorttitle{Fluid-solid Floquet stability analysis of self-propelled foils}
\shortauthor{L. Benetti Ramos, O. Marquet, M. Bergmann and A. Iollo}

\title{Fluid-solid Floquet stability analysis of self-propelled heaving foils}

\author{Luis Benetti Ramos\aff{1}
  \corresp{\email{luisbenettiramos@gmail.com}},
  Olivier Marquet\aff{1},
  Michel Bergmann \aff{2,3}
 \and Angelo Iollo\aff{2,3}}

\affiliation{\aff{1}ONERA -- The French Aerospace Lab, F-92190 Meudon, France
\aff{2} Université de Bordeaux, IMB, UMR 5251, F-33400 Talence, France 
\aff{3} Equipe-projet Memphis, Inria Bordeaux-Sud Ouest, F-33400 Talence, France}

\begin{document}

\maketitle

\begin{abstract}
We investigate the role of linear mechanisms in the emergence of nonlinear horizontal self-propelled states of a heaving foil in a quiescent fluid. Two states are analyzed: a periodic state of unidirectional motion and a quasi-periodic state of slow back \& forth motion around a mean horizontal position. The states emergence is explained through a fluid-solid Floquet stability analysis of the non-propulsive symmetric base solution. Unlike a purely-hydrodynamic analysis, our analysis accurately determine the locomotion states onset. An unstable synchronous mode is found when the unidirectional propulsive solution is observed. The obtained mode has a propulsive character, featuring a mean horizontal velocity and an asymmetric flow that generates a horizontal force accelerating the foil. An unstable asynchronous mode, also featuring flow asymmetry and a non-zero velocity, is found when the back \& forth state is observed. Its associated complex multiplier introduces a slow modulation of the flapping period, agreeing with the quasi-periodic nature of the back \& forth regime. The temporal evolution of this perturbation shows how the horizontal force exerted by the flow is alternatively propulsive or resistive over a slow period. For both modes, an analysis of the velocity and force perturbation time-averaged over the flapping period is used to establish physical instability criteria. The behaviour for large solid-to-fluid density ratio of the modes is thus analyzed. The asynchronous fluid-solid mode converges towards the purely-hydrodynamic one, whereas the synchronous mode becomes marginally unstable in our analysis not converging to the purely-hydrodynamic analysis where it is never destabilised.

\end{abstract}


%%%%%%%%%%%%%%%%%%%%%%%%%%%%%%%%%%%%%%%%%%%%%%%%%%%%%%%%%%%%%%%%%%%%%%%%%%%%%%%%%%%
\section{Introduction}

A common locomotion strategy adopted by aquatic or flying animals \citep{Gray1933,Wu2010,Shyy2010},
and more recently employed in the conception of large and small-scale artificial swimmers \citep{Barrett1996,Ramananarivo2013,Williams2014}, is the flapping motion of appendages such as wings, tails and fins. A fundamental question that impacts the design of micro-swimmers and aerial-vehicles \citep{Williams2014,Faux2018} is the critical size above which flapping-based propulsion remains efficient and applicable. 
Indeed, micro-organisms of very small scales, as cells or sperm, are known to exploit other locomotion strategies \citep{Lauga2011a}, ciliar or flagelar  propulsion, respectively. As first stated by \citet{Purcell1977} in the so-called scallop theorem, a reciprocal motion of appendages, for which the paths during the two half-strokes are identical but time-reversal, does not allow to generate a net thrust at those very small scales. This is due to the linearity and timeless nature of the surrounding flows which are entirely dominated by the viscous effects. An emblematic observation of the transition from ciliar to flapping propulsion has been achieved for the mollusc \emph{Clione Antartica} \citep{Childress2004} that disposes both of cilia and wings attached to its body. Whereas its cilia are always employed, the wings remain retracted to his body, being flapped exclusively after a critical velocity is reached. This switch of locomotion strategy was related to the evolution of the dynamical response of the surrounding flow as the Reynolds number, based on the swimming velocity, increases. The present paper aims at better understanding the emergence of flapping locomotion based on reciprocal motion.

To that aim, \citet{Vandenberghe2004,Vandenberghe2006} designed an experiment where flapping propulsion emerges exclusively from the flow and not from the motion asymmetry, which is generally explored in the animal world as to achieve a more efficient propulsion \citep{WeisFogh1973,Spagnolie2010}. The experiment consists of a horizontal flat rectangular foil immersed in still water filling a cylindrical tank and attached in its mid-span to a shaft. This shaft is vertically flapped with a sinusoidal motion and the foil is allowed to rotate, together with the shaft, around the vertical axis in the horizontal direction. Note that the foil is only heaving, not simultaneously heaving and pitching like in the experiments of \citet{Spagnolie2010} where the foil besides the imposed heaving was allowed to passively pitch around its leading edge. For a small enough frequency, the flow induced by a heaving motion of fixed amplitude is left-right symmetric. Thus, no hydrodynamic force is generated over the foil in the horizontal direction for every instants, and the foil does not rotate. However, once a critical Stokes number $\beta=f^*(c^*)^2/\nu$ is attained (a non-dimensional number similar to the Reynolds number that uses the dimensional flapping frequency $f^*$ and the foil chord $c^*$ as characteristic time and length scales, as well as the fluid kinematic viscosity $\nu$), the surrounding flow breaks its initial symmetry and generates horizontal forces. The foil then achieves locomotion and eventually reaches a permanent forward regime in equilibrium with the fluid. Subsequent numerical studies were dedicated to understand how the transient dynamics and the self-propelled regimes of this model problem evolve with respect to its control parameters. These studies simplified \citet{Vandenberghe2004} configuration, working with a two-dimensional cross-section of the experiment (imposed heaving and horizontally self-propelled foils), thus neglecting its rotational flow effects. Investigating the self-propulsion of elliptical foils in a two-dimensional incompressible flow under a fixed nondimensional chord-based flapping amplitude $A=0.5$, \citet{Alben2005} revealed that as the flapping frequency (equivalently the Stokes number) is increased, the foil motion transition between different self-propelled regimes that are a unidirectional propulsion (as in the experiments of \citet{Vandenberghe2004}), a quasi-periodic back \& forth motion around a fixed point in space (with a frequency remarkably lower than the flapping one) and even a chaotic motion. These authors have also shown that these self-propelled regimes are greatly impacted by the thickness-to-chord aspect ratio $h$ and the solid-to-fluid density ratio $\rho$. In one hand, thinner ellipses of aspect ratio $h=0.1$ are able to break symmetry at lower flapping frequencies and present thus a greater exponential growth of their horizontal velocity than thicker foils. On the other hand, for foils of greater density ratio ($\rho>10$) the existence of non-coherent and chaotic regimes is greatly reduced or even suppressed when compared to their lighter equivalents. \cite{Lu2006} for instance have shown for a fixed flapping amplitude and frequency that a non-coherent state of motion can be suppressed thanks to the increase of the density ratio. A similar observation was made by \cite{Zhang2009} while decreasing the aspect ratio of elliptical and rectangular foils, where in both cases smaller aspect ratios were found to be more prone to unidirectional locomotion than thicker ones. \cite{Lu2006} have equally mapped in the plan flapping amplitude/ frequency the transition between symmetric and non-symmetric flows (thus propelled ones) revealing that the transition occurs for smaller frequencies at higher flapping amplitudes. Later on, \citet{Deng2016a} established the same frontier for different aspect ratios, revealing that thinner foils break symmetry earlier in flapping amplitude and frequency. The authors also compared the frontier between symmetric and non symmetric flows for propelled and non-propelled foils for two-dimensional ellipses \citep{Deng2016a} or three-dimensional oblate spheroids \citep{Deng2016,Deng2017,Deng2018a}, indicating that the frontier of flow symmetry breaking is obtained in both two and three-dimensional cases for smaller frequency and amplitude for self-propelled foils. Using two-dimensional numerical simulations, we revisit in this work the nonlinear regimes of locomotion for a thin rectangular foil ($h^*=0.05 c^*)$, of density $\rho_{s}= 100 \rho_{f}$ and flapping with a fixed maximal amplitude  $A^{*}=0.5 c^{*}$. By varying the Stokes numbers in the range $1 \leq \beta \leq 20$, we first aim at carefully characterizing and identifying the transition between various self-propelled regimes of the foil motion: the non-propulsive, the unidirectional propulsive and the back \& forth motion. This parametric investigation is useful to accurately determine critical values of the Stokes numbers for which transition between these nonlinear regimes are observed. They will be used as basis of comparison for the second objective of this work, i.e. predicting the emergence of these regimes using linear stability analysis of the periodic flows generated by the flapping foil. \\

Floquet analysis allows to investigate the linear stability of periodic solutions \citep{Floquet1883}. In hydrodynamics,  \cite{Barkley1996} first performed this analysis on a two-dimensional time-periodic wake flow to explain the onset of three-dimensional structures in the wake of a \textit{fixed} circular cylinder. The Floquet analysis of time-periodic flows generated by flapping bodies has then been considered by \citet{Elston2004,Elston2006} for two-dimensional oscillating cylinder flows. They successfully explained the emergence of two and three dimensional flow asymmetries observed in experiments and simulations. More recently, \cite{Jallas2017} performed a Floquet analysis of time-periodic propulsive wake generated by a pitching wing. They identify an unstable synchronous mode that successfully explains the lateral deviation of the propulsive vortex street observed when increasing the flapping frequency. To investigate the emergence of the self-propelled regimes described above,  \cite{Deng2016a,Deng2016,Deng2017,Deng2018a} first proposed to consider a Floquet analysis for various flapping foil configurations. Based on the observation that flow symmetry breaking occurs prior to the self-propulsion of the foil in temporal simulations, \citet{Deng2016a,Deng2018a} applied a purely-hydrodynamic analysis, that does not consider  a perturbation of the foil speed in the propulsion direction. For a certain range of control parameters, results of this purely hydrodynamics analysis are in good agreement with nonlinear results. In particular,
they identify unstable asynchronous  modes at flapping frequencies where 
the foil exhibit a slow quasi-periodic back \& forth motion.
However, some disagreements between results of the purely-hydrodynamic analysis and the nonlinear results of self-propelled simulations were also reported, as in the case of low aspect ratio $h=0.1$ ellipses, where linear analysis fails to predict the onset of unidirectional forward locomotion \citep{Deng2016a}. In the present work, we introduce the so-called fluid-solid Floquet analysis that considers the foil speed as a perturbation variable and takes into account the inherent coupling between the flow perturbation and the rigid motion of the foil at the perturbation level. We will demonstrate, by comparison with nonlinear results, that this fluid-solid coupling is essential to correctly represent and predict the emergence of self-propelled regimes. \\

The importance of the fluid-solid coupling in linear stability analysis has a long history in aeroelasticity (see the review by \cite{dowell1989modern}) that investigates the infinitesimal motion of structures immersed in high Reynolds number flows. Fluid-solid stability analyses for lower Reynolds number flows are more recent. To our knowledge, \cite{Cossu2000} first performed the fluid-stability analysis of the steady wake cylinder flow to explain the sub-critical vortex-induced vibration of the cylinder when mounted on a spring. The path of bodies freely rising or falling in fluids under the effect of gravity (see \cite{Ern2012} for a review) is another example where fluid-solid linear stability analyses successfully explained the emergence of various trajectories. \cite{Tchoufag2014} first elucidate the path instability of buoyancy-driven disks/thin cylinders and then of freely rising spheroidal bubble \citep{Tchoufag2014a}. Recently, 
\cite{negi2019global} proposed a simplified formulation to handle the linearized fluid-structure interaction for rigid bodies. 
Fluid-solid stability analysis has also been extended to deformable (elastic) structures, to explain the dynamics of inverted flags in uniform flows \citep{Goza2018} and of flexible splitter plates  clamped to the rear of a cylinder \citep{Pfister2020}. Note also that \cite{tammisola2012surface} investigated the global instability of planar jets and wakes in two immiscible fluids, focusing on the effect of surface tension. In all of these studies, the temporal evolution of perturbations over a steady base flow solutions was considered. To our knowledge, the fluid-solid stability analysis of time-periodic flow solutions has never been addressed in the context of fluid-solid interaction. In the present study, we introduce the mathematical formalism of such analysis and apply it to explain the emergence of self-propelled flapping states. Additionally, a time-averaged analysis is proposed to highlight the role of the fluid-solid coupling in the destabilization of the Floquet modes. Such connections between linear modes and thrust efficiency have been for instance highlighted in the literature as key factors for an optimal frequency selection in flapping wings \citep{Triantafyllou1993,Moored2012}.\\

This article is organized in two parts. In 
\S \ref{sec:nonlinear}, we investigate numerically the nonlinear regimes of locomotion for a self-propelled heaving foil. The configuration and non-dimensional parameters are introduced before describing the governing nonlinear equations and numerical methods. The self-propelled solutions obtained for a fixed flapping amplitude and density ratio are then carefully described for three values of the Stokes number. The transition between regimes of non-propulsive, unidirectional propulsive and back \& forth motions are finally identified by varying the Stokes number in the range $2 \le \beta \le 20$.  In \S \ref{ref:FluidSolidFloquet}, we introduce first the fluid-solid Floquet stability analysis of self-propelled foils and then the time-averaged analysis that allows to establish instability criteria based on the velocity and force of the Floquet mode. 
Results of this fluid-solid analysis, performed at $\rho=100$ for symmetric non-propulsive solutions, are first described by analysing the synchronous and asynchronous modes found unstable at different Stokes numbers. Those results are then compared to those obtained with the purely-hydrodynamic Floquet analysis and with the nonlinear temporal simulations. The effect of the density ratio on the two unstable Floquet modes is finally described. 

\begin{figure} 
   \begin{center}
       \includegraphics[width=0.6\linewidth]{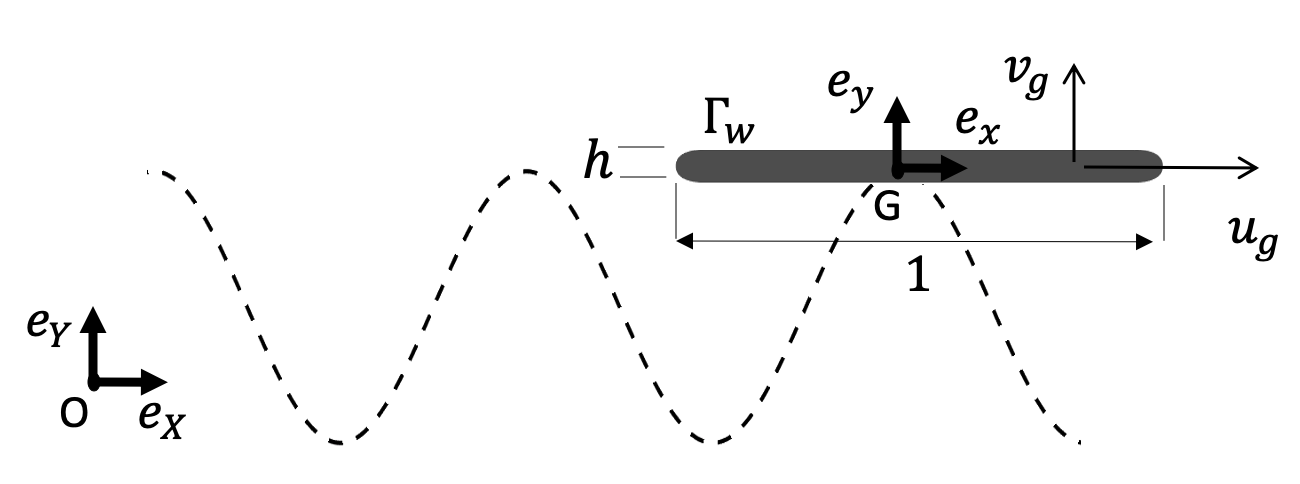}
    \end{center}
    \caption{Sketch of the foil configuration under sinusoidal vertical motion and horizontal translation with velocity $\mathbf{u}_g$. The foil chord (trailing to leading edge) and its thickness, the frames of reference and the solid/fluid interface $\Gamma_w$ are indicated.}
    \label{Configuration}
\end{figure}

%%%%%%%%%%%%%%%%%%%%%%%%%%%%%%%%%%%%%%%%%%%%%%%%%%%%%%%%%%%%%%%%%%%%%%%%%%%%%%%%%%%%%%%%%%%%%%%%%%%%%%%%%%
\section{Problem formulation and self-propelled nonlinear solutions} \label{sec:nonlinear}

We investigate the horizontally constrained locomotion of a vertically heaving foil of density $\rho_s$ immersed in an initially  quiescent fluid of density $\rho_f$ and viscosity $\nu$. The foil, shown in figure \ref{Configuration}, is  similar to the one used in the experimental studies by \citet{Vandenberghe2004,Vandenberghe2006}. Its rectangular shape is characterized by the thickness $h^*$ and chord $c^*$ with rounded corners of diameter equal to the foil thickness. The periodic displacement imposed along the vertical axis $e_y$ is
\begin{equation}\label{eqn:solidmotion}
    y_g^{*}(t)= - A^{*} \cos(2\pi f^{*} t^{*}) \, ,
\end{equation}
where the superscript $^*$ is used to indicate dimensional variables. $A^{*}$ is the maximal vertical amplitude and $f^{*}$ is the flapping frequency, and $T^{*}=1/f^{*}$ is the flapping period. The foil is free to move along the horizontal axis $e_x$ as a result of hydrodynamic forces acting on the solid-fluid interface $\Gamma_w$. This rigid-body fluid-structure interaction is characterized by four non-dimensional parameters, namely the frequency-based Stokes number $\beta$, the non-dimensional amplitude $A$, the solid-fluid density ratio $\rho$ and the non-dimensional thickness $h$, defined respectively as
\begin{equation}
     \beta = \frac{f^{*} (c^*)^2}{\nu}, \, A=\frac{A^{*}}{c^*}, \, \rho=\frac{\rho_s}{\rho_f} \mbox{ and } h=\frac{h^*}{c^*} \,.
\end{equation}
These parameters are obtained by choosing the chord $c$ as characteristic length scale, the fluid density $\rho_{f}$ as characteristic mass scale and the flapping period $1/f^*$ as characteristic time. In the following, all variables are thus made non-dimensional using these scales. Note that the non-dimensional flapping period $T$ is thus equal to $1$ whatever the  values of the Stokes number $\beta$, which is the only parameter or variable containing the dependency to the dimensional frequency $f^{*}$. Other choices of characteristic scale are also possible and made in the literature. For instance, \cite{Alben2005} used  the flapping velocity $ A^{*} f^{*} $ as characteristic velocity, thus introducing the flapping amplitude based Stokes number $\beta_A = A^{*} f^{*} c
^*/\nu = A \beta$.

\smallskip
\smallskip

In the present study, the foil geometry and the flapping amplitude are fixed to $h=1/20$ and $A=0.5$, respectively. This aspect ratio 
is close to the experimental devices of \citet{Vandenberghe2004}, and 
this flapping amplitude equals the one adopted by \cite{Alben2005}. A discussion of the influence of these two parameters can be found in \cite{Zhang2009} and \cite{Deng2016a}. In this section, we will investigate numerically  the nonlinear dynamics of the foil for the fixed density ratio $\rho=100$ in the range of Stokes number $ 1 \le \beta \le 20$. 

%%%%%%%%%%%%%%%%%%%%%%%%%%%%%%%%%%%%%%%%%%%%%%%%%%%%%%%%%%%%%%%%%%%%%%%%%%%%%%%%%%%%%%%%%%%%%%%%%%%%%%%%%%

\subsection{Governing non-linear equations}\label{subsec:nonlinearfom}
%% ----------------------------------------------------------------------------------------------%%
The dynamics of the foil interacting with the surrounding fluid is described by
the non-dimensional variable $\mathbf{q}=(\mathbf{u},p,u_g)^{T}$ where $\mathbf{u}=(u,v)$ is the two-dimensional fluid velocity field, $p$ is the pressure field and $u_g$ is the foil horizontal velocity. The fluid-solid variable is governed by the evolution equation 
\begin{equation} \label{AllEqs-Compact}
\mathcal{B}\frac{\partial\mathbf{q}}{\partial t} = \mathcal{R}(\mathbf{q},v_g) \, ,
\end{equation}
where $v_g(t)=2 \pi A \sin(2 \pi t)$ is the non-dimensional foil vertical velocity and the operators $\mathcal{B}$ and $\mathcal{R}$ are defined as
\begin{equation}\label{eqn:residual}
     \mathcal{B}=\begin{bmatrix}{}
                \mathcal{I} & 0 & 0 \\
                0 & 0 & 0 \\
                0 & 0 & 1
        \end{bmatrix}
      , \, \mathcal{R}(\mathbf{q},v_g)=\begin{bmatrix}{}
      - \left( [\mathbf{u}-\mathbf{u}_g]\cdot \boldsymbol{\nabla} \right) \mathbf{u}  - \nabla p + \beta^{-1} \Delta \mathbf{u}
                \\
                -\nabla \cdot \mathbf{u} \\
                \displaystyle (\rho S)^{-1} \displaystyle F_x(\mathbf{u},p)
        \end{bmatrix}  \, .
 \end{equation}{}
The first and second lines are the incompressible Navier-Stokes equations written in a non-inertial frame of reference, denoted $(G,\mathbf{e}_x,\mathbf{e}_y)$ in figure \ref{Configuration}, that translates at the foil speed $\mathbf{u}_g=(u_g,v_g)$ in the laboratory frame of reference $(O,\mathbf{e}_X,\mathbf{e}_Y)$. Note that both solid and fluid velocities are absolute velocities \citep{Mougin2003a,Jenny2004}, the relative flow velocity $(\mathbf{u}-\mathbf{u}_g)$ appearing in the non-linear term of the momentum equations. While the solid vertical velocity $v_g$ is imposed, the temporal evolution of the foil horizontal velocity $u_g$ is governed by the Newton's second law, as stated by the third line in (\ref{AllEqs-Compact},\ref{eqn:residual}). The horizontal acceleration is equal to the horizontal hydrodynamic force $F_x(\mathbf{u},p)$ 
weighted by the non-dimensional mass of the foil $\rho S$ ($S=h(1-h)+\pi h^2/4$ being its non-dimensional surface). This hydrodynamic force depends on the fluid velocity and pressure as
\begin{equation}\label{eqn:fluidforce}
F_x = \int_{\Gamma_w}([-p \mathbf{\mathcal{I}} + \beta^{-1} (\nabla\mathbf{u}+\nabla\mathbf{u}^T)] \cdot \mathbf{n}) \cdot \mathbf{e_x} \, \mathrm{d} \Gamma_w \,.
\end{equation}
where $\Gamma_w$ denotes the fluid-solid boundary. An additional coupling between the fluid and solid variables is due to the equality of velocities at the fluid-solid interface, i.e. 
\begin{equation}\label{eqn:fluidsolid}
    \mathbf{u}(\Gamma_w,t)=\mathbf{u}_g(t)= [u_g(t), 2\pi A \sin(2\pi t)]^T \,.
\end{equation}{}
The fluid is at rest sufficiently far away from the foil.

\begin{figure}
\vspace{0.4cm}
    \centering
    \begin{tabular}{ll}
    (a) & (b)\\
    \includegraphics[width=0.4\linewidth]{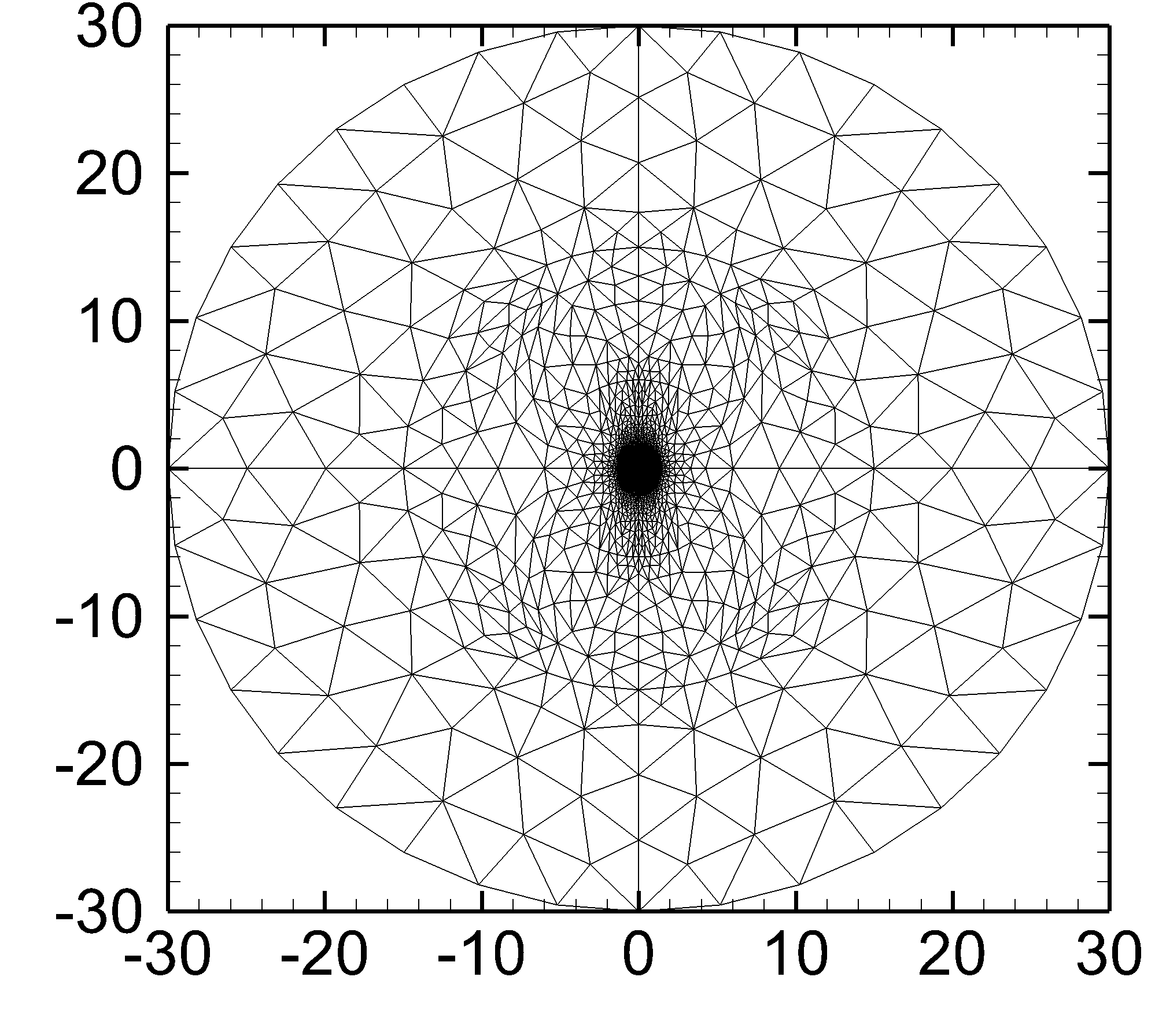}
    & \includegraphics[width=0.4\linewidth]{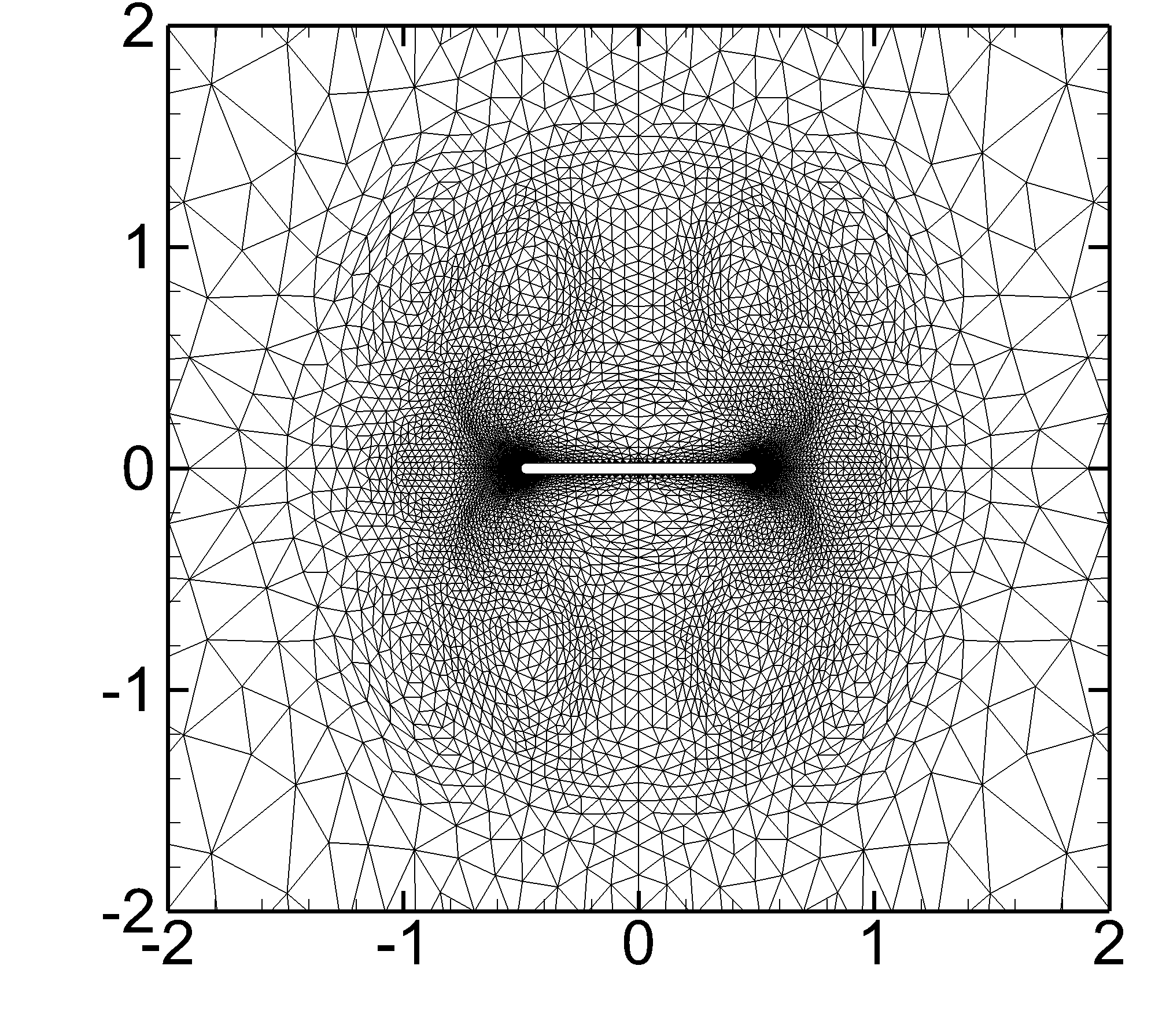}
    \end{tabular}
    \caption{Computational domain and mesh. (a) Full and (b) close-up views of a typical mesh  adapted to the flow solution}
    \label{Mesh}
\end{figure}{}

\subsection{Numerical methods}\label{subsec:nonlinearmethod}

The system of equations \eqref{AllEqs-Compact}, \eqref{eqn:residual}, \eqref{eqn:fluidforce} and  
\eqref{eqn:fluidsolid} is discretized in time using the following $r$-order semi-explicit scheme
\begin{eqnarray} \label{eqn:NS-Implicit}
\frac{\alpha_0}{\Delta t} \mathbf{u}^{n+1} +
\nabla{p}^{n+1} -\beta^{-1} \Delta \mathbf{u}^{n+1} &=&  \mathbf{f}^{n+1} \nonumber \\
\nabla \cdot \mathbf{u}^{n+1} &=& 0 \, \nonumber \\
    (u^{n+1},v^{n+1})(\Gamma_{w}) &=& \left(u_g^{n+1}, 2 \pi \,A \, \sin(2 \pi t^{n+1}) \right) \\
\frac{\alpha_0}{\Delta t} u_g^{n+1} &=& (\rho S)^{-1}F_x(\mathbf{u}^{n+1},p^{n+1}) - \sum_{k=1}^r \frac{\alpha_k}{\Delta t} u_{g}^{n+1-k} \, \nonumber \\
(u^{n+1},v^{n+1})(\Gamma_{e}) &=& (0,0) \nonumber,
\end{eqnarray}
where the right-hand side forcing term $\mathbf{f}^{n+1}$ in the momentum equation is defined as 
 \begin{eqnarray*} \label{LinearOperator}
\mathbf{f}^{n+1} &=& - \sum_{k=1}^r \gamma_{k} (\mathbf{u}^{n+1-k}- \mathbf{u_g}^{n+1-k})\cdot \nabla \mathbf{u}^{n+1-k} - \sum_{k=1}^r \frac{\alpha_k}{\Delta t} \mathbf{u}^{n+1-k} \,.
\end{eqnarray*}
with $\Delta t$ the time step and $(\mathbf{u}^{n+1},p^{n+1})$ the velocity and pressure at time $t_{n+1}=(n+1) \Delta t$. A quiescent fluid condition is applied in the external boundary $\Gamma_e$ of the computational domain, typically far away from the foil. The time derivatives are approximated by $r$-order backward differential formulae. The linear diffusion and pressure gradient terms are implicit, while the nonlinear convection terms are extrapolated with $r$-order formulae. A first-order scheme ($r=1$, $\alpha_0=1$, $\alpha_1=-1$ and $\gamma_1=1$) is used for the first two temporal iterations ($n \le 1$), before switching in the subsequent iterations ($n>1$) to a second-order scheme ($r=2$, $\alpha_0=3/2$, $\alpha_1=-2$, $\alpha_2=1/2$, $\gamma_1=2$ and $\gamma_2=-1$). To avoid severe time-step restrictions for small values of density ratio induced by an explicit coupling \citep{Causin2005}, the equality of fluid and solid velocity is treated implicitly. To allow the use of an existing fast implementation to solve the flow equations \cite{Jallas2017}, we use a segregated approach, proposed by \citet{Jenny2004} and detailed in Appendix  \ref{Appendix:TimeMarchingScheme}, to solve the coupled fluid-solid problem.
Typically, the time step is set to $\Delta t=10^{-2}$ for small values of the Stokes number ($\beta=2$) and is decreased to $\Delta t=5\cdot10^{-4}$ for larger values ($\beta=19$), so as to ensure the numerical stability of this semi-explicit temporal scheme \citep{Kress2006}.\\

The linear equations \eqref{eqn:NS-Implicit} are discretized in space using a classical finite-element method. The flow velocity is discretized with quadratic elements (P2) while the pressure is discretized with linear element (P1). The implementation is based on the FreeFEM software \citep{Hecht2012}. The computational domain, displayed in figure  \ref{Mesh}(a), is a circle of (non-dimensional) diameter $60$ centered at the foil center of mass,
the external boundary of this circular domain being $\Gamma_{e}$. A Delaunay triangulation of the computational domain results in mesh with typically $1.2 \times 10^{4}$ triangles. As spatially symmetric solutions are expected, a particular attention was given to create a symmetric mesh and not artificially insert asymmetries in the flow.
To create a mesh that is symmetric with respect to the $x$ and $y$-axis and refined in flow regions exhibiting large velocity gradients (see figure \ref{Mesh}-b), we have proceed as follows. Once a first solution has been computed, we adapt a mesh of a quarter domain to several instants of the periodic flow, using the hessian-based mesh adaptation implemented in FreeFEM. We refer to \citet{Fabre2018} for a practical review. After duplicating and appropriately rotating this quarter-mesh, the full mesh can finally be assembled. The triangle size is typically of order $\mathcal{O}(10^{-2})$ close to the foil, and $1$ in the external part of the domain. Mesh refinement and domain size were chosen based on the convergence of the foil horizontal velocity and the vertical hydrodynamic force. Greater domains or mesh refinement have exhibited little influence over these values. The validation of this numerical method is detailed in Appendix \ref{Appendix:SectionValidation} by comparison with results of \citet{Spagnolie2010}.

%%%%%%%%%%%%%%%%%%%%%%%%%%%%%%%%%%%%%%%%%%%%%%%%%%%%%%%%%%%%%%%%%%%%%%%%%%%%%%%%%%%%%%%%%%%%%%%%%%%%%%%%%%
\subsection{Results}

\begin{figure}
   \centering
          \begin{subfigure}{0.24\textwidth}
    \includegraphics[width=\linewidth]{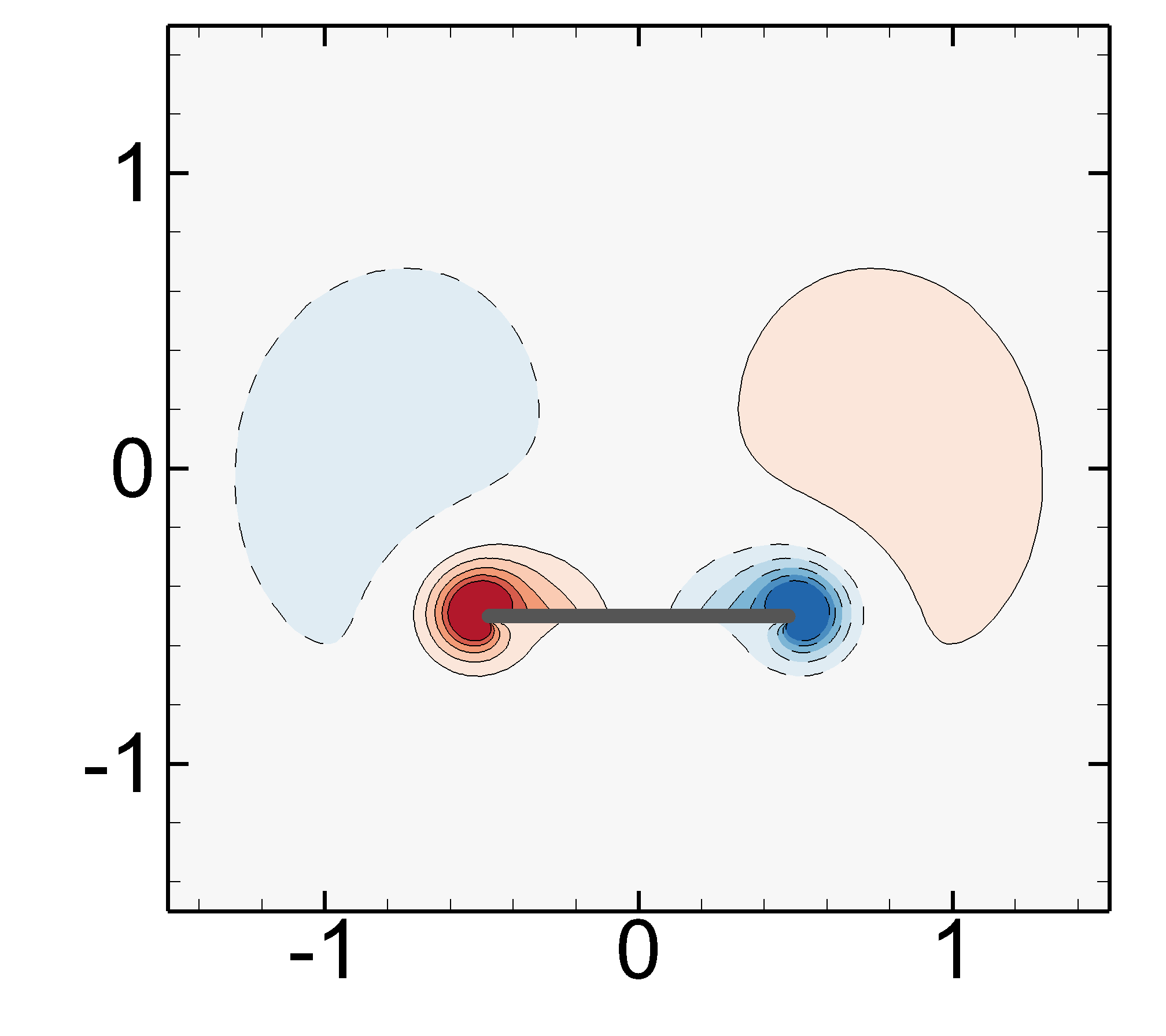}
    \caption{$t_0$}
    \end{subfigure}
       \begin{subfigure}{0.24\textwidth}
    \includegraphics[width=\linewidth]{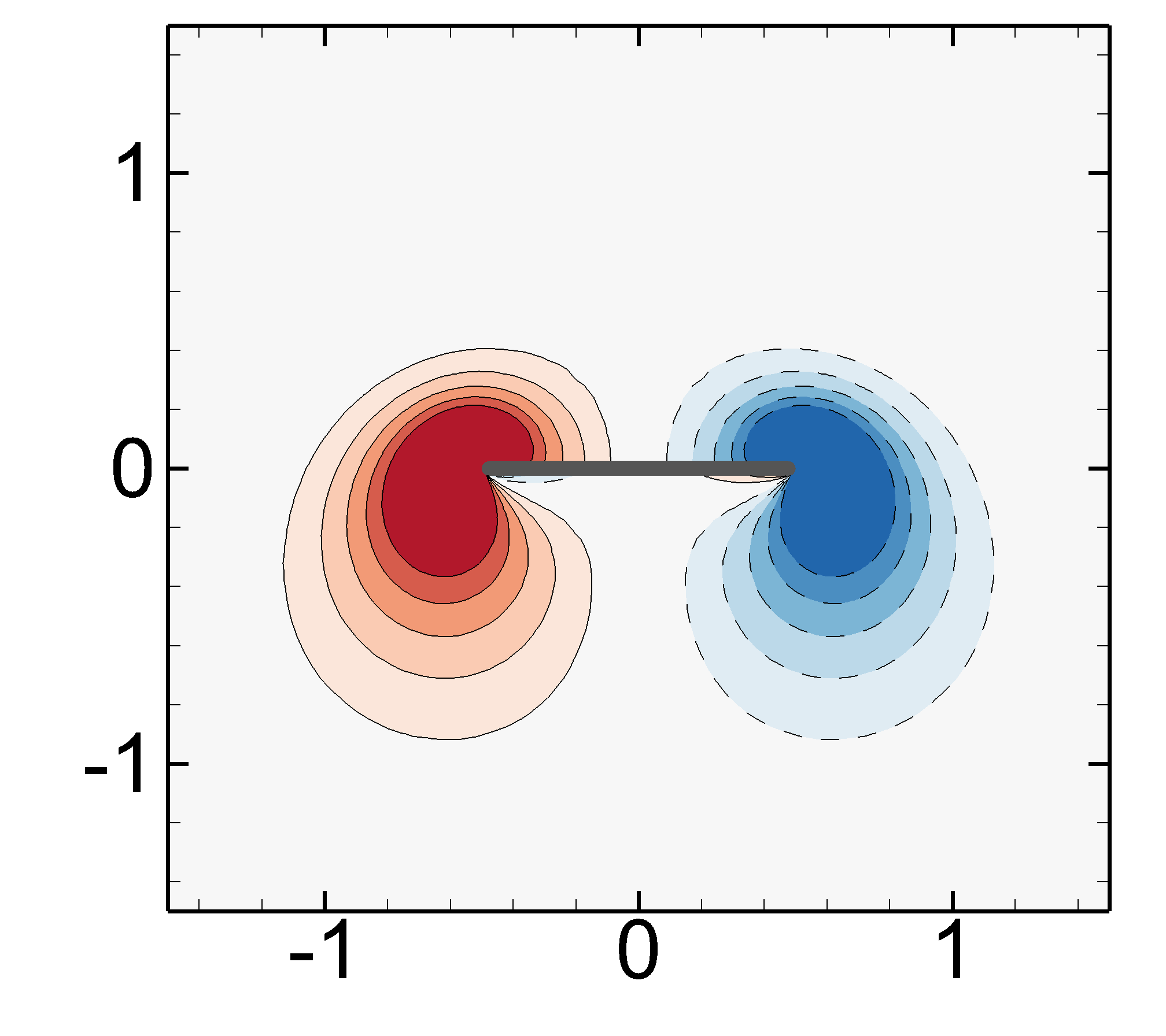}
    \caption{$t_0+1/4$}
    \end{subfigure}
       \begin{subfigure}{0.24\textwidth}
    \includegraphics[width=\linewidth]{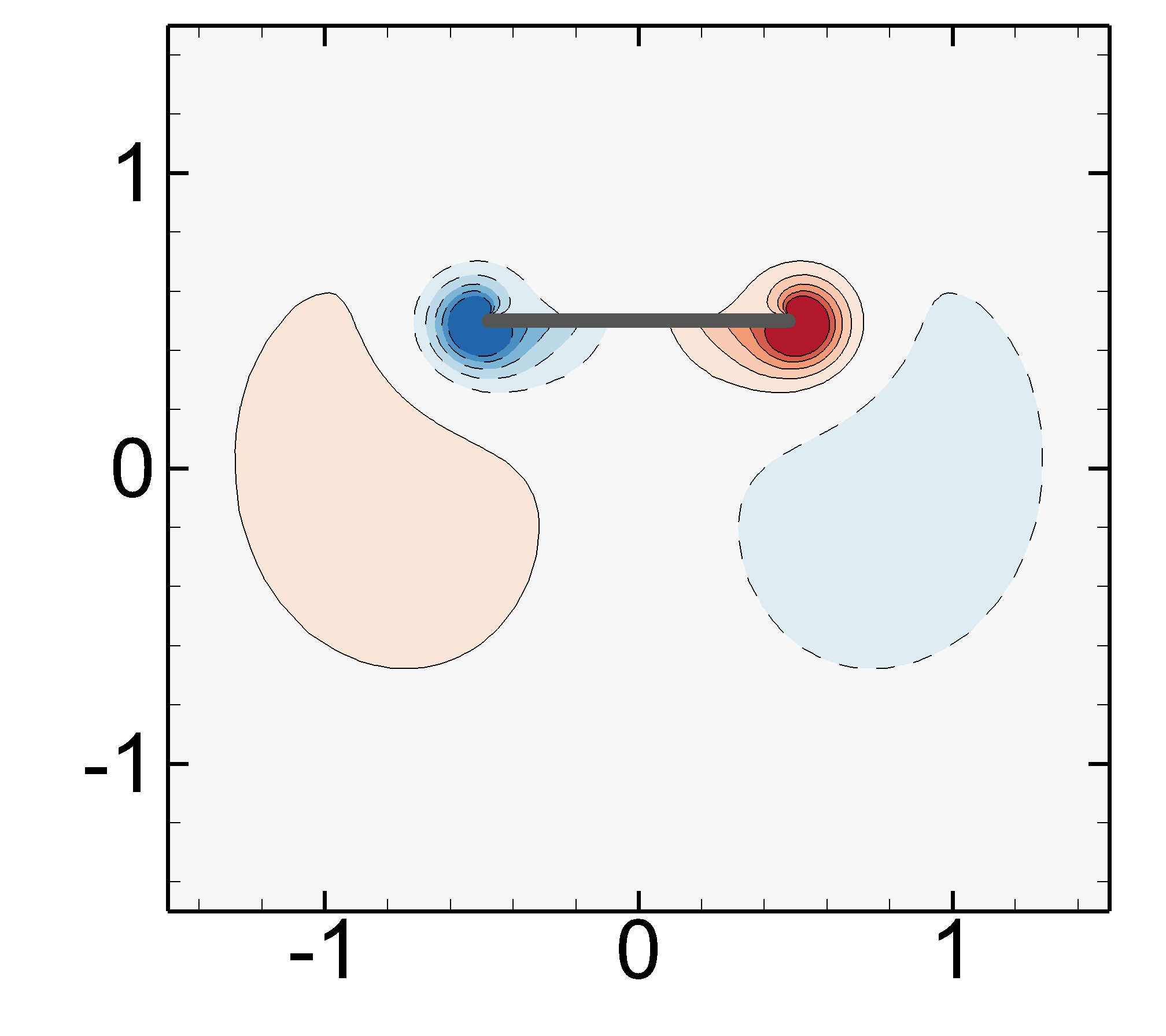}
    \caption{$t_0+1/2$}
    \end{subfigure}
    \begin{subfigure}{0.24\textwidth}
    \includegraphics[width=\linewidth]{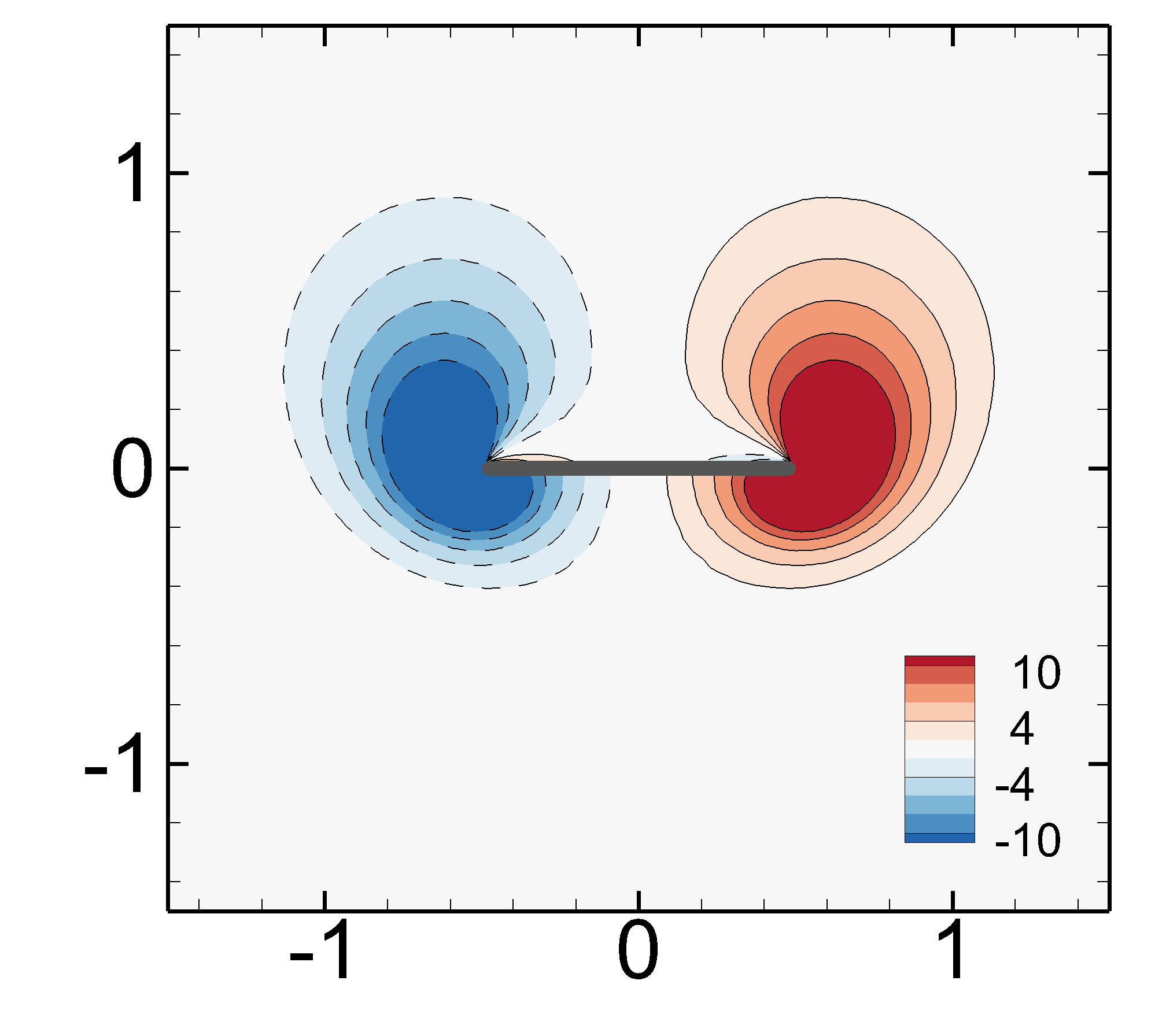}
    \caption{$t_0+3/4$}
    \end{subfigure}
   \begin{subfigure}{0.24\textwidth}
    \includegraphics[width=\linewidth]{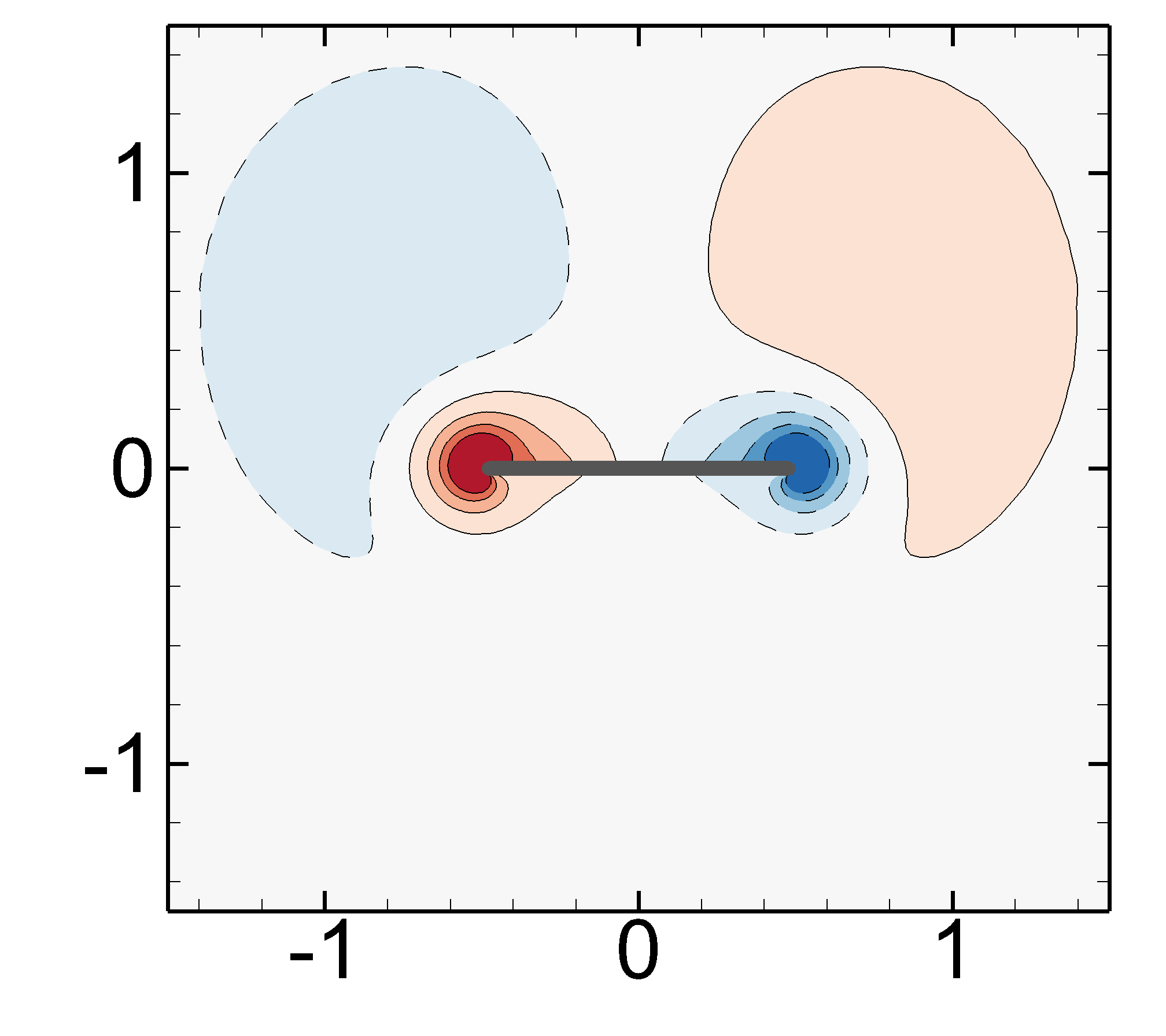}
    \caption{$t_0$}
    \end{subfigure}
       \begin{subfigure}{0.24\textwidth}
    \includegraphics[width=\linewidth]{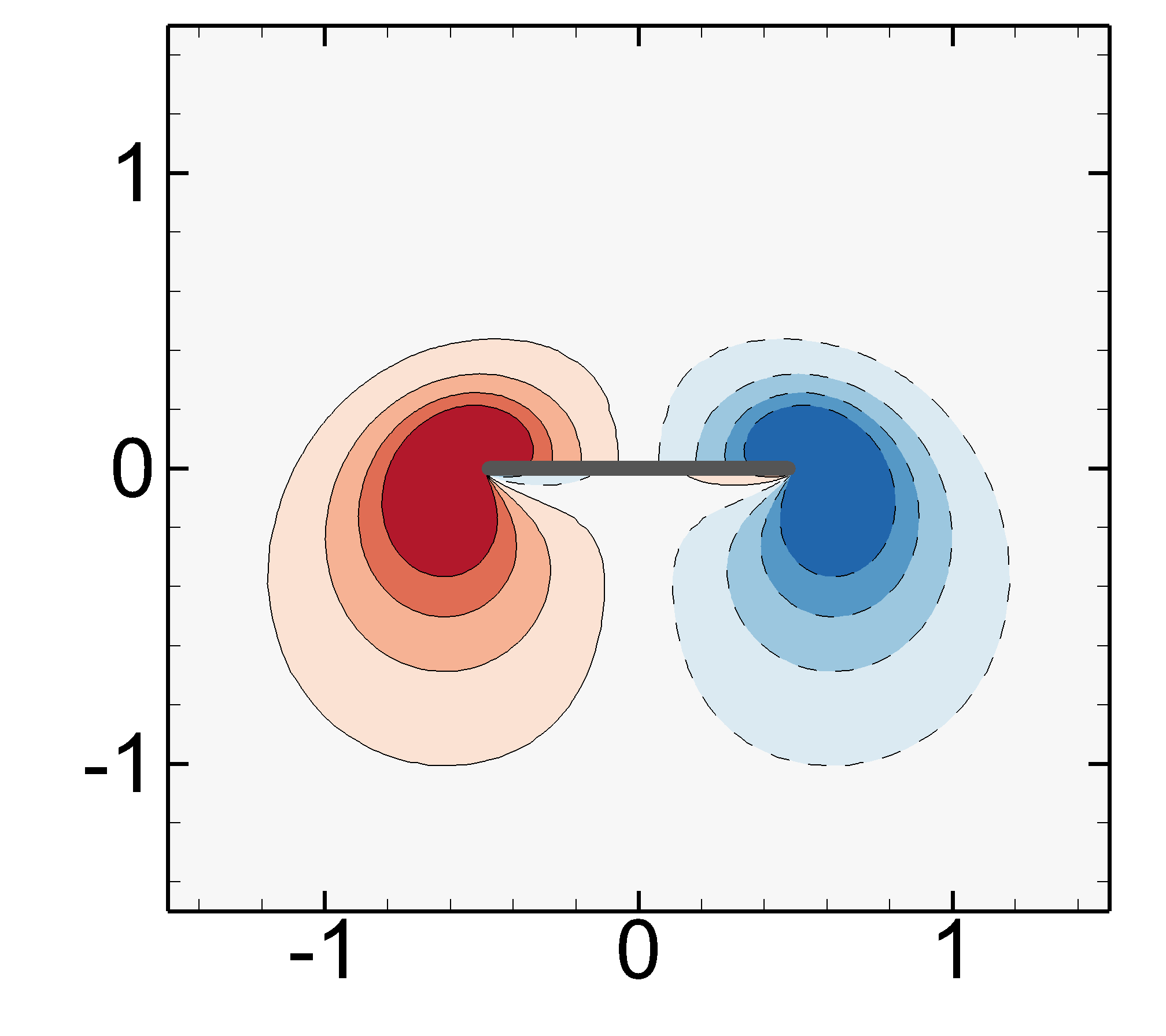}
    \caption{$t_0+1/4$}
    \end{subfigure}
       \begin{subfigure}{0.24\textwidth}
    \includegraphics[width=\linewidth]{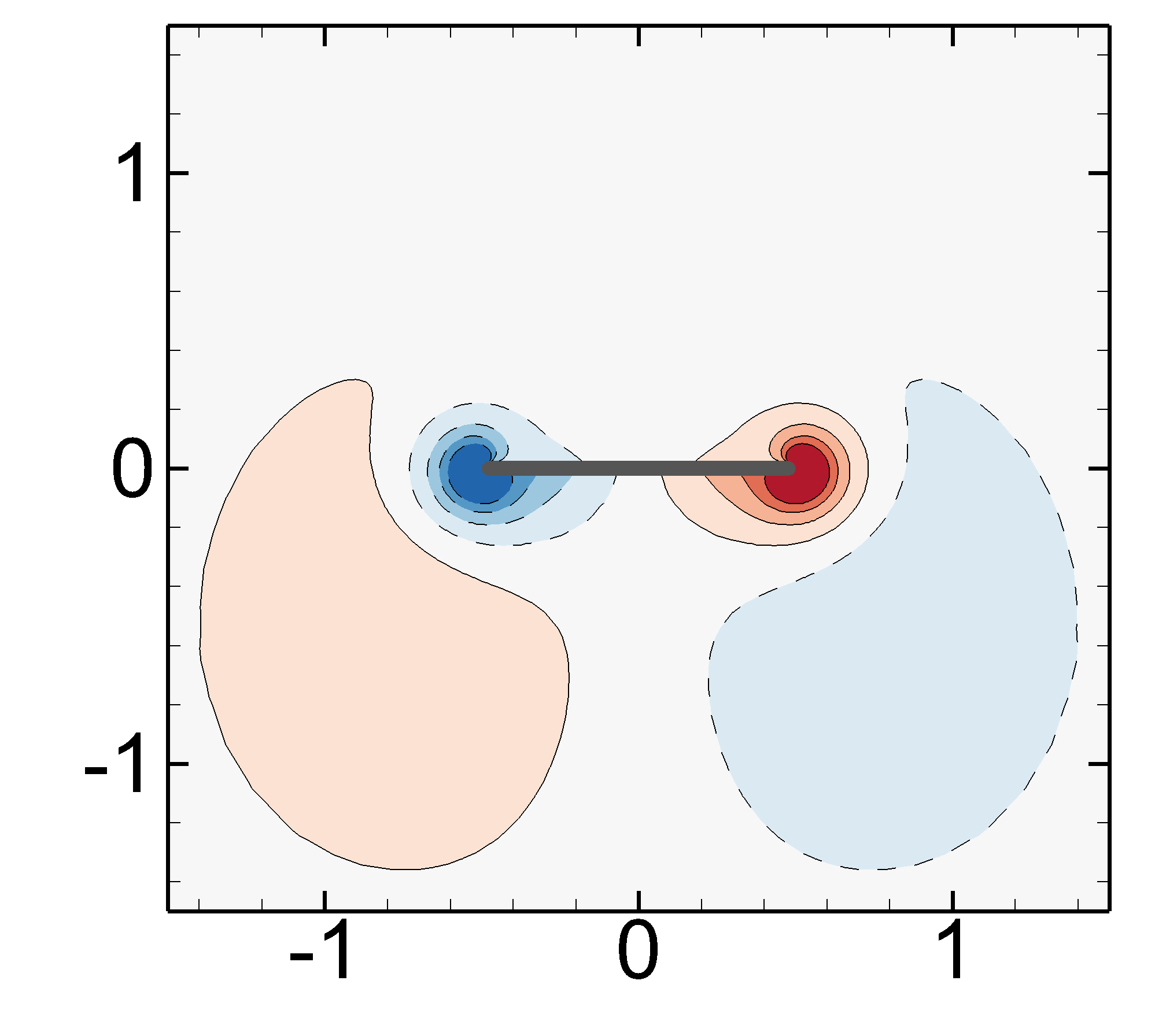}
    \caption{$t_0+1/2$}
    \end{subfigure}
    \begin{subfigure}{0.24\textwidth}
    \includegraphics[width=\linewidth]{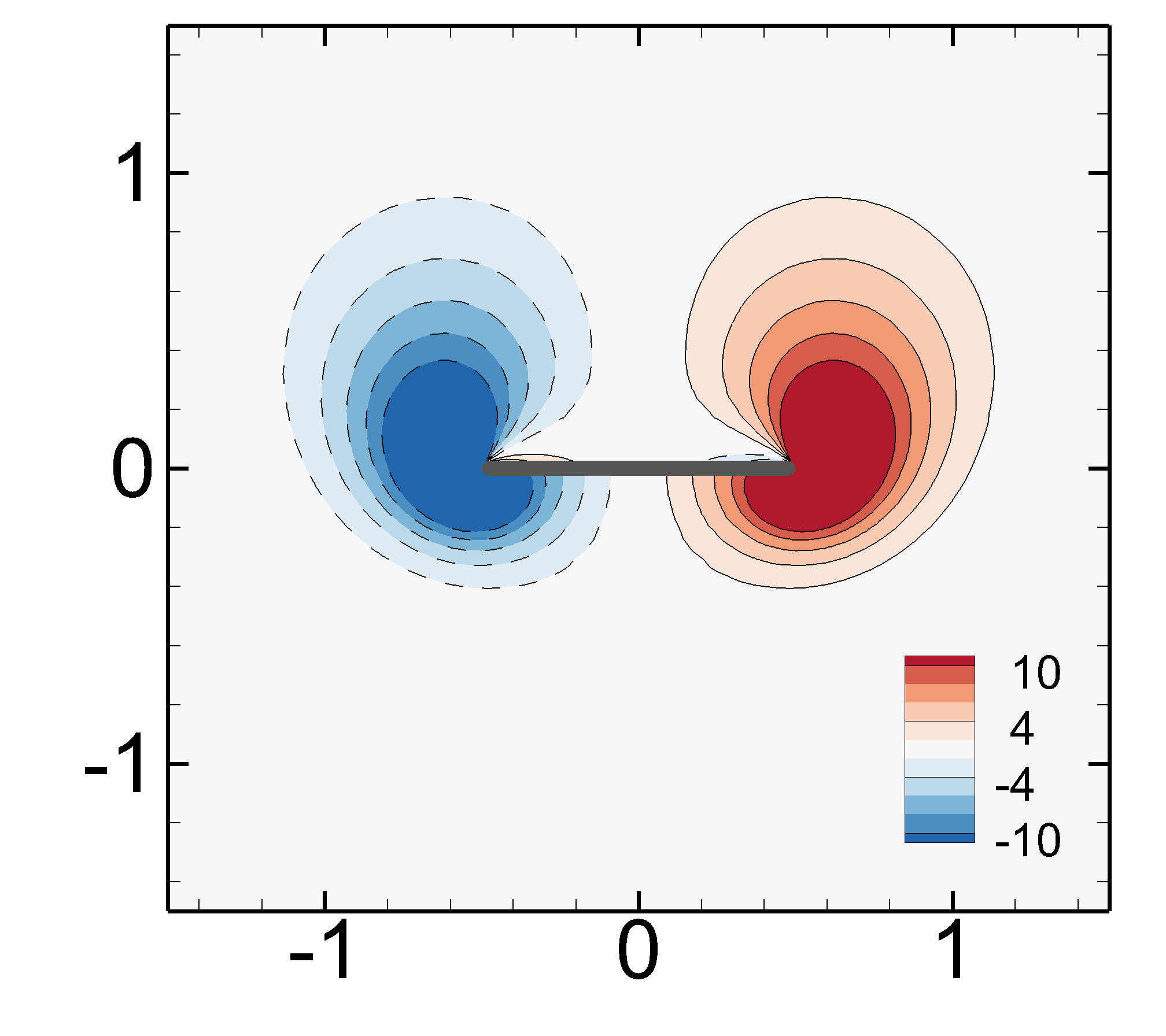}
    \caption{$t_0+3/4$}
    \end{subfigure}
    \caption{Symmetric solution for $\beta=2$. The vorticity field is depicted in the laboratory frame of reference $(X,Y)$ (a-d) and in the non-inertial frame of reference $(x,y)$ (e-h) for four equally-spaced instant of the unitary period. The initial time $t_0$ corresponds to the lowest vertical position of the foil.}
    \label{2-NonLinear-Sym}
\end{figure}

Unsteady nonlinear simulations are performed for values of the Stokes numbers in the range $1 \le \beta \le 20$. The amplitude $A=0.5$ and the foil aspect ratio $h=1/20$ are kept fixed throughout this study. When increasing the Stokes number, three different types of solution are successively observed, called hereinafter the (1) symmetric non-propulsive, (2) unidirectional propulsive and (3) back \& forth solutions. In \S \ref{Subsec:RegimesRho100} we first describe these solutions for three representative values of Stokes number and for a fixed density ratio $\rho=100$, concluding the section by a summary of the Stokes numbers range for which these solutions are obtained. In \S \ref{Subsec:Foil_Shape_Rho} these results are compared to the ones of a smaller density ratio closer to aquatic swimming ($\rho=1$).
These different type of solutions have already been experimentally or numerically observed in previous studies.
The transition from non-propulsive to unidirectional propulsive solutions was investigated in the works of \citet{Vandenberghe2004,Vandenberghe2006}, while back \& forth solutions have been computed numerically in \citep{Lu2006,Alben2005,Deng2018a}.
Self-propelled regimes presented in this section are thus not new but aim to establish the transition route for comparison with the linear Floquet stability analysis performed in the next section.

% \subsubsection{Symmetric non-propulsive solutions}
% \label{Subsec:Symmetric}

\subsubsection{Self-propelled regimes for $\rho=100$}
\label{Subsec:RegimesRho100}

A typical solution obtained for small values of the Stokes number is displayed for $\beta=2$ in figure \ref{2-NonLinear-Sym}. The flow induced by the flapping foil inherits the spatial symmetry of the foil and the temporal symmetry of the imposed vertical motion \citep{Elston2004}. It satisfies the $x$-reflection spatial symmetry in the non-inertial frame of reference, i.e. 

\begin{equation} \label{spatialsymmetry}
(u,v,p,\omega_{z})(x,y,t)=(-u,v,p,-\omega_{z})(-x,y,t) \, ,
\end{equation}

and the spatio-temporal symmetry 

\begin{equation} \label{spatiotemporalsymmetry}
(u,v,p,\omega_{z})(x,y,t)=(u,-v,p,-\omega_{z})(x,-y,t+T/2) \, ,
\end{equation}

which is the combination of the $y$-reflection symmetry and the $T/2$ time-reciprocal translation. The vorticity $\omega_{z}$, used to display the solution, in figure \ref{2-NonLinear-Sym}(a-d) and (e-h) respectively at the inertial and non-inertial frames of reference for four equally-spaced instants of the period $T$, is clearly an odd function of the $x$ variable for every time instants. Physically, the spatial symmetry is seen by the vortices of equal shape but different sign shed one each side of the foil during its vertical motion. The spatio-temporal flow symmetry is observed by the inversion of the vorticity sign in opposite foil strokes. A direct consequence of the spatial flow symmetry is the absence of instantaneous hydrodynamic forces acting in the horizontal direction, i.e. $F_x(t)=0$. Consequently, the foil is not accelerated in that direction and its velocity remains equal zero, hence the name of  symmetric non-propulsive solution.

% \subsubsection{Unidirectional propulsive solutions}

\smallskip
\smallskip

\begin{figure}
   \centering
   \begin{subfigure}{0.24\textwidth}
    \includegraphics[width=\linewidth]{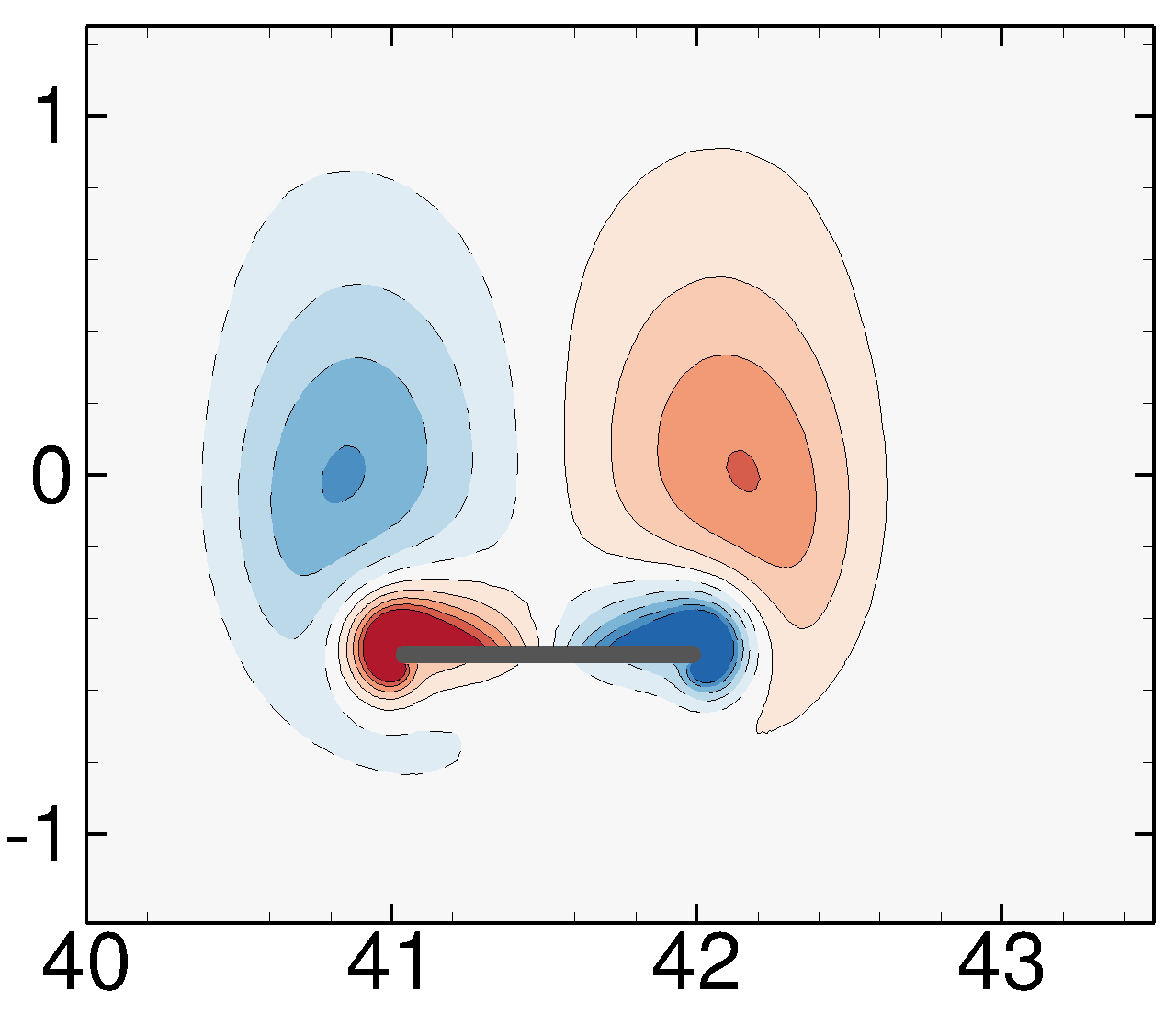}
    \caption{$t_0=324$}
    \end{subfigure}
       \begin{subfigure}{0.24\textwidth}
    \includegraphics[width=\linewidth]{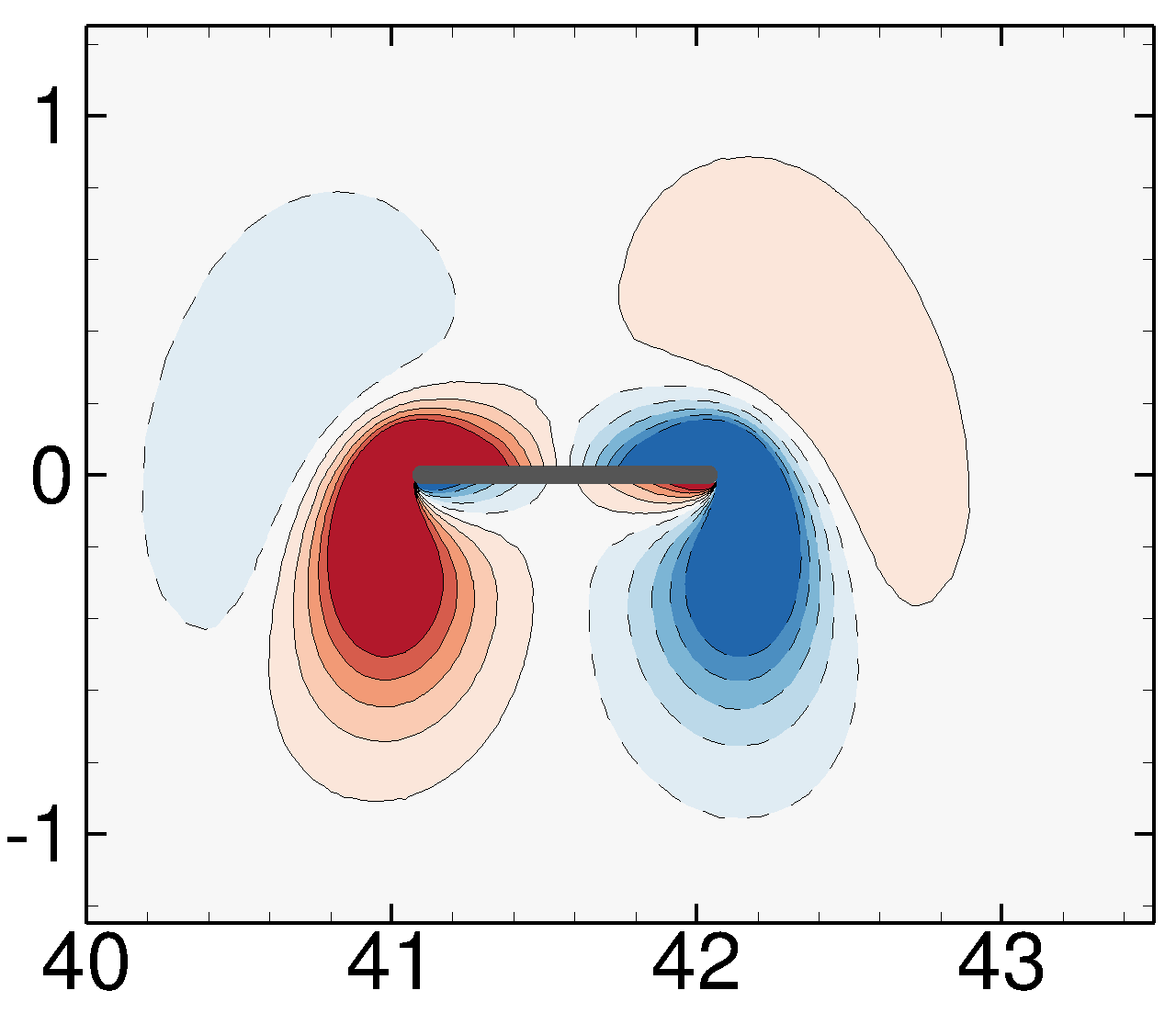}
    \caption{$t_0+1/4$}
    \end{subfigure}
       \begin{subfigure}{0.24\textwidth}
    \includegraphics[width=\linewidth]{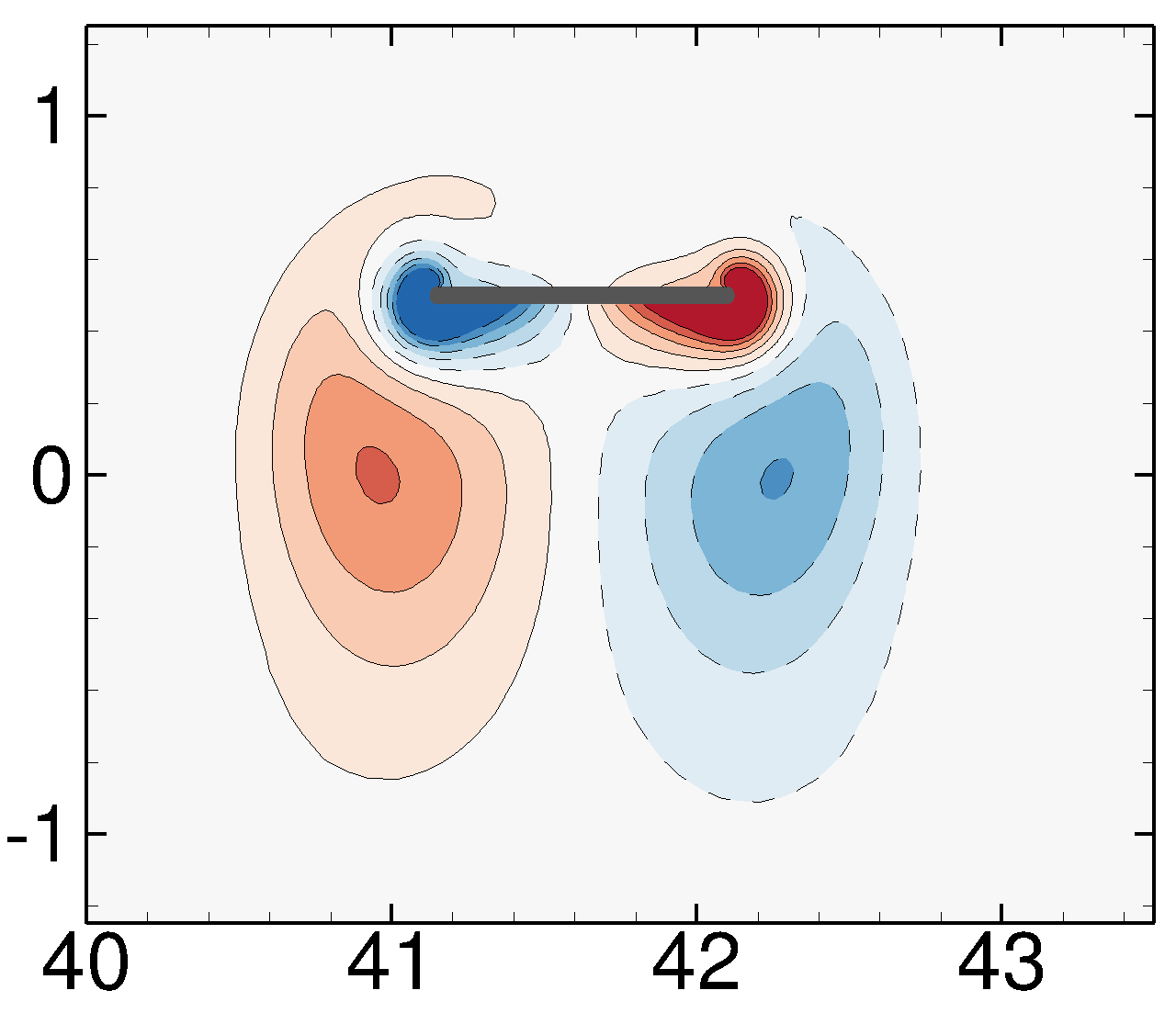}
    \caption{$t_0+1/2$}
    \end{subfigure}
    \begin{subfigure}{0.24\textwidth}
    \includegraphics[width=\linewidth]{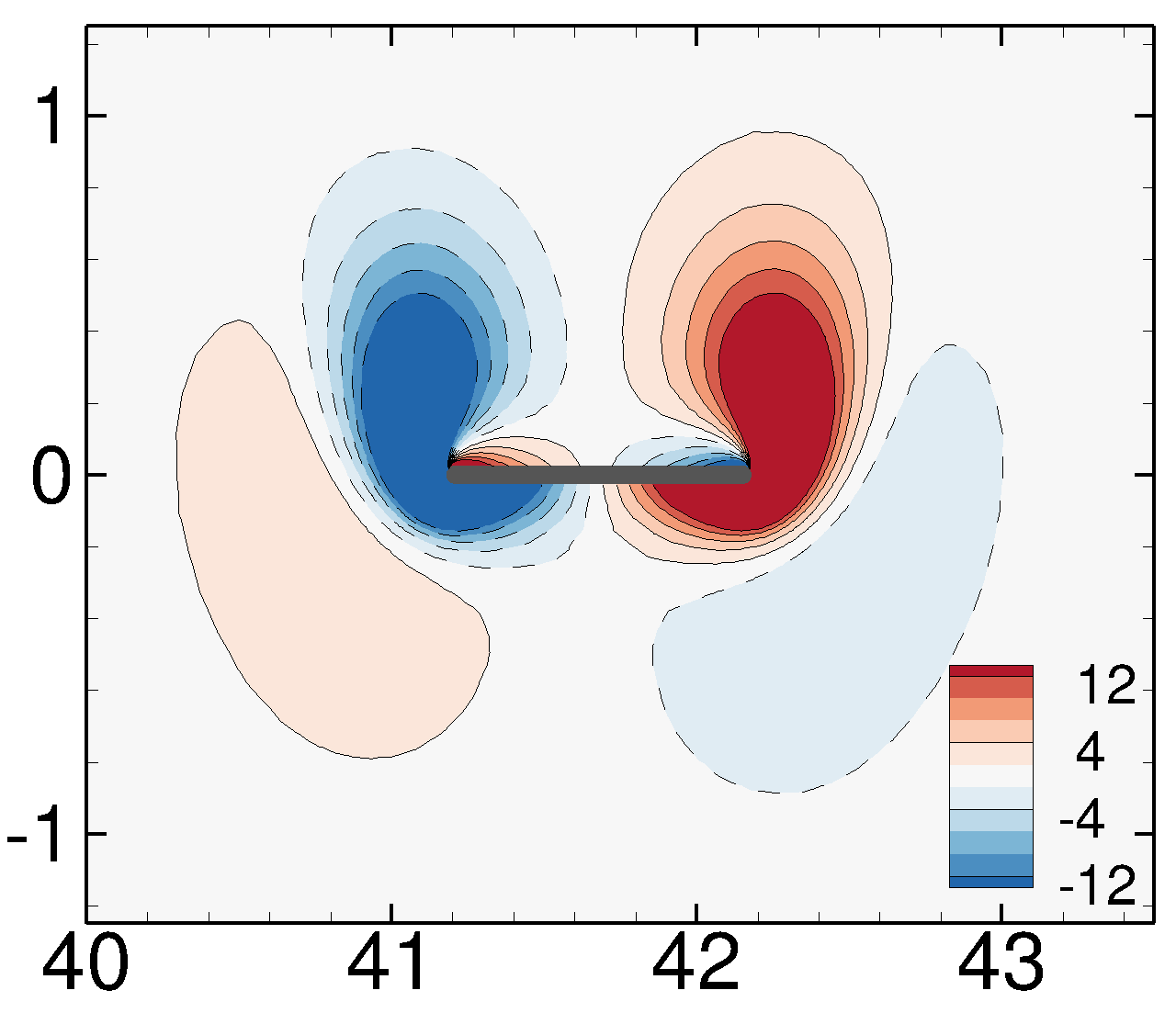}
    \caption{$t_0+3/4$}
    \end{subfigure}
    \begin{subfigure}{0.45\textwidth}
       \includegraphics[width=\linewidth]{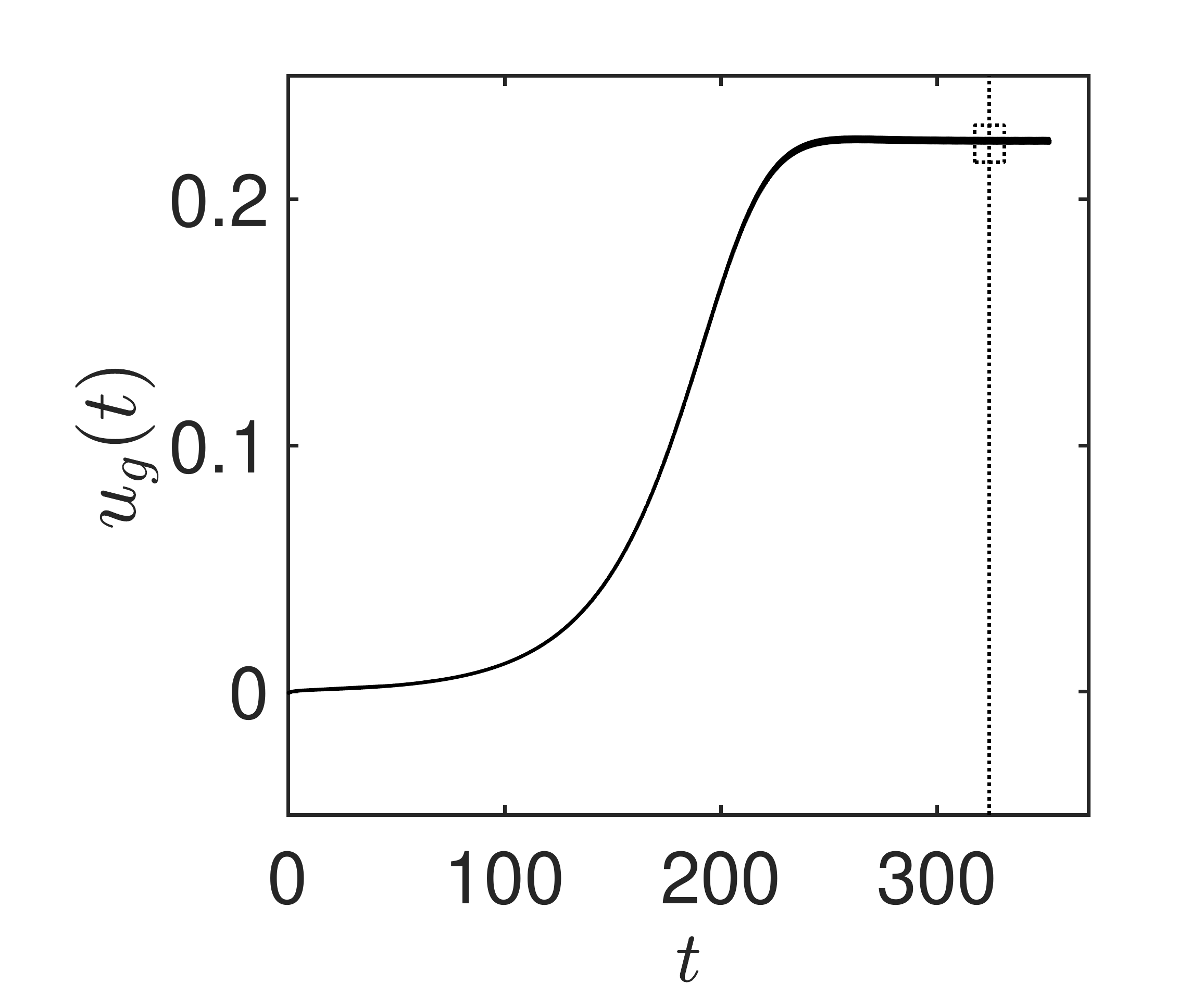}
    \caption{}
    \end{subfigure}
    \begin{subfigure}{0.45\textwidth}
        \includegraphics[width=\linewidth]{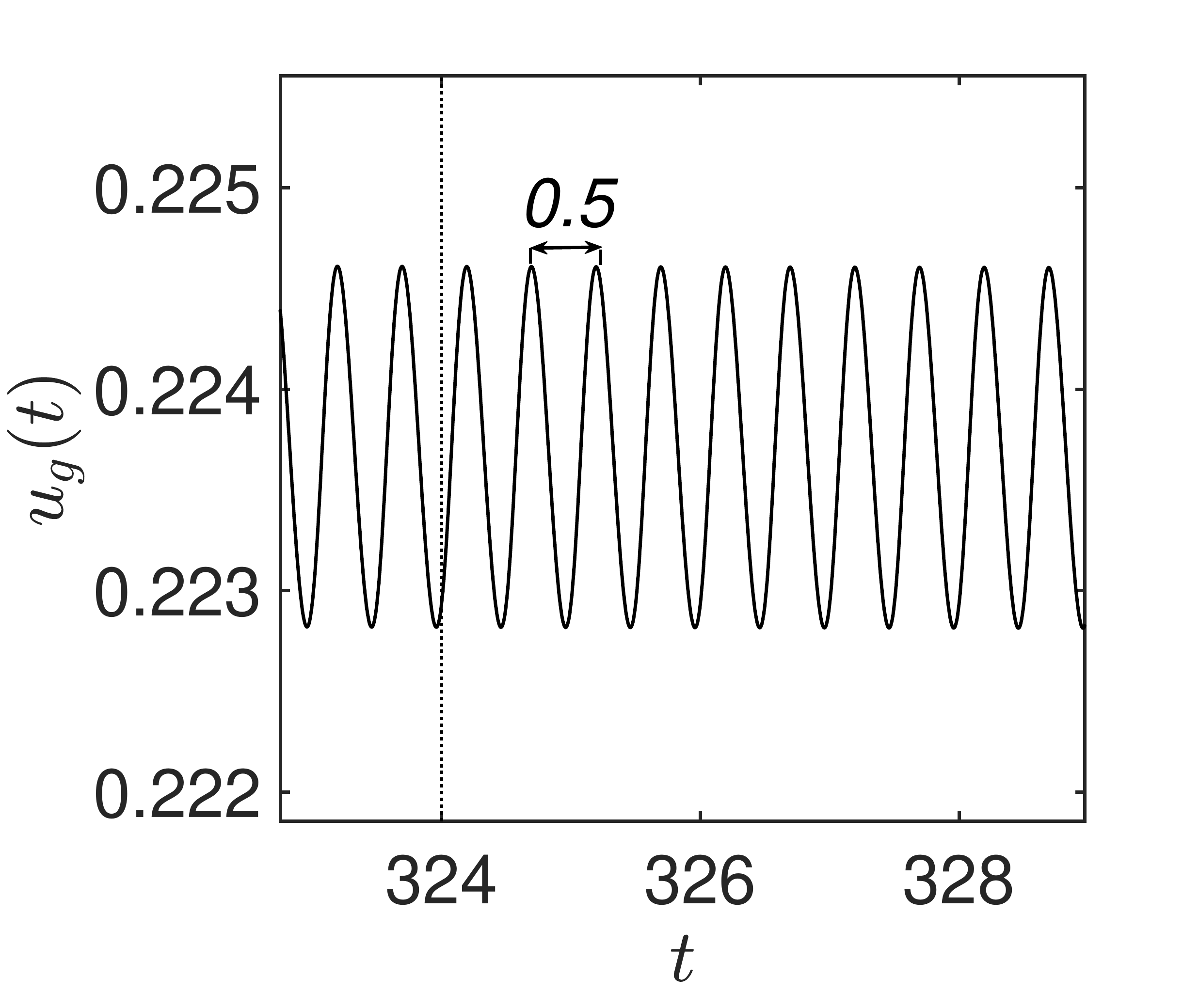}
    \caption{}
    \end{subfigure}

    \caption{Unidirectional propulsive solution for $\beta=6$: (a-d) Vorticity flow field along a flapping period. Time evolution of  the horizontal velocity $u_g$ (initially equal to zero) (e) over the whole simulated time and (f) restricted to time window indicated by the rectangle in (e). The instant $t_0$ is depicted with a vertical line in (e) and (f). The period of the horizontal velocity is $0.5$, half the vertical flapping period $1$.}
            \label{2-NonLinear-Prop}
\end{figure}

\begin{figure}
\vspace{0.5cm}
   \centering
    \begin{tabular}{ll}
       (a) & (b) \\
       \includegraphics[width=0.45\linewidth]{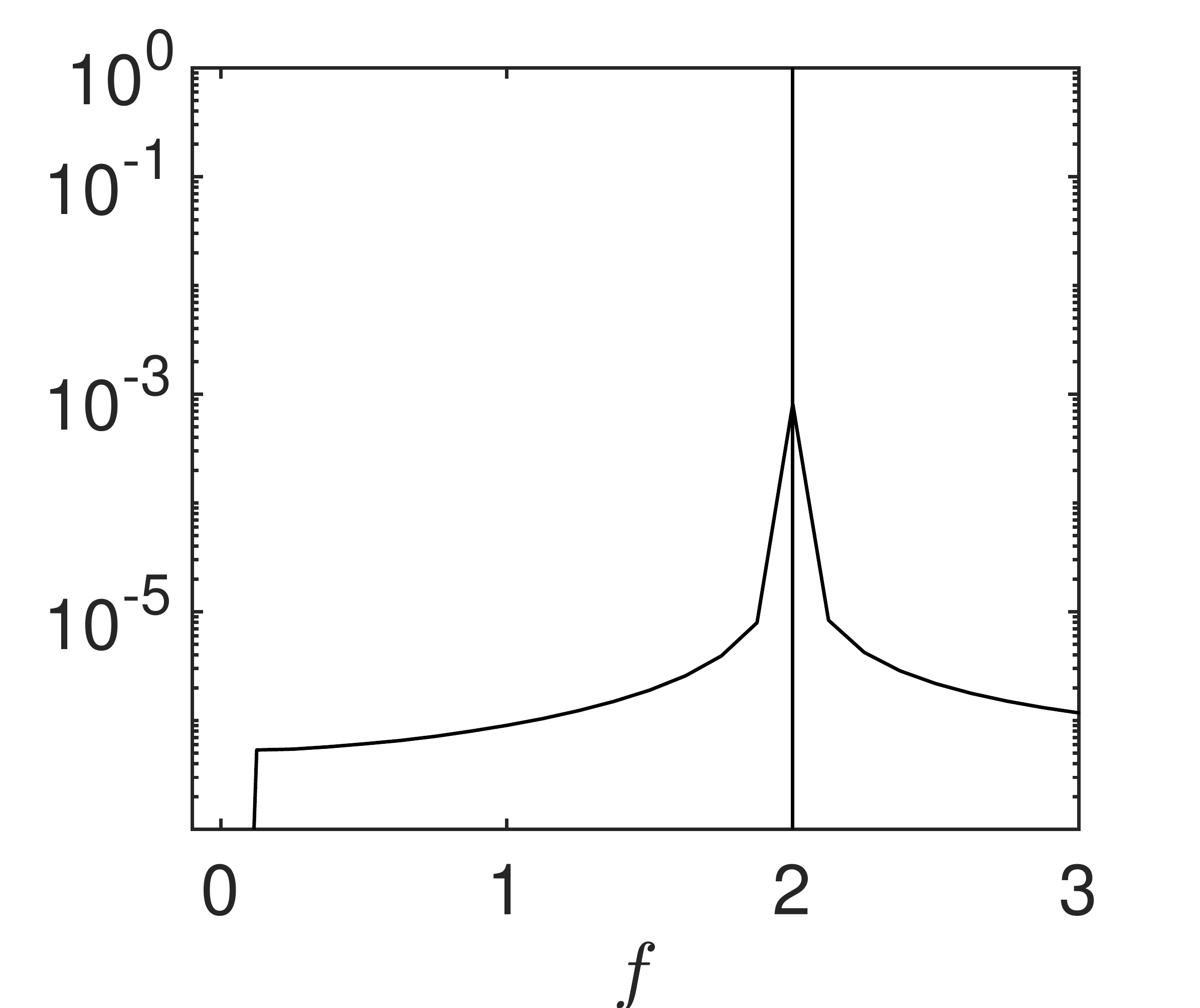}
         & \includegraphics[width=0.45\linewidth]{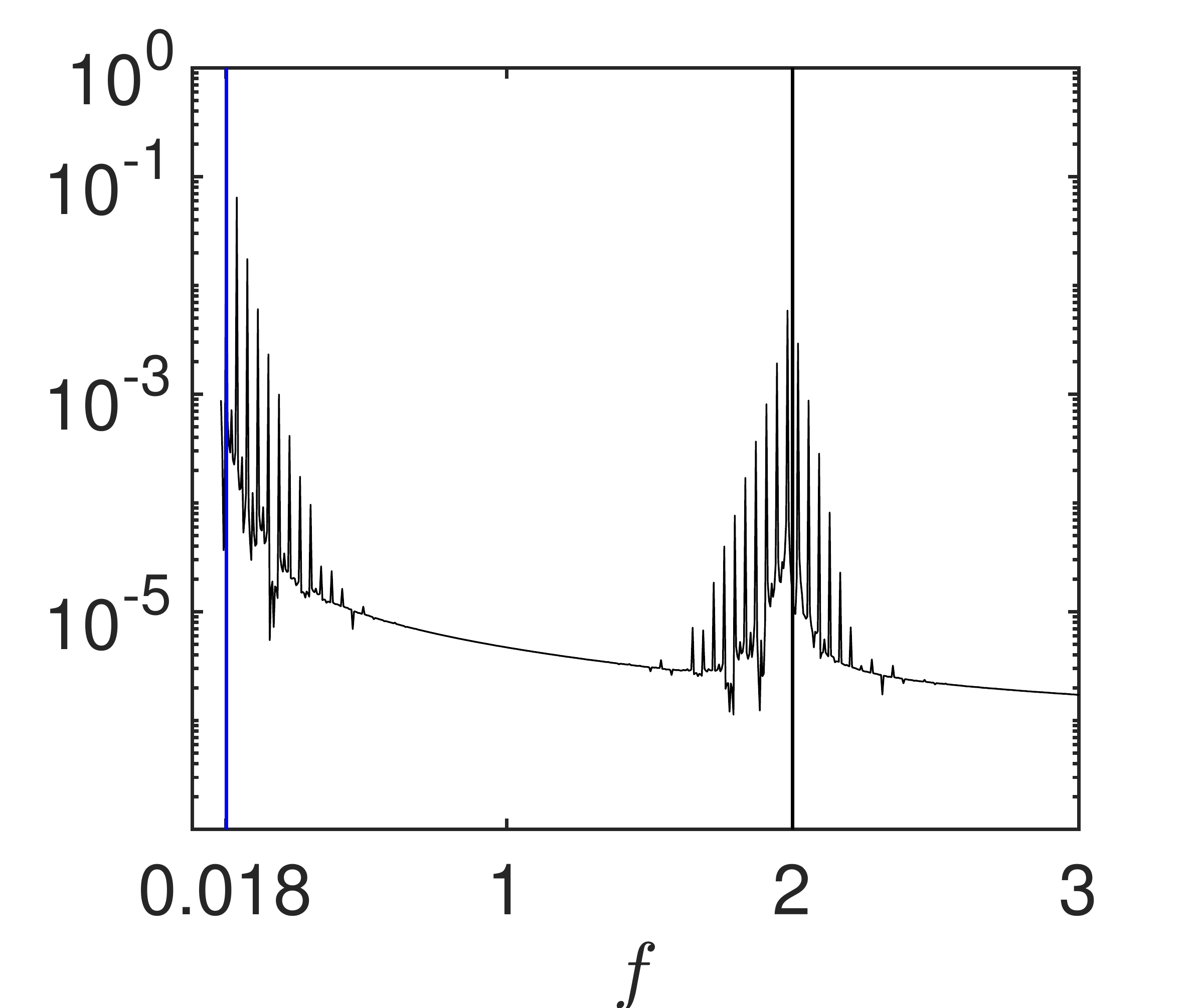}
    \end{tabular}
    \caption{Frequency spectra of the horizontal foil speed for (a) the periodic unidirectional propulsive solution ($\beta=6$) and (b) the quasi-periodic back \& forth propulsive solution ($\beta=13$). The fundamental frequency of the horizontal speed (vertical black lines)  is twice the vertical flapping frequency. The low frequency of the quasi-periodic solution (figure \ref{2-NonLinear-BandF}-e) is identified by the blue line.}
            \label{Fig:Fourier-NonLinear}
\end{figure}

As the Stokes number (dimensional frequency) is increased, the flow breaks the spatial symmetry \eqref{spatialsymmetry}, as seen in figure \ref{2-NonLinear-Prop}(a-d) for $\beta=6$. Vortices shed on each side of the foil are of slightly different shape and intensity for all time instants. This asymmetric flow then induces an instantaneous horizontal force accelerating the foil. Figure \ref{2-NonLinear-Prop}(e) shows that an initial small perturbation of the horizontal velocity $u_g$ grows exponentially in time, before saturating for $t>200$ towards a periodic state, as shown in the close-up view displayed in figure \ref{2-NonLinear-Prop}(f). The amplitude of oscillation of the horizontal velocity is very weak compared to its time-averaged value, denoted hereinafter  $<u_g>$. Being positive, the flapping foil self-propels 
in the positive $x$-direction. Solutions self-propelling in the negative $x$-direction can also be found by modifying the initial horizontal velocity. The effect of the initial condition on the symmetry breaking direction was investigated in \citet{Jallas2017}. The Fourier spectrum of the foil horizontal velocity, displayed in figure \ref{Fig:Fourier-NonLinear}(a), shows that it oscillates at the (non-dimensional) frequency $f=2$, i.e. twice the (non-dimensional) flapping frequency equal to $f=1$ independently from the Stokes number. This doubling-frequency of the horizontal velocity is related to the spatio-temporal flow symmetry (Eq. \ref{spatiotemporalsymmetry}). Over one flapping period, the horizontal force acting on the foil is identical during upward and downward strokes.

% \subsubsection{Back \& Forth propulsive solutions}
% \label{Subsec:BF}

\smallskip
\smallskip

\begin{figure}
   \centering
      \begin{subfigure}{0.24\textwidth}
       \includegraphics[width=\linewidth]{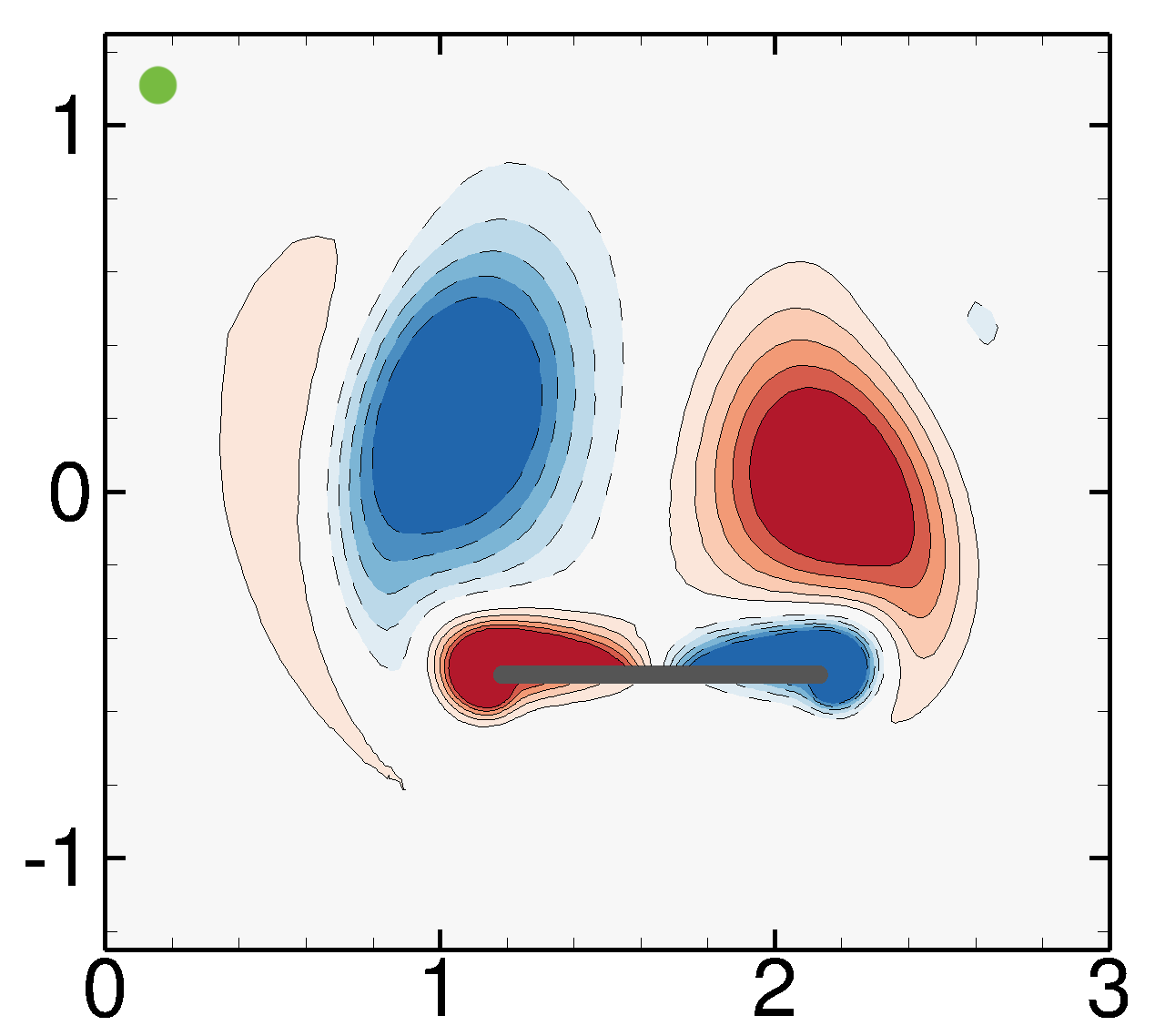}
    \caption{$t_0=178$}
    \end{subfigure}
       \begin{subfigure}{0.24\textwidth}
       \includegraphics[width=\linewidth]{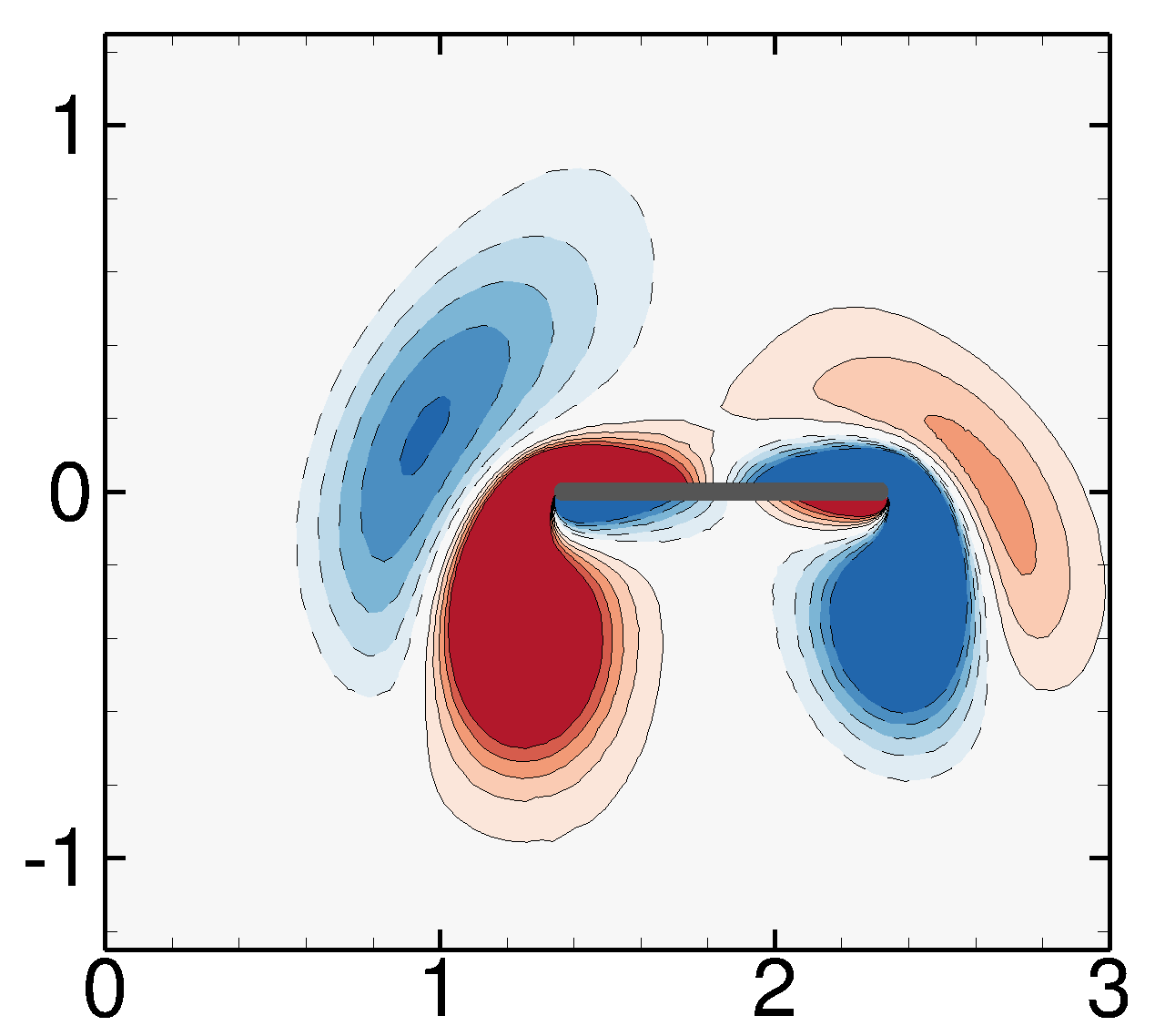}
    \caption{$t_0+1/4$}
    \end{subfigure}
          \begin{subfigure}{0.24\textwidth}
       \includegraphics[width=\linewidth]{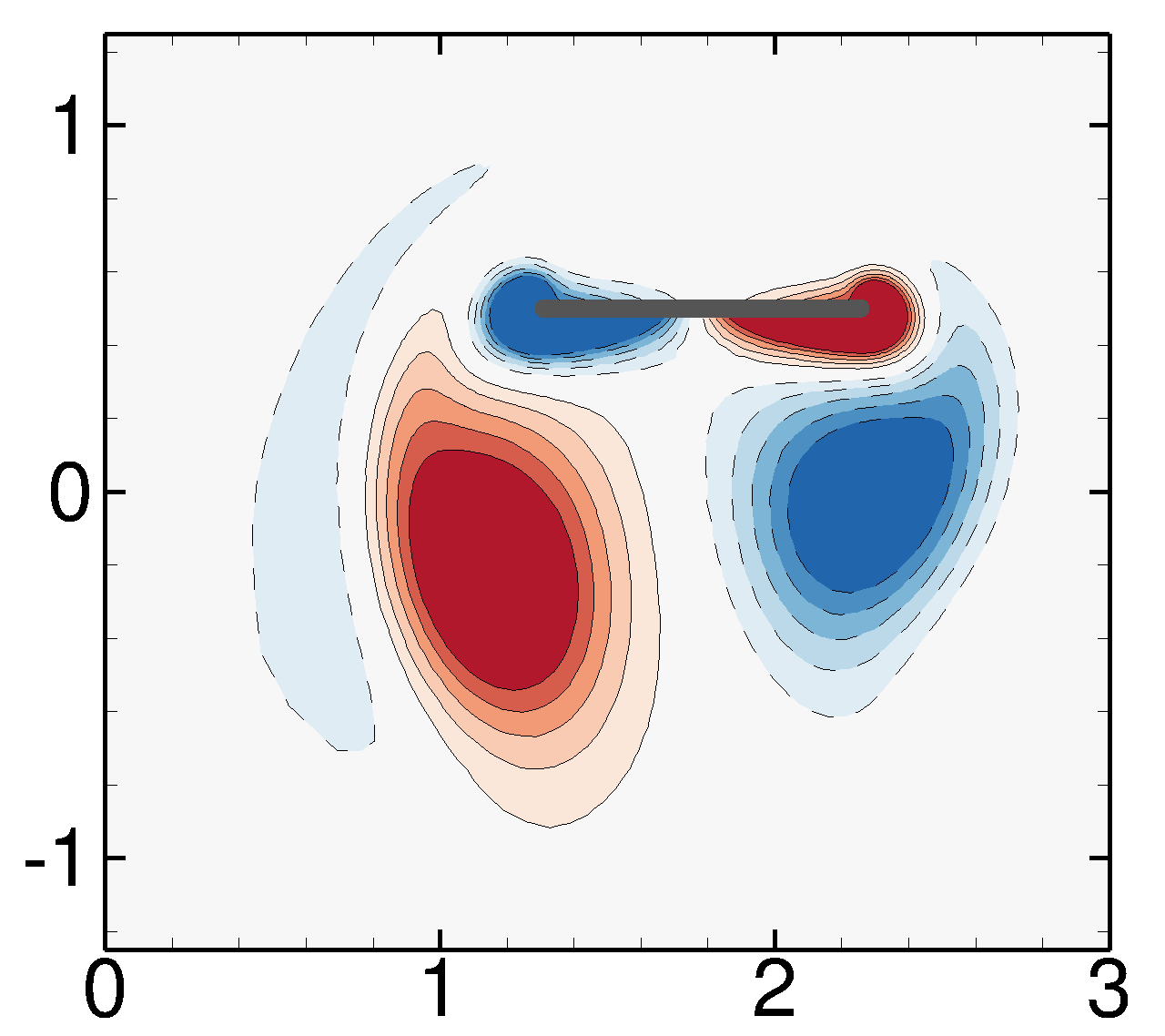}
    \caption{$t_0+1/2$}
    \end{subfigure}
       \begin{subfigure}{0.24\textwidth}
       \includegraphics[width=\linewidth]{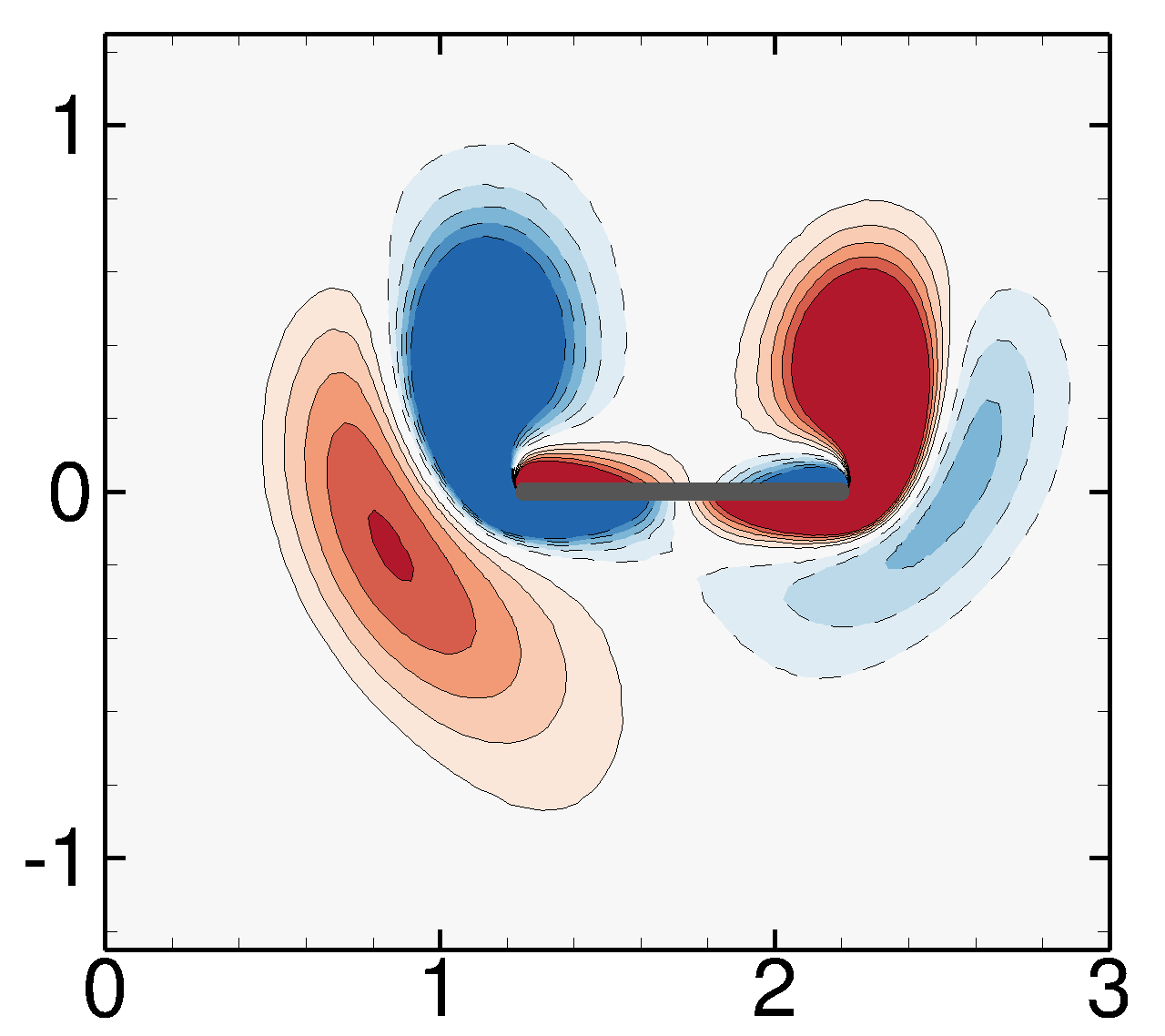}
    \caption{$t_0+3/4$}
    \end{subfigure}
    \begin{subfigure}{0.48\textwidth}
       \includegraphics[width=\linewidth]{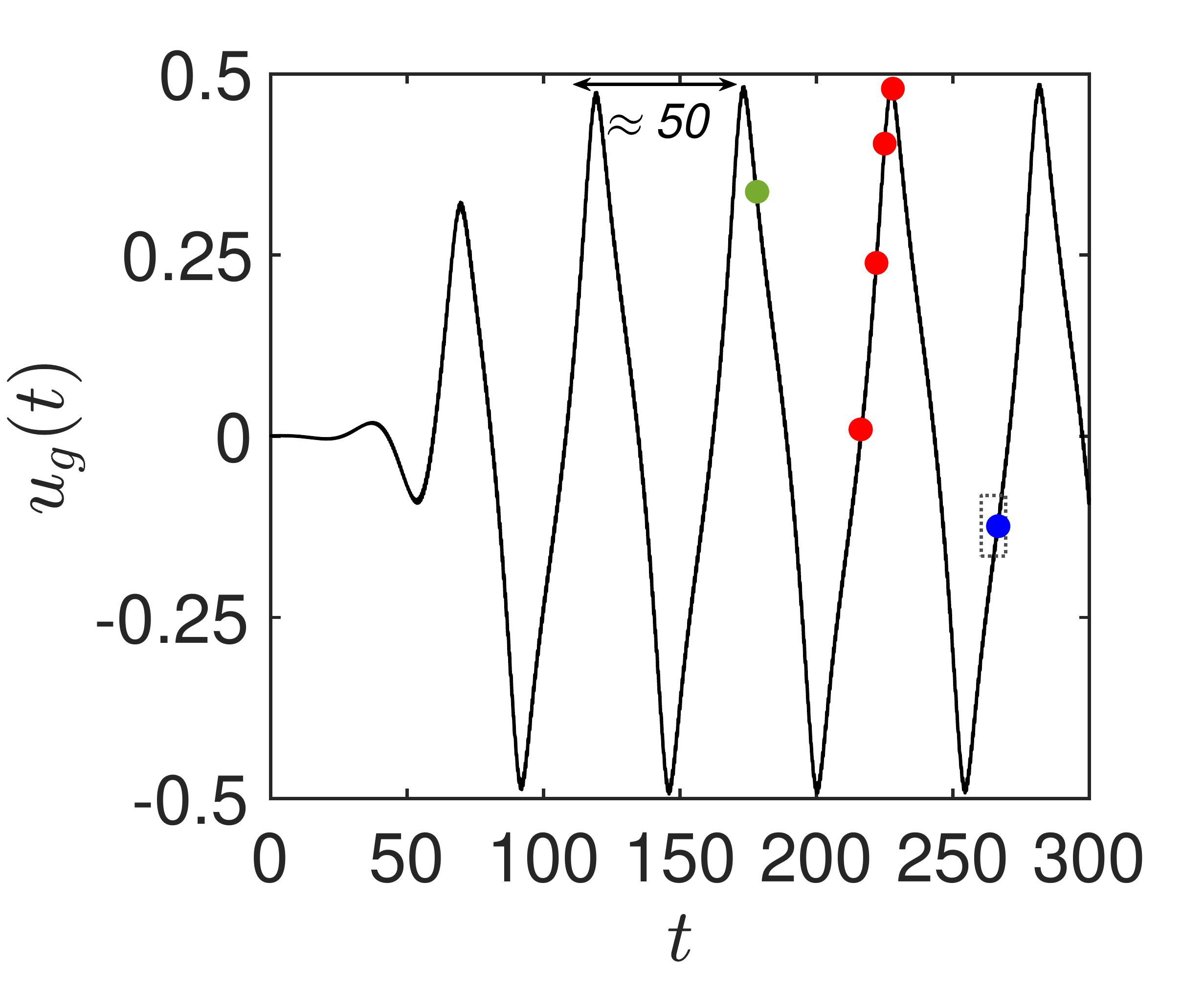}
    \caption{}
    \end{subfigure}
    \begin{subfigure}{0.48\textwidth}
        \includegraphics[width=\linewidth]{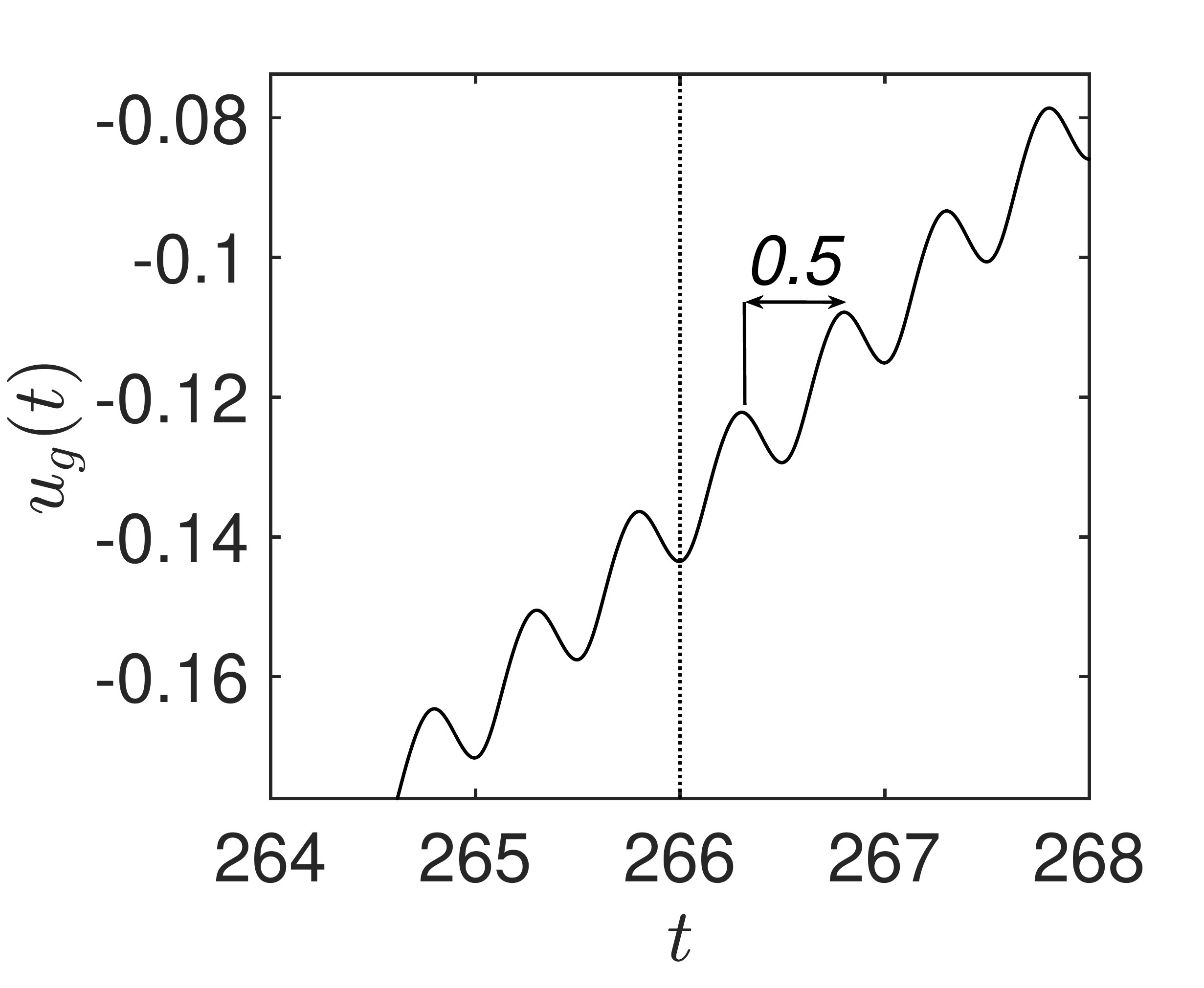}
    \caption{}
    \end{subfigure}
           \begin{subfigure}{0.24\textwidth}
       \includegraphics[width=\linewidth]{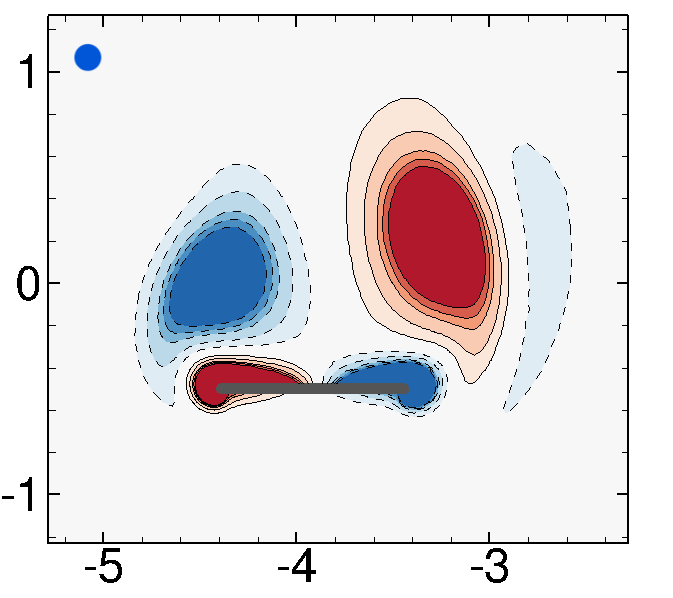}
    \caption{$t_1=266$}
    \end{subfigure}
    \begin{subfigure}{0.24\textwidth}
       \includegraphics[width=\linewidth]{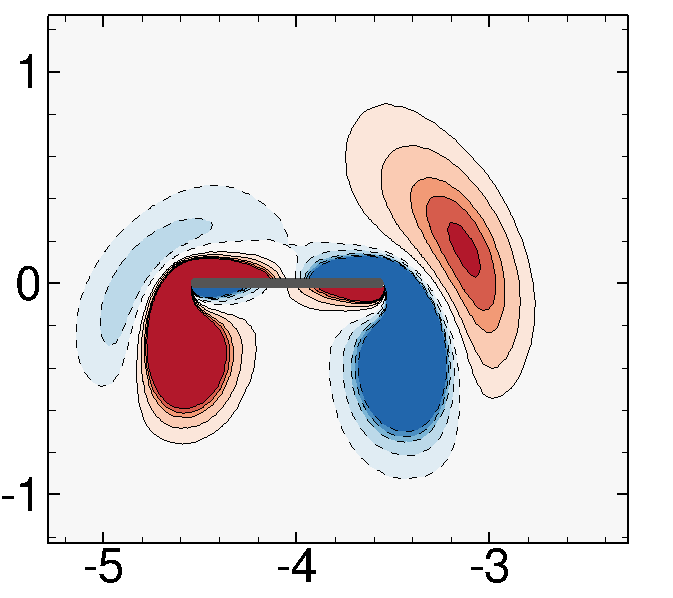}
    \caption{$t_1+1/4$}
    \end{subfigure}
           \begin{subfigure}{0.24\textwidth}
       \includegraphics[width=\linewidth]{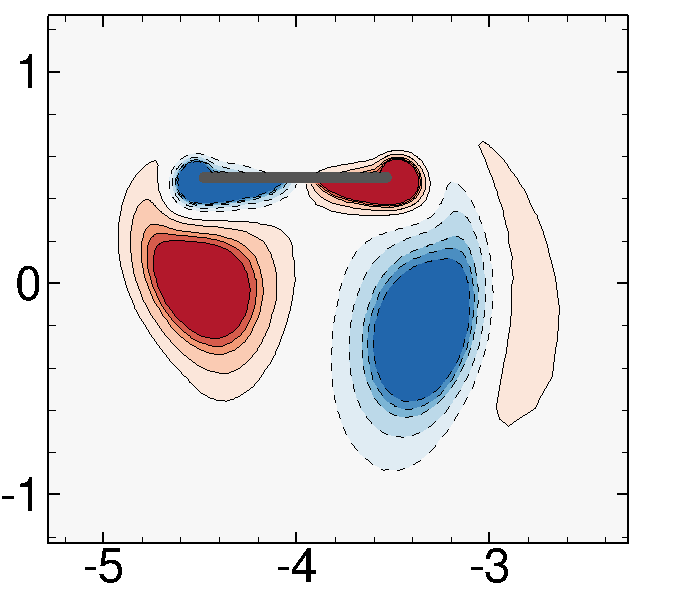}
    \caption{$t_1+1/2$}
    \end{subfigure}
    \begin{subfigure}{0.24\textwidth}
       \includegraphics[width=\linewidth]{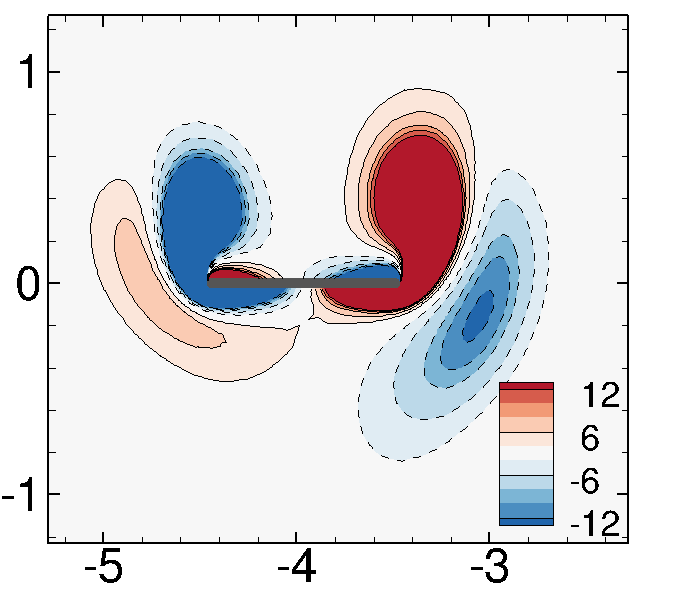}
    \caption{$t_1+3/4$}
    \end{subfigure}
    \caption{Back \& Forth solution for $\beta=13$: (a-d) and (g-j) Vorticity contours for a flapping period -- starting from different time instants in (a) and (g). (e-f) Horizontal speed $u_g$ time evolution with dotted rectangle close-up in (f). Instants (a) and (g) are indicated by filled green and blue dots in (e). Long and small periods of $u_g(t)$ are respectively indicated in (e) and (f).}
            \label{2-NonLinear-BandF}
\end{figure}

\begin{figure}
\vspace{0.5cm}
   \centering
    \begin{tabular}{llll}
    (a) & (b) & (c) & (d) \\
       \includegraphics[width=0.24\linewidth]{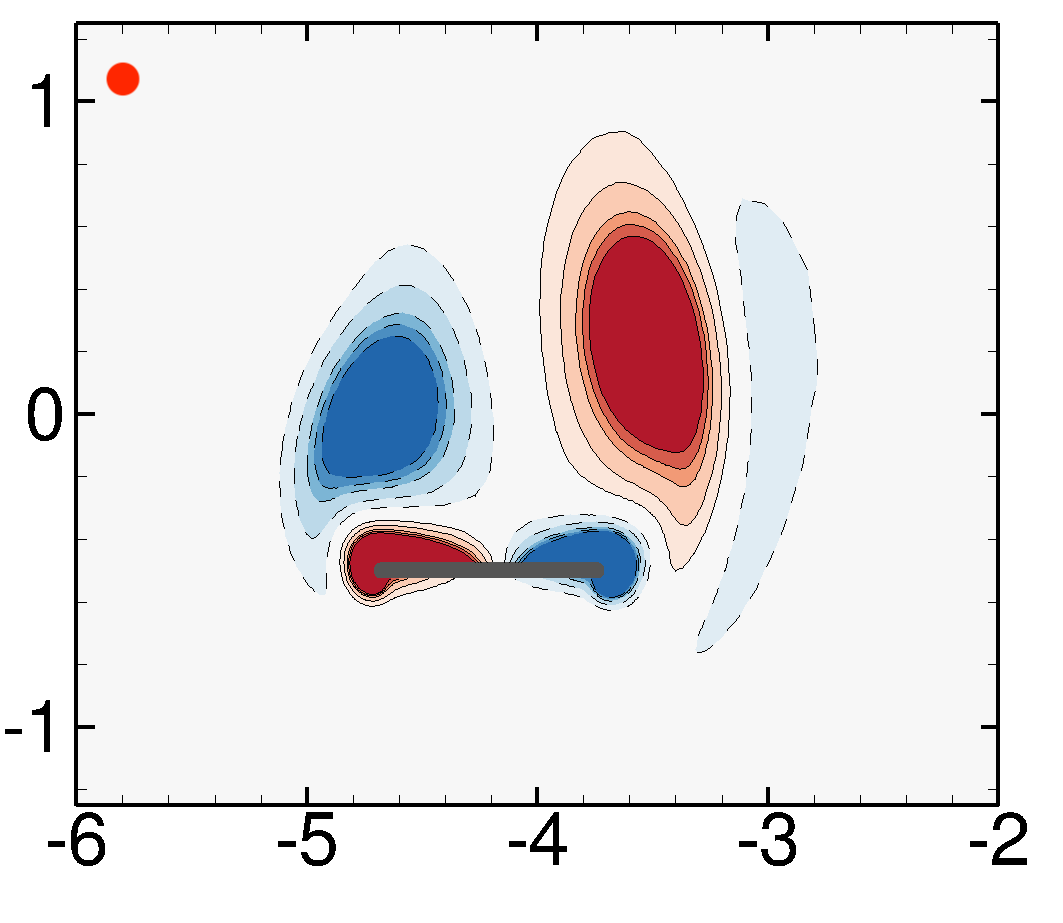}
       & \includegraphics[width=0.24\linewidth]{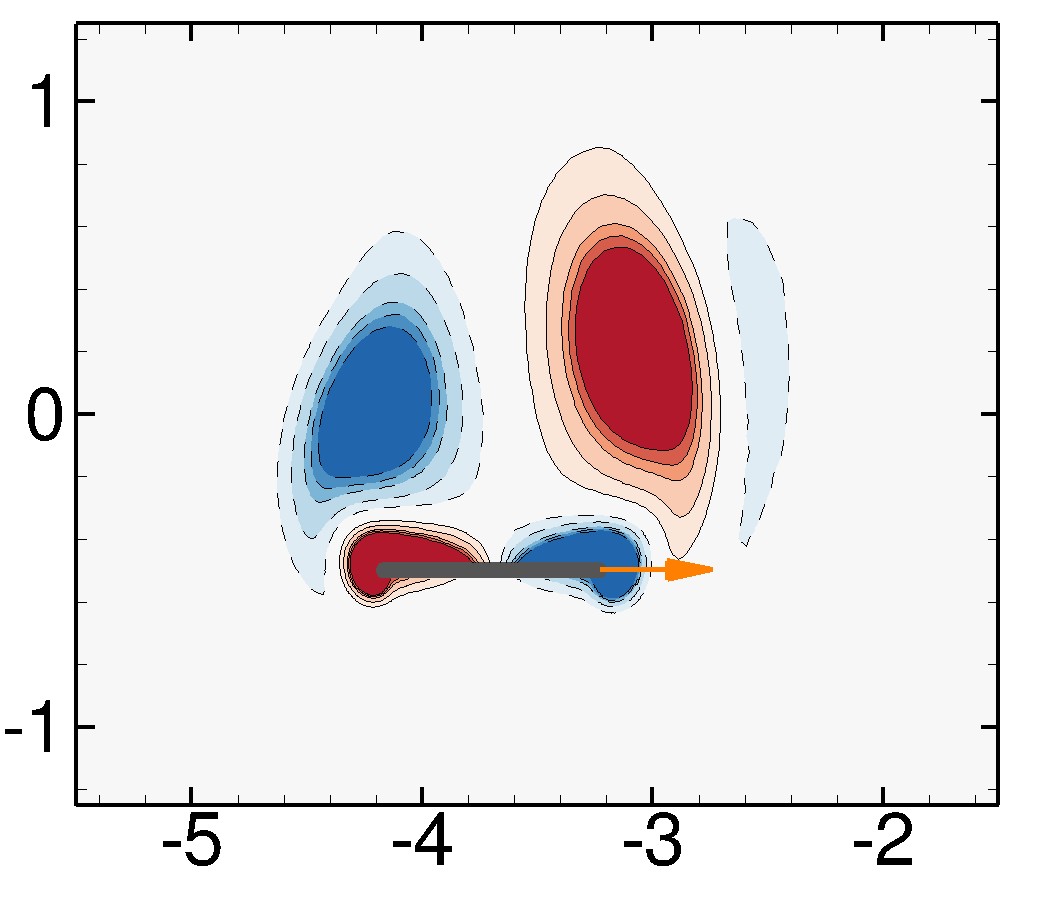}
       & \includegraphics[width=0.24\linewidth]{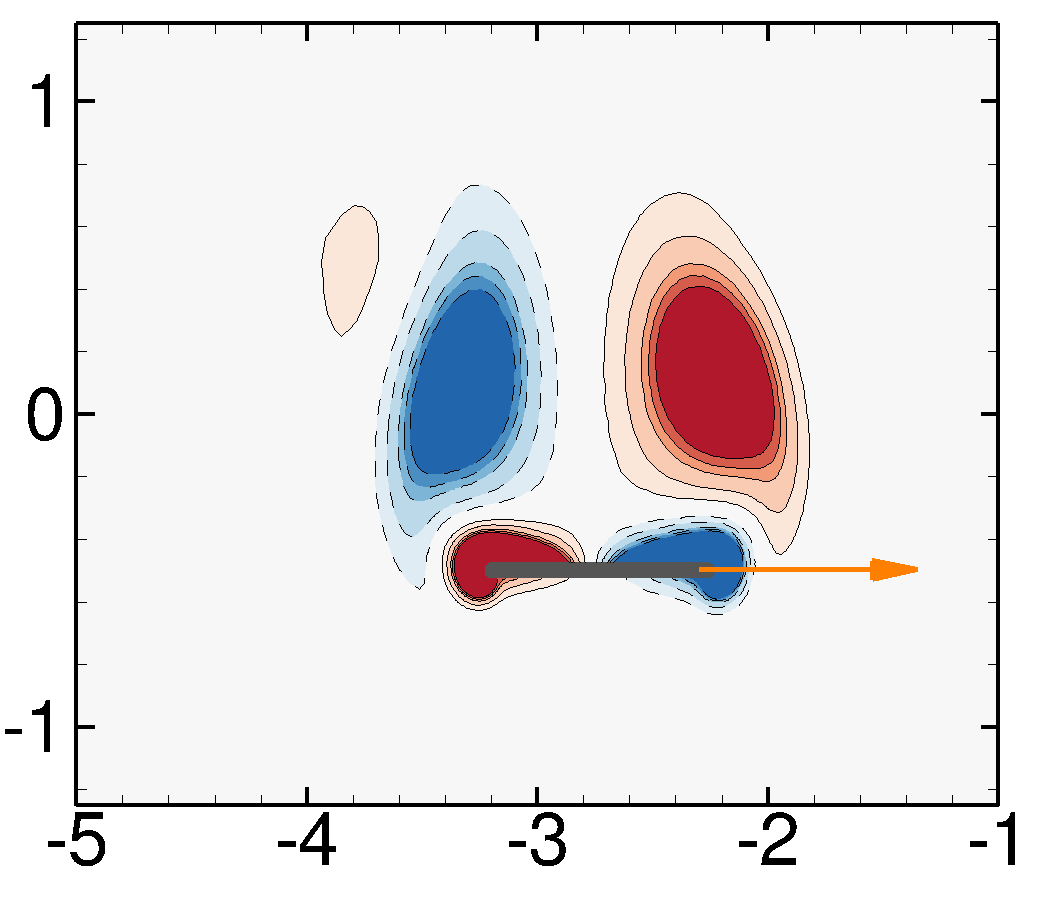}
       & \includegraphics[width=0.24\linewidth]{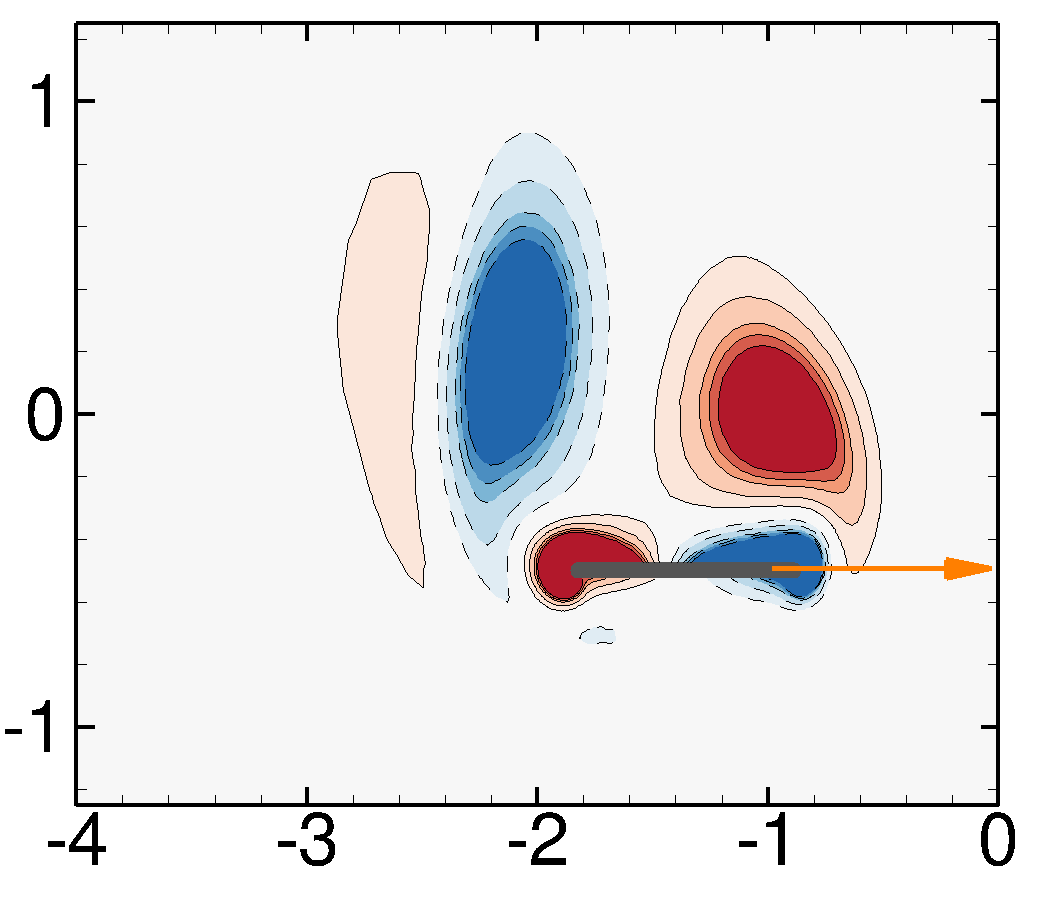}
    \end{tabular}
    \caption{Back \& Forth solution for $\beta=13$: (a-d) Vorticity contours of the time instants represented in figure \ref{2-NonLinear-BandF}(e) by filled red dots. An orange arrow indicates the horizontal velocity of these time instants. These time instants are equally spaced of 3 flapping periods. }
            \label{Fig:BF-Stroboscopic}
\end{figure}

For higher values of the Stokes number, the spatial flow symmetry is still broken but the propulsion is no longer unidirectional. The foil periodically reverses its propulsive direction, as if a restoring force was at play. A typical solution obtained for $\beta = 13$ is displayed in figure \ref{2-NonLinear-BandF}. The time evolution of the horizontal velocity shown in \ref{2-NonLinear-BandF}(e) clearly indicates that, after an initial exponential growth in time, the foil velocity slowly oscillates between positive and negative values, over a period about $50$ times larger than the flapping period. This solution is no longer periodic, but quasi-periodic, as clearly shown by the Fourier spectrum 
of the foil horizontal velocity displayed in figure \ref{Fig:Fourier-NonLinear}(b). Two fundamental frequencies are obtained, one at $f=2$ corresponding to twice the flapping frequency, and one around $f=0.018$ corresponding to the slow period. The multiple peaks observed around each fundamental frequency are induced by non-linear interactions. Coming back to the horizontal velocity of the foil, its time-average over the long period is zero. Thus, this solution is not a coherent (unidirectional) propulsive state \citep{Alben2005}. The foil oscillates back \& forth around a fixed point in space. Nevertheless, the horizontal velocity, time-averaged  along the (short) flapping period, is either positive or negative, as seen in figure \ref{2-NonLinear-BandF}(f). A propulsive effect is thus obtained at this time-scale. The instantaneous vorticity fields displayed in figures \ref{2-NonLinear-BandF} (a-d) correspond to a flapping period (marked with the green dot in figure \ref{2-NonLinear-BandF}-e) where the velocity of the foil is positive, while those shown in figure \ref{2-NonLinear-BandF}(g-j) correspond to a negative foil velocity (blue dot). In both cases, the leading-edge vortex is of smaller size and closer to the foil than the trailing-edge vortex. Interestingly, such vortex pattern is not observed for all phases of the long period, and in particular in the acceleration phases, marked with red dots in figure \ref{2-NonLinear-BandF}(e). The corresponding instantaneous vorticity fields are depicted in figure \ref{Fig:BF-Stroboscopic}. In between the instants corresponding to figure \ref{Fig:BF-Stroboscopic}(a) and (b), the foil accelerates and self-propels in the right direction but the leading-edge vortex (right) is now of larger size and further away from the foil, compared to the trailing-edge vortex (left). This suggests that the foil motion is induced by a suction of the leading-edge vortex. As the foil further accelerates, the leading and trailing-edge vortices are progressively convected downstream until the more classical propulsive pattern is recovered when the foil reaches its maximal velocity (see figure \ref{Fig:BF-Stroboscopic}-d).

\begin{figure}
   \begin{center}
   \begin{tabular}{ll}
   (a) & (b) \\
      \includegraphics[width=0.45 \linewidth]{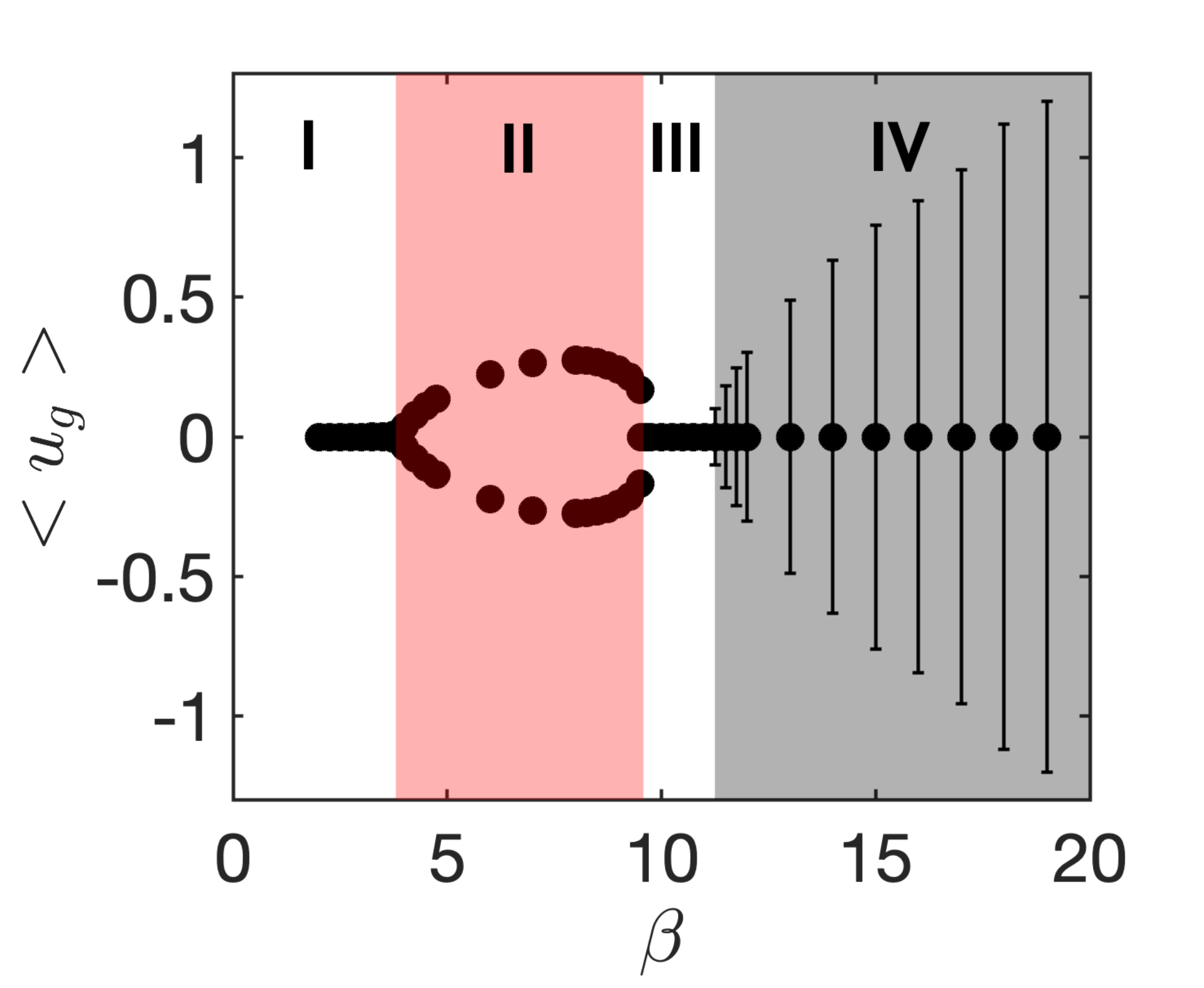}
    &  \includegraphics[width=0.45 \linewidth]{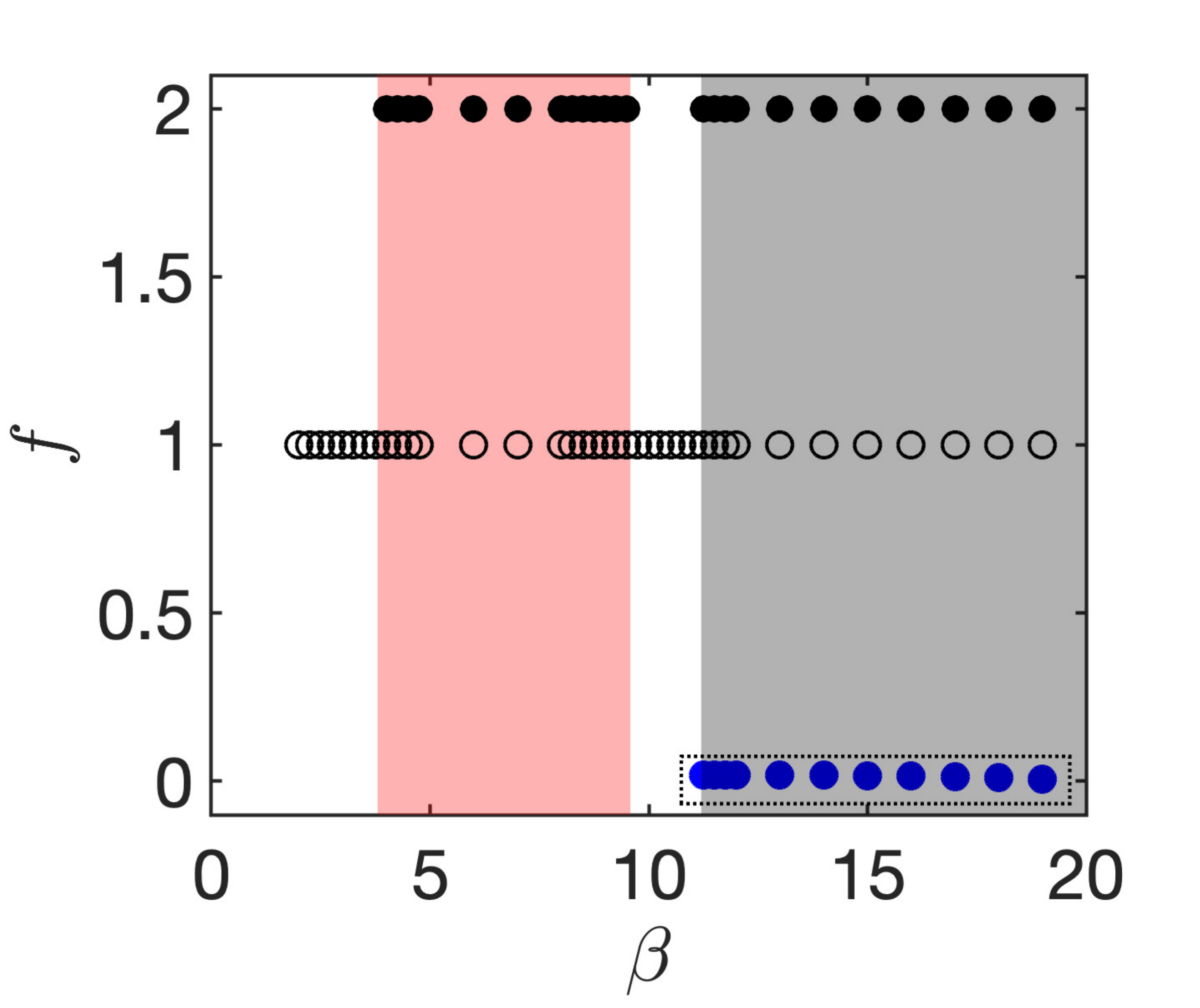} \\
   (c) & (d) \\
      \includegraphics[width=0.45 \linewidth]{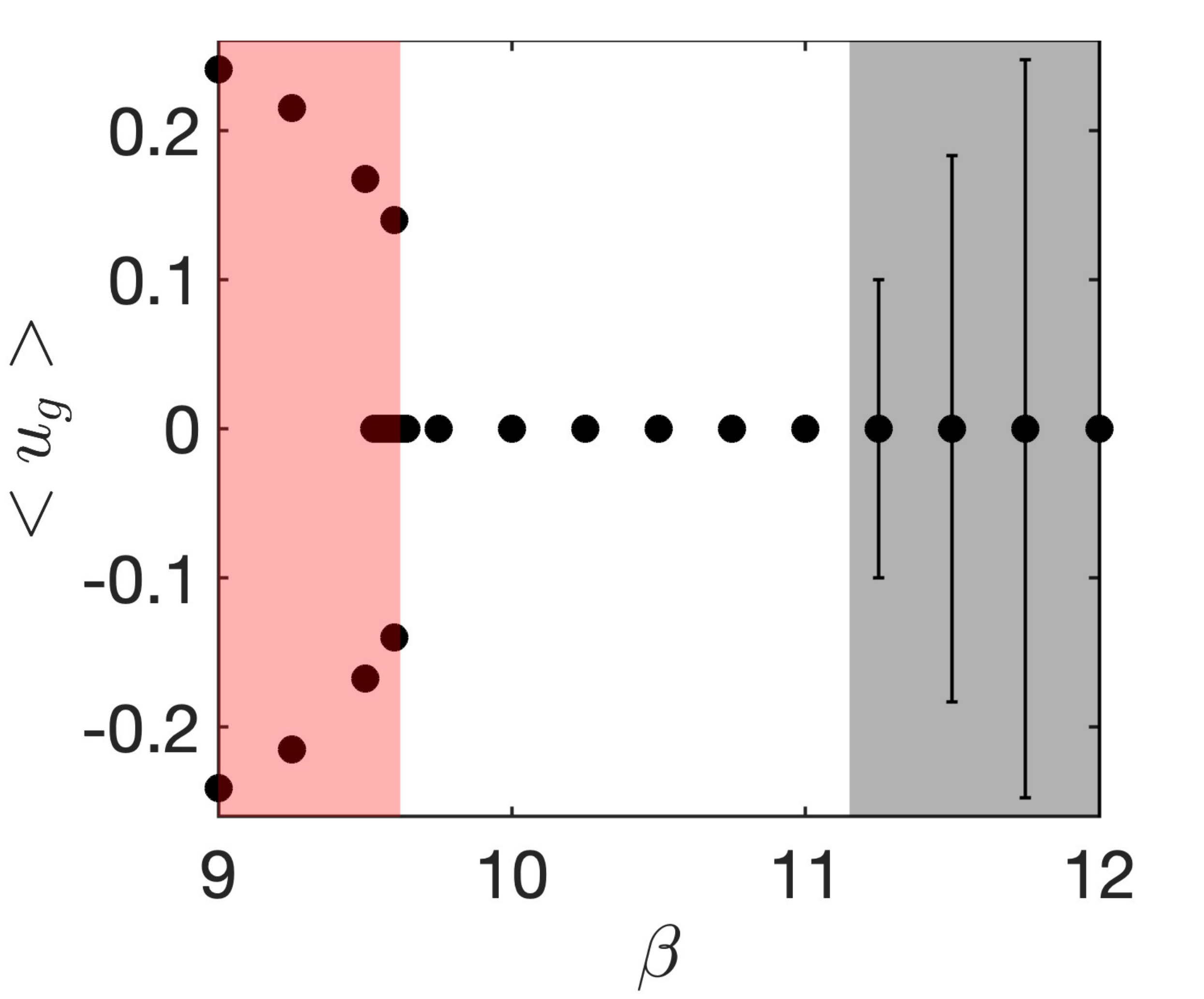}
    &  \includegraphics[width=0.45 \linewidth]{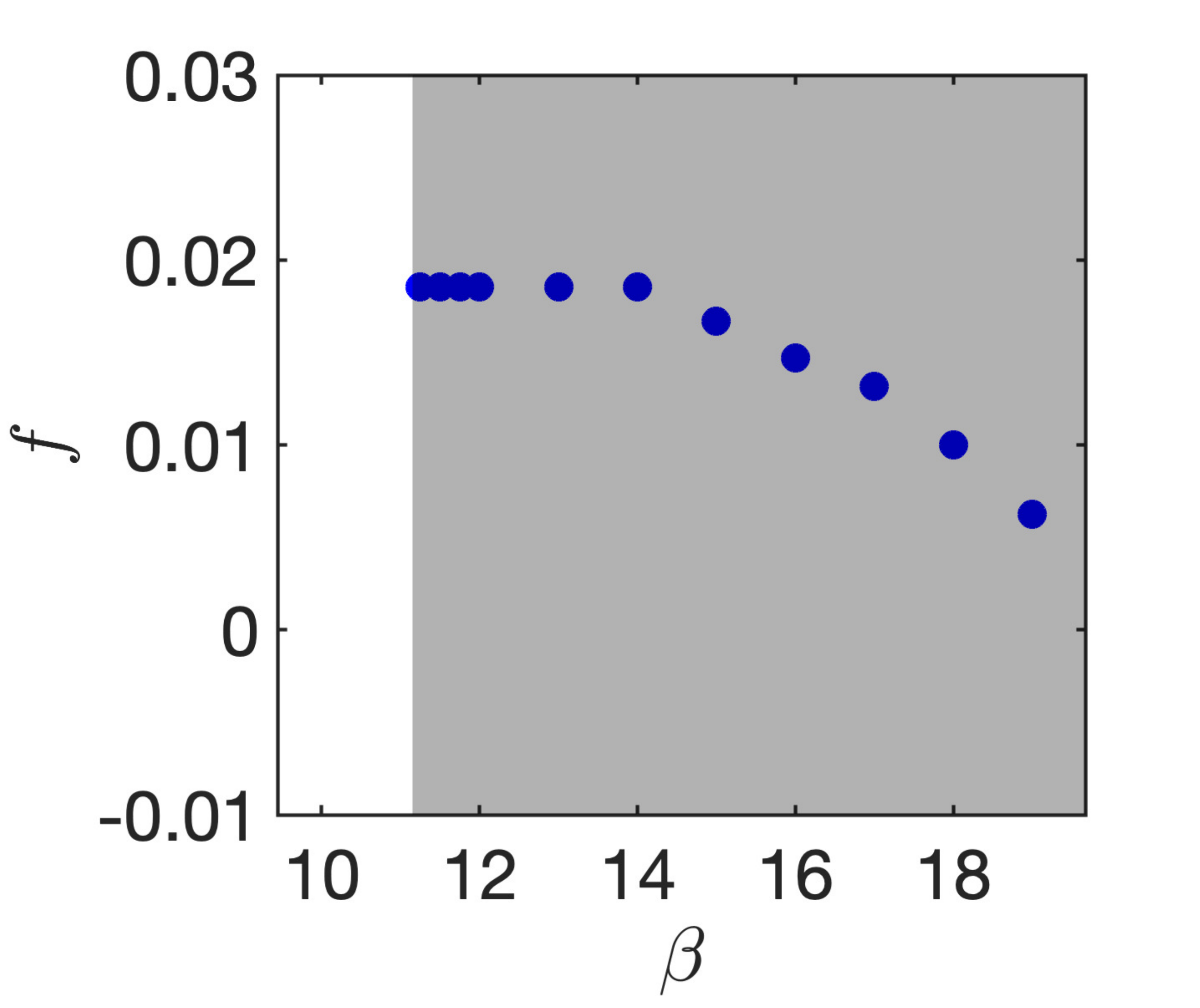} 
   \end{tabular}
    \end{center}
    \caption{Non-propulsive (white), unidirectional propulsive (red) and back \& forth (grey) regimes as a function of the Stokes number $\beta$. (a,c) Time-averaged (circles) and oscillation amplitude (error bars) of the foil horizontal velocity. (b,d) Forcing frequency of the foil vertical velocity (open circles) and frequency of the horizontal foil velocity (filled circles). (c) is a close-up view of (a) highlighting the transition between unidirectional propulsive, non-propulsive and back \& forth solutions. (d) is a close-up view of (b) showing the evolution of the low-frequency in the back \& forth solution as a function of the Stokes number. Parameters: $A = 0.5$ and $\rho = 100$.}
    \label{2-NonLinear-MeanU}
\end{figure}

% \subsubsection{Regimes of self-propelled solutions when varying the Stokes number}

\smallskip
\smallskip

The existence and characterization of the three propulsive solutions is displayed in figure \ref{2-NonLinear-MeanU} 
as a function of the Stokes number in the range $ 2\le \beta \le 20 $. The regimes of non-propulsive, propulsive and back \& forth solutions are identified with white, red and grey background colors, respectively. The horizontal velocity is depicted in figure \ref{2-NonLinear-MeanU}(a) and (c), with black dots for the time-averaged value $<u_g>$ and vertical bars for the fluctuation amplitude. For the back \& forth solution, the long period is used for time-averaging. The frequencies $f$ of the foil velocities are shown in figure \ref{2-NonLinear-MeanU}(b) and (d), the open circles denoting the vertical flapping frequency, while the filled circles correspond to the frequency of the horizontal velocity. Symmetric non-propulsive solutions exist for small Stokes numbers $\beta<4$ (region I) and for intermediate values in the range $9.53 \leq \beta \leq 11.25$ (region III). For these Stokes numbers, no locomotion is achieved and the foil remain in its position, producing a spatially symmetric periodic flow. Propulsive solutions appear for $\beta=4$. They are characterized by non-zero time-averaged horizontal velocities --- both negative and positive depending on initial conditions --- with very small amplitudes of fluctuations ad a frequency of oscillation equal to $f=2$. As the Stokes number is increased, this frequency remains constant while the (absolute value of) time-averaged propulsive velocity continuously increases until $\beta \sim 8.5$. The mean velocity then decreases and abruptly (discontinuously) falls to zero for $\beta=9.58$. By decreasing again the Stokes number, we have identified a small range of Stokes number ($9.53 \le \beta \le 9.58$), visible in Figure \ref{2-NonLinear-MeanU}(c),  where non-propulsive and propulsive solutions co-exist. Therefore, the bifurcation from propulsive to non-propulsive solution is sub-critical around $\beta=9.5$, unlike the transition from non-propulsive to propulsive solution at $\beta=4$, which is super-critical. Finally, back \& forth solutions are observed when increasing the Stokes number $\beta \ge 11.25$ (region IV). They are characterized by zero time-averaged horizontal speed with large amplitude of fluctuations. These quasi-periodic solutions are characterized by two fundamental frequencies, the high frequency (black dots) and the low-frequency (blue dots). As seen in figure \ref{2-NonLinear-MeanU}(d), the low-frequency decreases towards zero when increasing the Stokes number.

\subsubsection{Self-propelled regimes for $\rho=1$}
\label{Subsec:Foil_Shape_Rho}

\begin{figure}
   \begin{center}
   \begin{tabular}{ll}
   (a) & (b) \\
      \includegraphics[width=0.45 \linewidth]{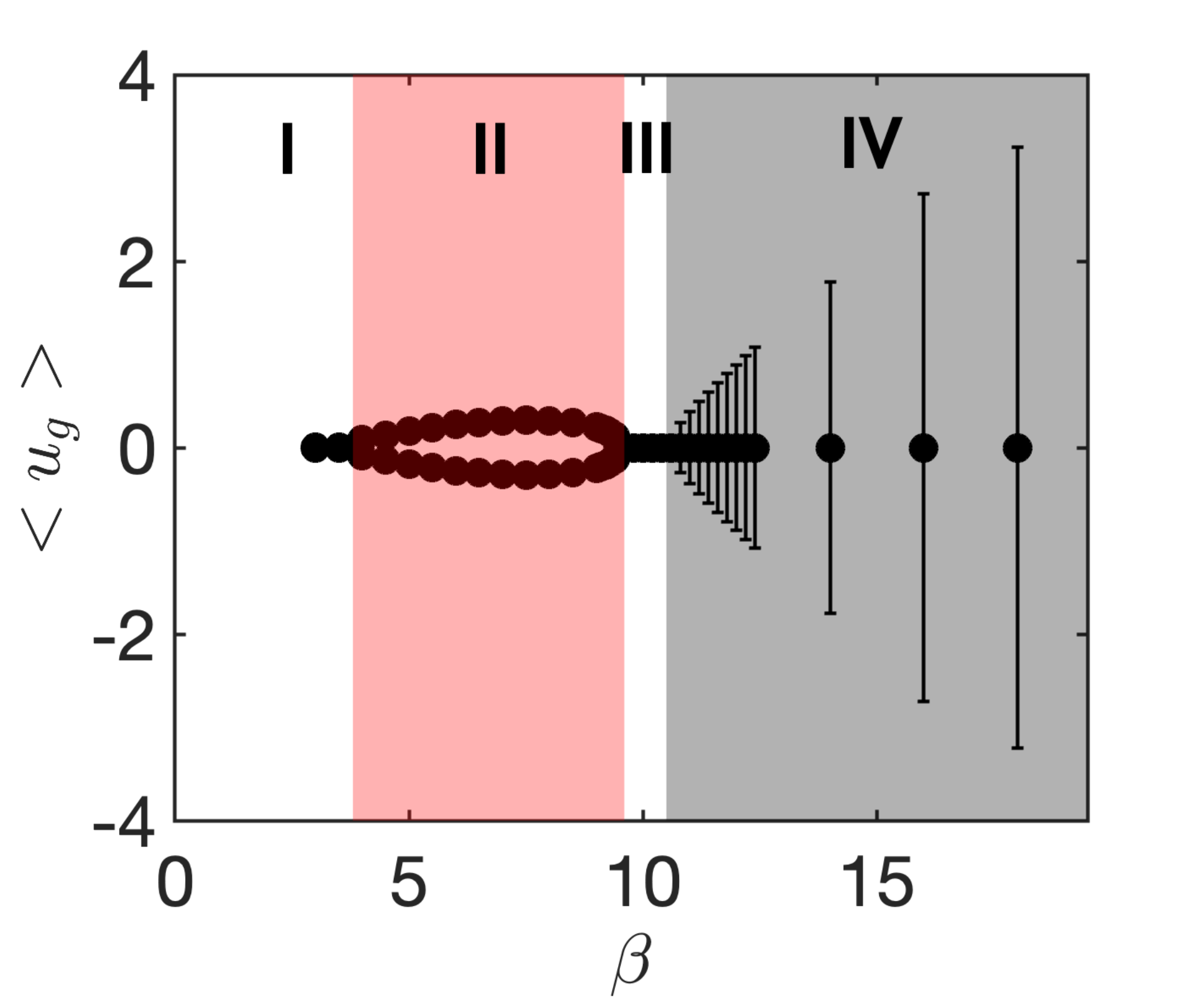}
    & \includegraphics[width=0.45 \linewidth]{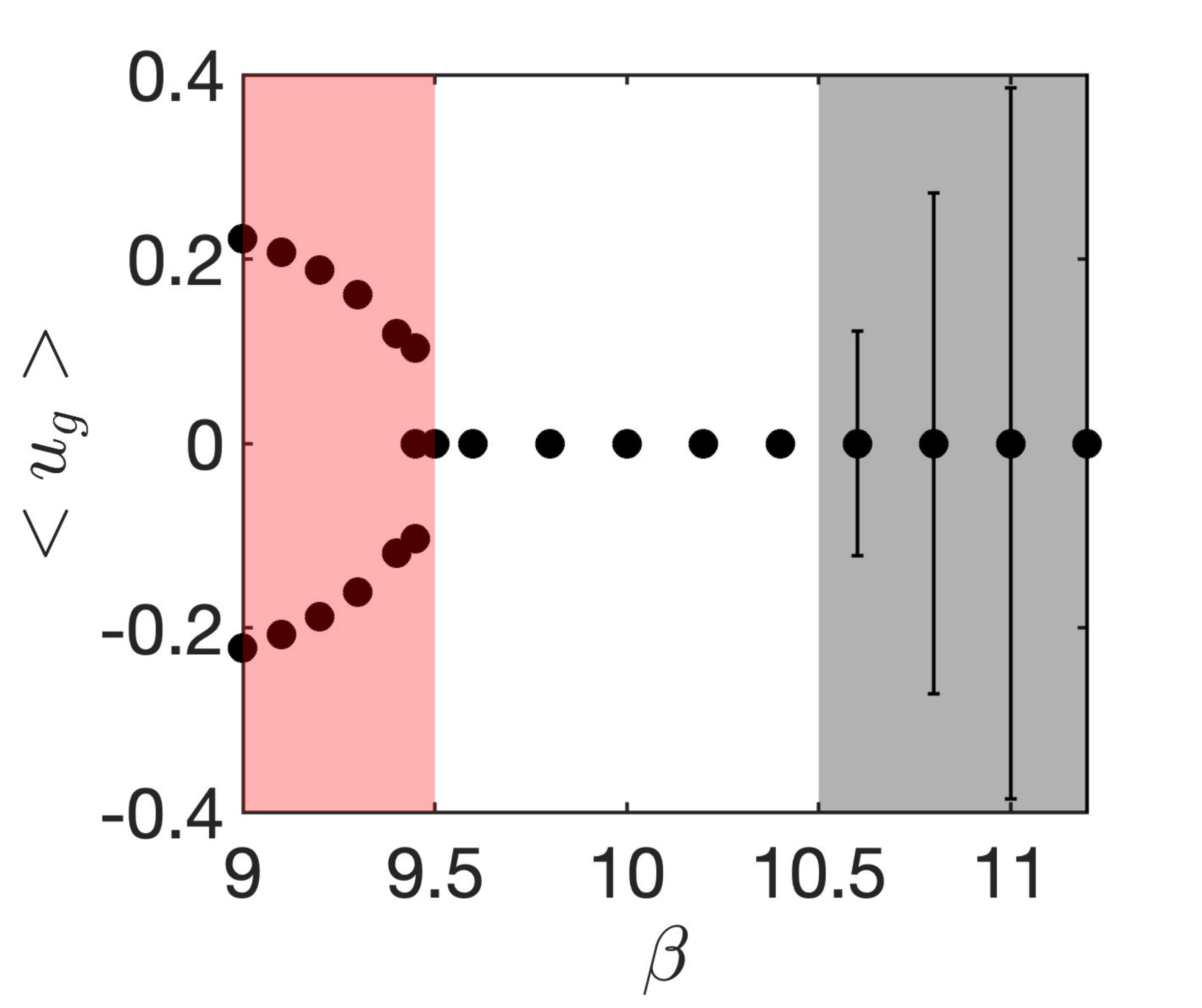}
   \end{tabular}
    \end{center}
    \caption{Self-propelled regimes for a density ratio $\rho = 1$. Background colors are the same as the previous figure. (a) Evolution of the time-averaged (circles) and amplitude (error bars) of the foil horizontal velocity with the Stokes number. (b) Close-up view highlighting the transition between the self-propelled regimes. Other control parameters: $A = 0.5$ and $h = 0.05$.}
    \label{2-NonLinear-MeanU-rho}
\end{figure}

Considering a density ratio closer to swimming organisms ($\rho=1$), figure \ref{2-NonLinear-MeanU-rho}, the same self-propelled regimes and transitions are identified. In figure \ref{2-NonLinear-MeanU-rho}(a), as the previously presented case of $\rho=100$, symmetric non-propulsive solutions become unidirectional propulsive ones for the critical Stokes number $\beta=4$. The time-averaged velocity thus increase with an average value slightly slower than $\rho=100$ up to $\beta \sim 8.5$, decreasing beyond this point and abruptly falling to zero for $\beta=9.5$, a small change in the re-stabilization $\beta$ of the previous density ratio. By decreasing the Stokes number, co-existing unidirectional propulsive and non-propulsive solutions were again obtained, this time for $9.42 \leq \beta \leq 9.5$, visible in figure \ref{2-NonLinear-MeanU-rho}(b). Back \& forth solutions are finally observed for $\beta \geq 10.6$. These solutions are significantly encouraged for smaller density ratios. Beside the smaller critical Stokes number, they present now a velocity four times greater in amplitude than for $\rho=100$. As presented by the two considered values of $\rho=1$ and $\rho=100$, the self-propelled regimes transition route appears to be robust to variations of the density ratio. \\

Self-propelled regimes and the transition route, for this Stokes number range, are also invariant with respect to the foil geometry. Simulations conducted for an elliptical foil of aspect ratio $h=0.1$, reported in \S \ref{Appendix:Alben}, have presented similar results and transition route as the ones described for the rectangular foil with rounded edges.

%%%%%%%%%%%%%%%%%%%%%%%%%%%%%%%%%%%%%%%%%%%%%%%%%%%%%%%%%%%%%%%%%%%%%%%%%%%%%%%%%%%%%%%%%%%%%%%%%%%%%%%%%%
\section{Fluid-solid stability analysis of non-propulsive periodic solutions}\label{ref:FluidSolidFloquet}

The emergence of the propulsive solutions identified in the previous section with nonlinear unsteady simulations is now investigated by analyzing the stability of non-propulsive periodic solutions. 
The Floquet stability analysis is introduced in \S \ref{subsec:floquet} by considering a perturbation of the horizontal foil velocity, in addition to the flow perturbation. The numerical method is then explained in \S \ref{subsec:numericalmethod}. Results of such stability analysis, that couples the fluid and solid perturbations, are presented in  \S \ref{subsec:unstablemodes} for the density ratio $\rho=100$. First the two synchronous and asynchronous modes found unstable are carefully described. Then these modes are discussed in light of the non-linear results previously described. Finally, the influence of the density ratio on the linear results is discussed in \S \ref{subsec:densityratio}.

%% ----------------------------------------------------------------------------------------------%%
\subsection{Fluid-solid Floquet stability analysis }\label{subsec:floquet}

The flow variables and the foil horizontal velocity are decomposed as 
\begin{align}
    (\mathbf{u},p,u_{g})=(\mathbf{u}_{b},p_{b},0) + \epsilon (\mathbf{u}',p',u_{g}'), 
\label{eqn:decomposition}
\end{align}
where $(\mathbf{u_b},p_b)$ denote the periodic base flow fields. Since it satisfies the spatial symmetry \eqref{spatialsymmetry} at every instant of the flapping period, the foil horizontal velocity of the periodic base solution is equal to zero. Infinitesimal perturbations ($\epsilon \ll 1$) are superimposed to the periodic base solution meaning that in addition to perturbing the base-flow field $(\mathbf{u}',p')$ as in \citet{Deng2016a}, the foil horizontal velocity $u_{g}'$ is perturbed. Note that no perturbation of the vertical velocity is considered since the flapping velocity $\mathbf{v}_{g}$ is imposed in the present analysis. By injecting the above decomposition into \eqref{AllEqs-Compact}-\eqref{eqn:fluidsolid} and retaining the first-order term in $\epsilon$, we obtain the following system of equations governing the linear dynamic around the non-propulsive periodic solution
\begin{equation} \label{eqn:fluid-solid-perturbation}
\mathcal{B} \; \frac{\partial}{\partial t} 
\underbrace{\begin{bmatrix}{}
\mathbf{u}' \\
p' \\
u_g'
\end{bmatrix}}_{\mathbf{q}'} = 
\underbrace{\begin{bmatrix}{}
- \left[ (\mathbf{u}_{b}(t)-v_g(t) \mathbf{e}_{y}) \cdot \boldsymbol{\nabla} \right] - \boldsymbol{\nabla} \mathbf{u}_{b}(t) + \beta^{-1} \boldsymbol{\Delta}& -\boldsymbol{\nabla} &  (\boldsymbol{\nabla} \mathbf{u}_{b}) \cdot \mathbf{e}_{x} \\
-\boldsymbol{\nabla}  & 0 & 0 \\
(\rho S)^{-1} \mathcal{F}_v & (\rho S)^{-1} \mathcal{F}_p & 0 
\end{bmatrix}}_{\mathcal{L}(\mathbf{u}_{b},p_{b})}
\begin{bmatrix}{}
\mathbf{u}' \\
p' \\
u_g'
\end{bmatrix}
\end{equation}
The first two rows are the linearized momentum and mass equations governing the fluid velocity and pressure perturbations. They are coupled to the foil velocity perturbation $u_g'$ via two terms: firstly, the bulk term $(\boldsymbol{\nabla} \mathbf{u}_{b}) \cdot \mathbf{e}_{x}$ (block $(1,3)$ in the right-hand side matrix) that modifies the production of fluid perturbation in the momentum equation, and secondly, the boundary conditions at the foil surface $\Gamma_{w}$, where the equality of fluid and solid perturbations holds
\begin{equation}\label{BC-Fluid-rho-fini}
 (u',v')(\mathbf{x}_w,t)=(u_g',0) \,. 
\end{equation}
The third row indicates that the horizontal acceleration of the foil is equal to the horizontal force exerted by the flow perturbation, which is  here separated into  viscous $\mathcal{F}_v$  and pressure $\mathcal{F}_p$   contributions, respectively defined as  
 \begin{eqnarray}
    \mathcal{F}_v (\mathbf{u'})  = \beta^{-1}  \displaystyle\int_{\Gamma_w} (\nabla\mathbf{u'}+\nabla\mathbf{u'}^{T})\cdot \mathbf{n}) \cdot \mathbf{e_x} \, \mathrm{d}\Gamma_w \;\;,\;\;
    \mathcal{F}_p (p')= \displaystyle\int_{\Gamma_w}(- p'\mathbf{n}) \cdot \mathbf{e_x} \, \mathrm{d}\Gamma_w \, .
 \end{eqnarray}
We note that, in \eqref{eqn:fluid-solid-perturbation}, the viscous and pressure forces are weighted by the invert of the foil mass $\rho S$ . Consequently, when the ratio of solid to fluid density increases, the effect of the flow on foil perturbations decreases. In the limit of infinite density ratio, i.e. $\rho \rightarrow \infty$, it even vanishes leading to the following one-way coupled fluid-solid system    
\begin{equation} \label{eqn:fluid-solid-infinitedensity}
\mathcal{B} \; 
\frac{\partial}{\partial t} 
\begin{bmatrix}{}
\mathbf{u}' \\
p' \\
u_g'
\end{bmatrix}
= 
\begin{bmatrix}{}
- \left[ (\mathbf{u}_{b}-v_g \mathbf{e}_{y}) \cdot \boldsymbol{\nabla} \right] - \boldsymbol{\nabla} \mathbf{u}_{b} + \beta^{-1} \boldsymbol{\Delta}& -\boldsymbol{\nabla} &  (\boldsymbol{\nabla} \mathbf{u}_{b}) \cdot \mathbf{e}_{x} \\
-\boldsymbol{\nabla}  & 0 & 0 \\
0 & 0 & 0 
\end{bmatrix}
\begin{bmatrix}{}
\mathbf{u}' \\
p' \\
u_g'
\end{bmatrix}
\end{equation}
In that limit case, the horizontal acceleration of the foil is zero, but not its horizontal velocity. This velocity might still affect the flow perturbation via the coupling terms described above. This one-way coupling analysis is thus different from the hydrodynamic Floquet analysis, performed for instance by \citet{Elston2004}, \citet{Elston2006} on a forced oscillating cylinder and more recently applied by \citet{Deng2016} and \citet{Deng2016a} respectively on the forced oscillation of an ellipsoid and on the self-propulsion of different aspect ratio oscillating ellipses.\\

In the hydrodynamic Floquet analysis, the horizontal perturbation velocity is assumed to be zero ($u_g'=0$). The perturbation equations  \eqref{eqn:fluid-solid-perturbation} then simplify to
\begin{equation} \label{eqn:fluid}
\mathcal{B}_f \; 
\frac{\partial}{\partial t} 
\begin{bmatrix}{}
\mathbf{u}_{f}' \\
p_{f}' 
\end{bmatrix}
= 
\begin{bmatrix}{}
- \left[ (\mathbf{u}_{b}-v_g \mathbf{e}_{y}) \cdot \boldsymbol{\nabla} \right] - \boldsymbol{\nabla} \mathbf{u}_{b} + \beta^{-1} \boldsymbol{\Delta}& -\boldsymbol{\nabla}  \\
-\boldsymbol{\nabla}  & 0 
\end{bmatrix}
\begin{bmatrix}{}
\mathbf{u}_{f}' \\
p_{f}' 
\end{bmatrix}
\end{equation}
where $\mathcal{B}_f$ is the portion of the operator $\mathcal{B}$ related to the fluid variable, and the subscript $f$ is introduced to indicate that the perturbation is purely hydrodynamic. At the foil boundary, they satisfy the no-slip boundary condition
\begin{equation}
 (u_{f}',v_{f}')(\mathbf{x}_w,t)=(0,0) \,. \label{BC-Fluid}
\end{equation}
All the above equations are closed by considering that the fluid perturbations vanish $\mathbf{u'}=0$ far away from the foil. \\

We would like to stress that the fluid-solid perturbation analysis encompasses the purely hydrodynamic perturbation analysis, since hydrodynamic perturbations should be retrieved in the fluid-solid analysis if the foil velocity perturbation is zero.  This will be further discussed when presenting results in \S \ref{subsec:densityratio} in the limit case $\rho \rightarrow \infty$.
 \\
Following \cite{Elston2004} or \cite{Jallas2017}, the perturbations are further decomposed in the form
\begin{equation} \label{Floquet}
\mathbf{q'}(\mathbf{x},t)= \sum \left( \mathbf{\hat{q}}_j (\mathbf{x},t) e^{\lambda_j t} + \mbox{c.c.} \right) \, ,
\end{equation}

where $\mathbf{\hat{q}}_j$ are T-periodic functions, called the Floquet modes, associated to the complex numbers $\lambda_j$, called the Floquet exponents. The Floquet multiplier, defined as $\mu_j=e^{\lambda_j T}$, is rather used in the following. It represents the complex amplitude gain of the periodic Floquet mode over one period, i.e. $\mathbf{\hat{q}}_j(\mathbf{x},T)=\mu_j \mathbf{\hat{q}}_j(\mathbf{x},0)$. The polar decomposition of the Floquet multiplier is $\mu_j=|\mu_j| e^{\rm{i} \phi_{j}} $, where the modulus $|\mu_j|$  quantifies the growth (or decay) of the corresponding Floquet mode over the period, and $\phi_{j}$ represents its phase shift over the same period. The stability of the periodic base solution is then addressed by considering the Floquet multiplier with largest modulus. If its absolute value, denoted $|\mu_0|$, is greater than one, the corresponding Floquet mode will grow over one period and the periodic base solution is thus unstable. When the leading Floquet multiplier is real ($\phi_{0}=0$), the Floquet mode is \textit{synchronous} as the perturbation evolves in time with the period of the base flow. When the leading Floquet multiplier is complex ($\phi_{0} \neq 0$), the Floquet mode is \textit{asynchronous} and a frequency, denoted $f'$ in the following, related to the phase of the Floquet mode as $f' = \phi_{0} /(2 \pi)$ is introduced. 

\subsection{Time-averaged analysis of fluid-solid Floquet modes}\label{subsec:timeaveragedanalysis}
To better understand how the periodic flow perturbation is related to the destabilisation of a fluid-solid Floquet mode, we examine the equation 
\begin{equation}
    \lambda \hat{u}_{g} +\frac{d \hat{u}_{g}}{dt} = \frac{1}{\rho S} F_x(\hat{\mathbf{u}},\hat{p})
\end{equation}
that expresses the instantaneous equilibrium between the horizontal force exerted by the flow component of the Floquet mode and the horizontal acceleration of the foil. The latter is composed of two terms, one related to the growth/decay of the mode, and one related to its instantaneous acceleration. Due to the periodicity of the Floquet mode, the latter disappears when time-averaging over a flapping period, yielding
\begin{equation}\label{eqn:acc-force}
    \lambda \left< \hat{u}_{g} \right>   = \frac{1}{\rho S} \left< \hat{F}_x \right>
\end{equation}
where $\left< \cdot \right>$ denotes the time-average over a flapping period. \\
For synchronous modes, the Floquet exponent and mode are real variables and the above expression gets 
\begin{equation}\label{eqn:growthrate-synchronous}
    \lambda_{r} = \frac{1}{\rho S} \frac{\left< \hat{F}_x\right>}{\left<\hat{u}_g\right>} \;.
\end{equation}
The growth rate of the Floquet mode is thus proportional to the ratio between the mean component of the force and the mean velocity of the Floquet mode. The Floquet mode is thus unstable (res. stable) when the force and velocity are of same (different) sign.  In the case of asynchronous modes, we introduce the polar decomposition of the time-averaged horizontal force and velocity in \eqref{eqn:acc-force} to obtain the simple relation 
\begin{equation}\label{eqn:growthrate-asynchronous}
    \lambda = \frac{1}{\rho S} \frac{|\left< \hat{F}_x\right>| }{|\left<\hat{u}_g\right>| } e^{\rm{i} \psi}
\end{equation}
showing that the growth rate (real part) is now also related to the phase difference $\psi$ between the time-averaged force and velocity, and not only to their ratio. The relations \eqref{eqn:growthrate-synchronous} and \eqref{eqn:growthrate-asynchronous} will be used in \S \ref{subsec:unstablemodes} for a physical discussion of the Floquet mode

\subsection{Numerical method}\label{subsec:numericalmethod}

The periodic non-propulsive solutions $(\mathbf{u_b},p_b)$ are computed by integrating in time the governing equations \eqref{AllEqs-Compact},\eqref{eqn:residual} using the same temporal and spatial discretization scheme as described in the previous section, but with the following boundary conditions
\begin{eqnarray}
(u_b,v_b)(x_{w},y_{w},t) &=& (0,2 \pi A sin(2 \pi t) )\\
(u_b,\partial_x v_b)(0,y,t) &=& (0,0) 
\end{eqnarray}
The first set of boundary condition, applied at any point $(x_w,y_w)$ of the foil surface, allows imposing the flapping motion of the foil in the vertical direction without any motion in the horizontal direction.  The second set of boundary condition, applied on the $y-$ axis, is used to enforce the spatial reflection symmetry of the flow characteristic of the non-propulsive solution. Typically, $50$ flapping periods are simulated to reach a periodic solution. Note that, for computational efficiency, the computational domain can be reduced to the left or right part of the full domain shown in figure \ref{Mesh}, but this is not mandatory. In that case, the periodic base solution on the full domain is retrieved by using the spatial symmetry relation \eqref{spatialsymmetry}. \\

The strategy to compute Floquet modes is the one proposed by \citet{Barkley1996}. The stability of an initial perturbation is assessed regarding the action of the propagator over one period $\mathbf{\Phi}$, also known as Monodromy matrix. The action of this Monodromy matrix over the perturbation at an arbitrary initial time $t_0$ is formally denoted $\mathbf{q'}(\mathbf{x},t_0+T)=\mathbf{\Phi} \, \mathbf{q'}(\mathbf{x},t_0) \, $. It is  actually obtained by time-integration along a period of the linearized equations  \eqref{eqn:fluid-solid-perturbation} with boundary conditions \eqref{BC-Fluid-rho-fini}, using the temporal and numerical discretization schemes previously described. An Arnoldi method with a modified Gram-Schmidt algorithm for the orthogonalization step 
\citep{Saad2011} is implemented in the FreeFEM software \citep{Hecht2012} to approximate the Monodromy matrix in a low-dimensional space. The eigenvalues of this reduced matrix approximate the Floquet multiplier and its eigenvectors allow reconstructing the Floquet modes at the initial time, i.e. $\hat{\mathbf{q}}(\mathbf{x},t_0)$. A minimal number of $30$ Arnoldi vectors is used in the following, this number being further increased in steps of $10$ when necessary in order for the dominant eigenvalue to converge to five significant digits. All computed modes are normalised by the kinetic energy of the coupled fluid-solid system. A validation of this method is detailed in Appendix \ref{Appendix:SectionValidation}. \\

Finally, as the Arnoldi method gives access to the Floquet mode at an initial time, the mode complete temporal evolution is obtained through time-integration of the following equation over one flapping period,  
\begin{equation} \label{AllEqs-Compact-Floquet} 
\mathcal{B} \, \frac{\partial \hat{\mathbf{q}}}{\partial t}  - \mathcal{L}(\mathbf{u}_b,v_g) \, \hat{\mathbf{q}} = - \lambda \, \mathcal{B} \, \hat{\mathbf{q}}, 
\end{equation}
using as initial condition the Floquet mode obtained with the Arnoldi algorithm. The Floquet exponent $\lambda$ being known,  the right-hand side term appropriately counteracts the temporal growth (resp. decay) of the unstable (resp. stable) Floquet mode.

\begin{figure}
   \centering
   \begin{subfigure}{0.24\textwidth}
    \includegraphics[width=\linewidth]{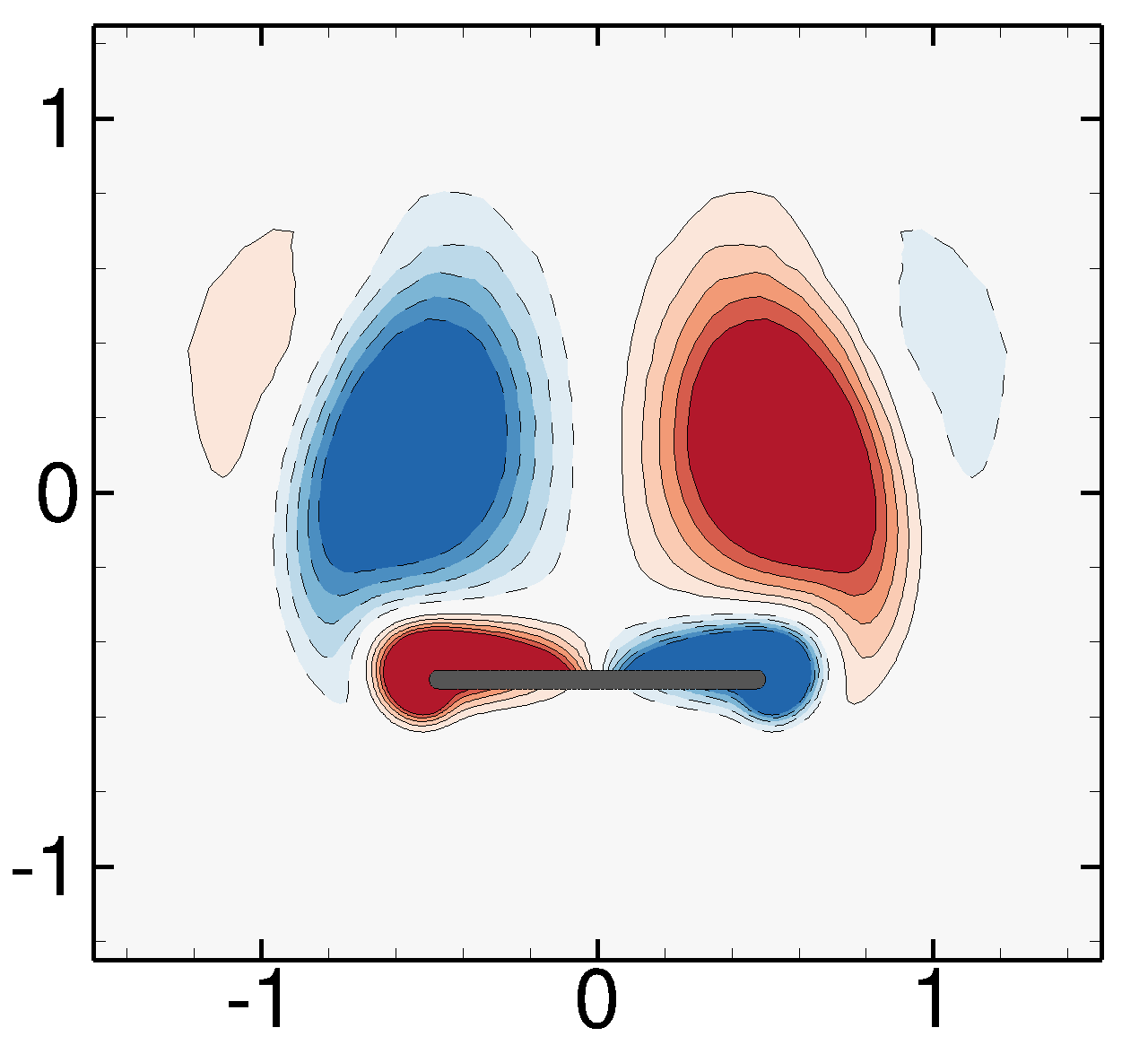}
    \caption{$t_0$}
    \end{subfigure}
       \begin{subfigure}{0.24\textwidth}
    \includegraphics[width=\linewidth]{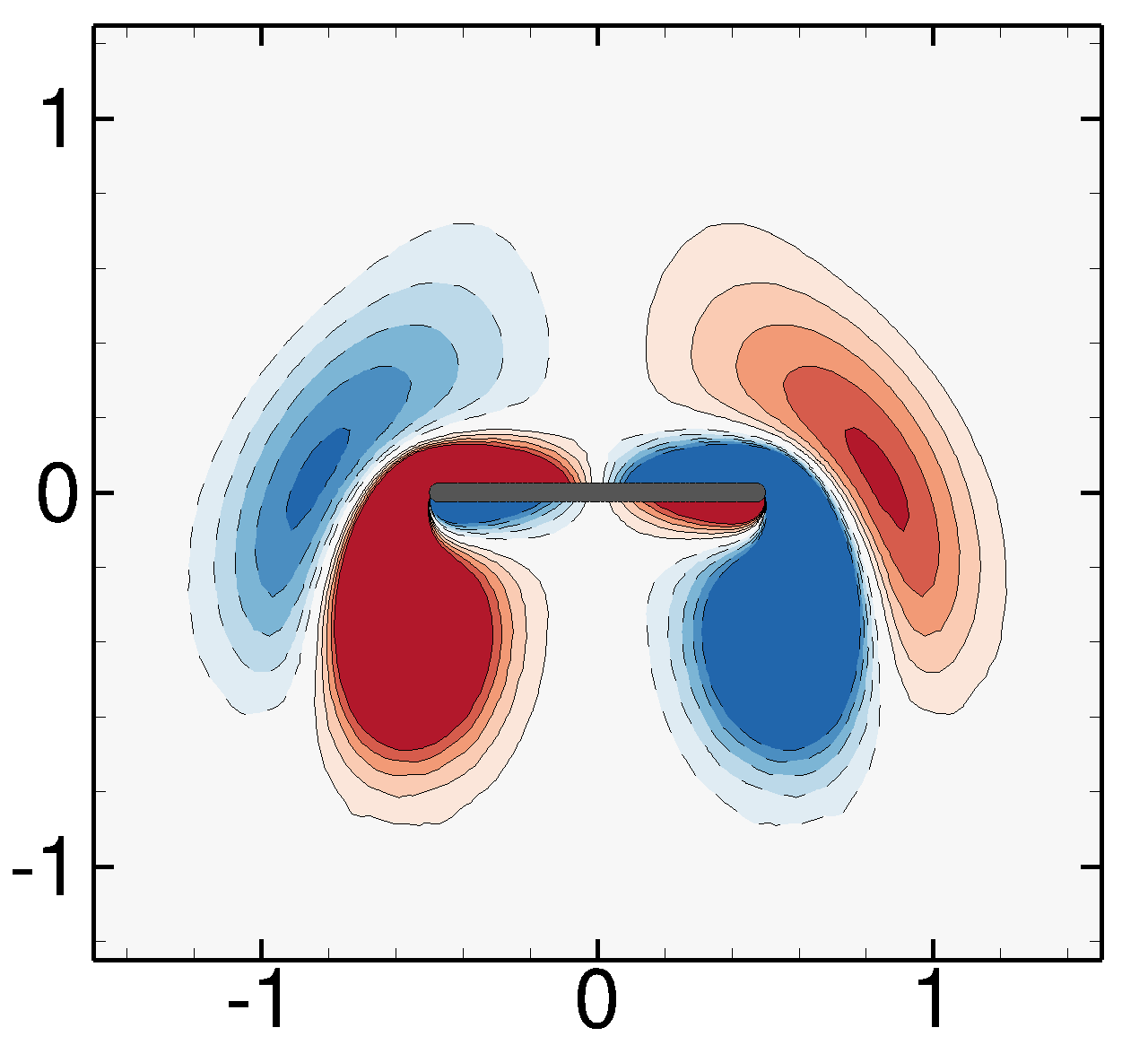}
    \caption{$t_0+1/4$}
    \end{subfigure}
       \begin{subfigure}{0.24\textwidth}
    \includegraphics[width=\linewidth]{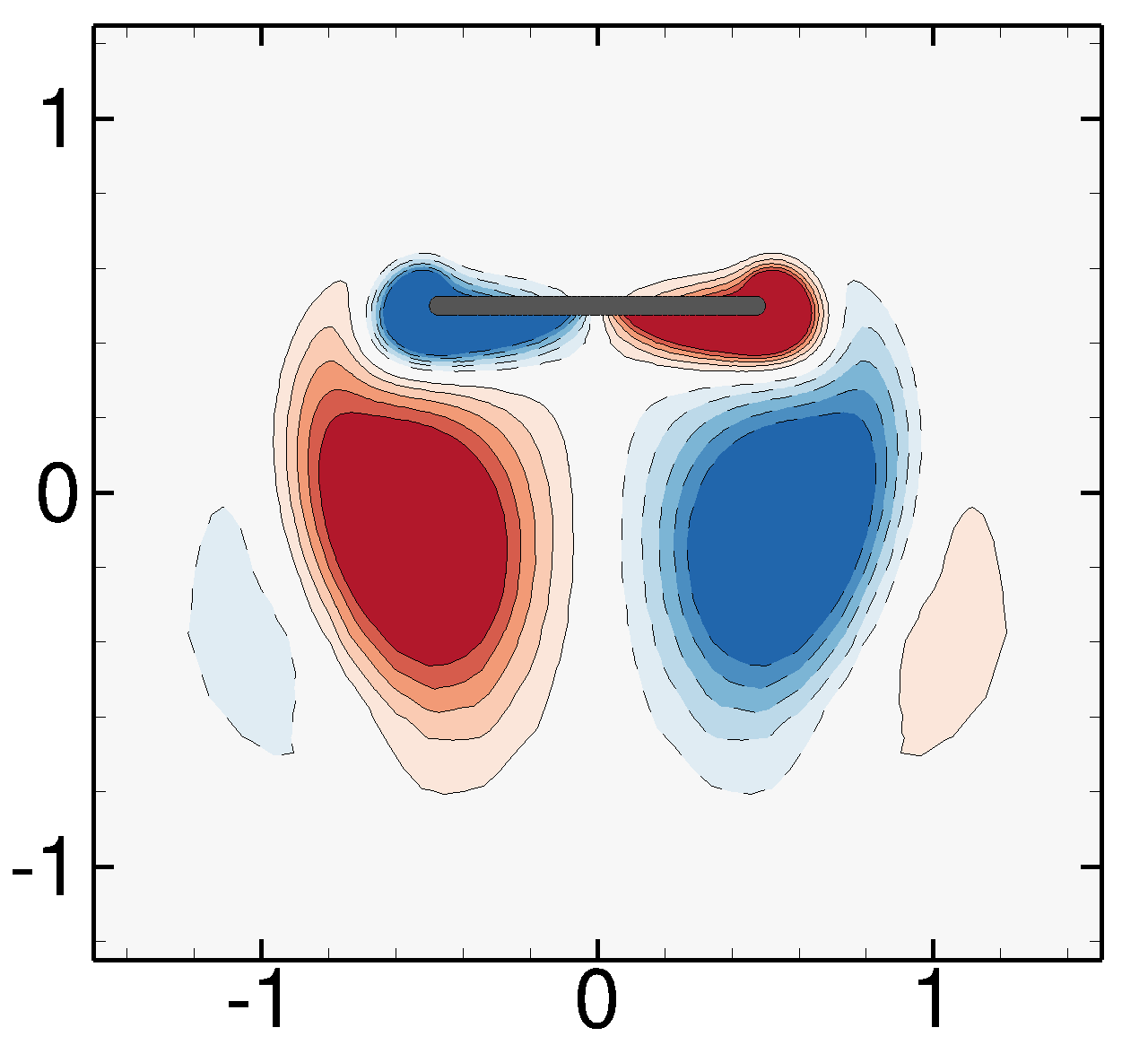}
    \caption{$t_0+1/2$}
   \end{subfigure}
    \begin{subfigure}{0.24\textwidth}
    \includegraphics[width=\linewidth]{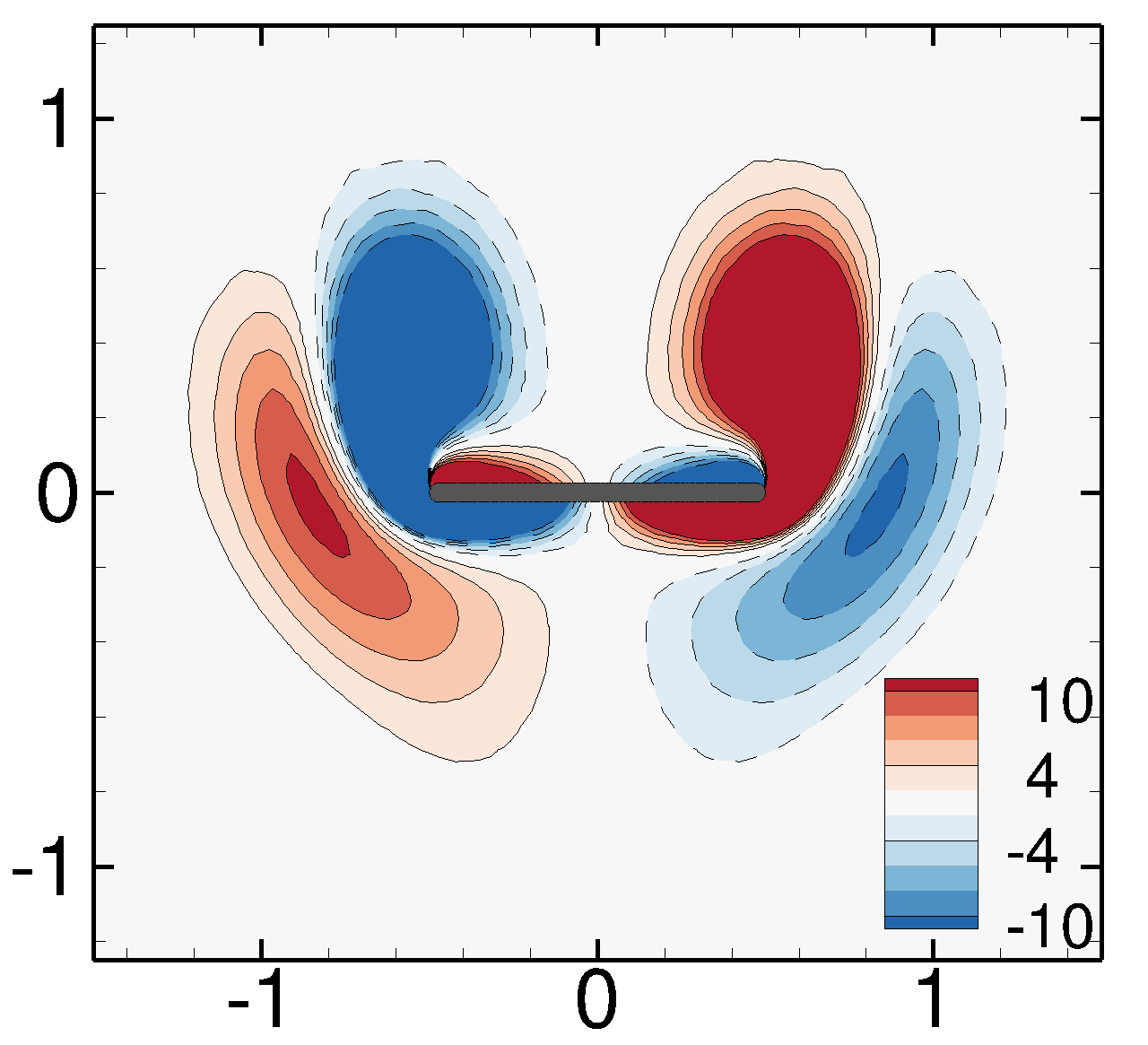}
    \caption{$t_0+3/4$}
    \end{subfigure}
    \caption{Non-propulsive solution obtained for $A=0.5$ and $\beta=13$. The instantaneous vorticity field is depicted at four instants of the (unitary) flapping period.}
    \label{4-Linear-SymPreserved}
\end{figure}

%% ----------------------------------------------------------------------------------------------%%
\subsection{Results of Floquet analyses for $\rho=100$}\label{subsec:unstablemodes}

The stability analysis of non-propulsive periodic solutions has been performed for the flapping amplitude $A=0.5$ and Stokes numbers in the range $1 \le \beta \le 20$.  The instantaneous vorticity fields of a typical base non-propulsive solution, obtained for $\beta=13$, are depicted in figure \ref{4-Linear-SymPreserved} at four instants in the flapping period. The spatial left-right symmetry \eqref{spatialsymmetry} along the $y$-axis is clearly satisfied at every instant of the flapping period. By comparing the dipolar structure at time $t_0+1/2$ and $t_0$, this solution also satisfies the spatio-temporal symmetry \eqref{spatiotemporalsymmetry}.  \\

\subsubsection{Floquet multipliers: fluid-solid versus hydrodynamic analysis}

\begin{figure}
   \begin{center}
    \begin{tabular}{ll}
    (a) & (b) \\
    \includegraphics[width=0.45\linewidth]{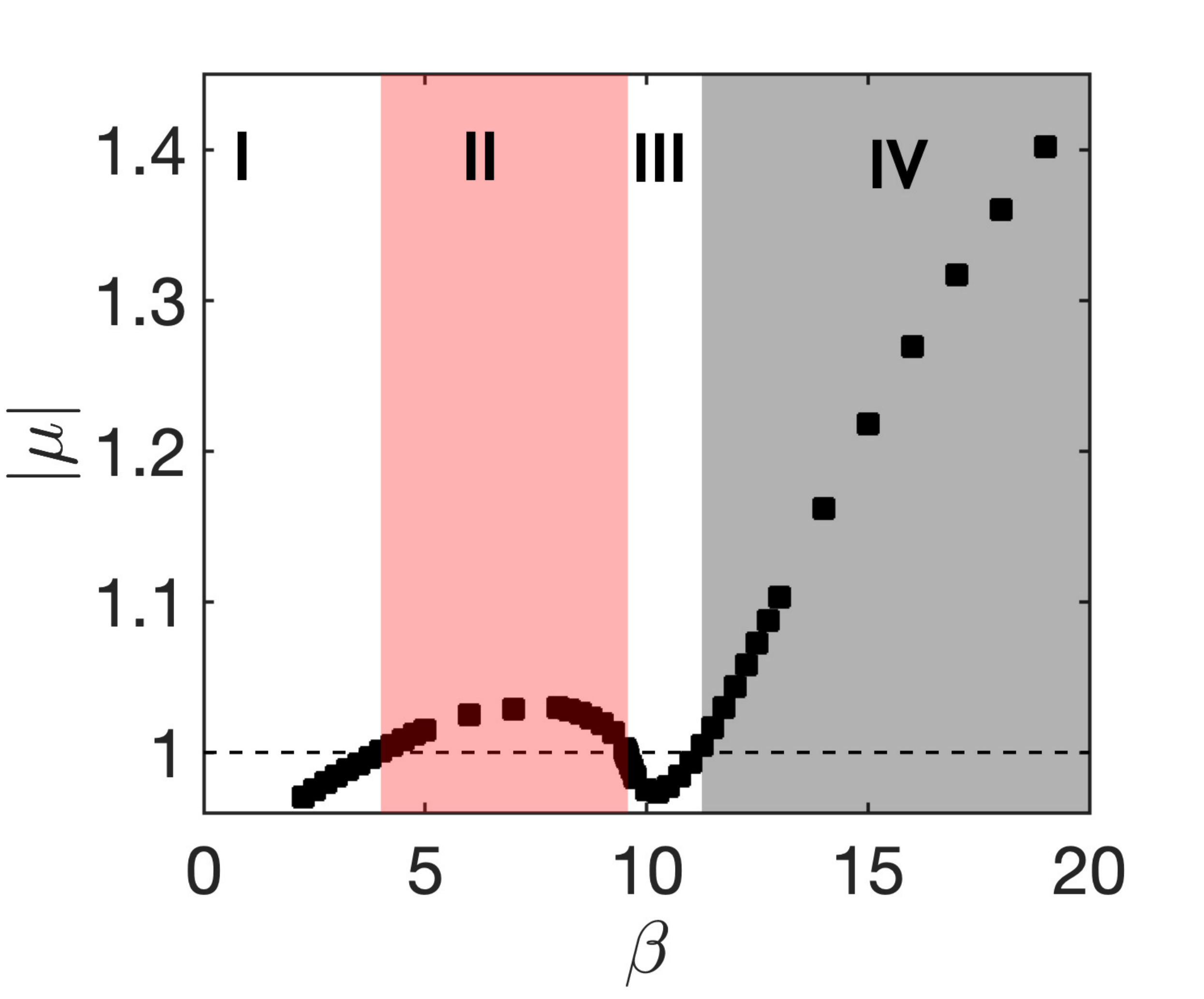}
    & \includegraphics[width=0.45\linewidth]{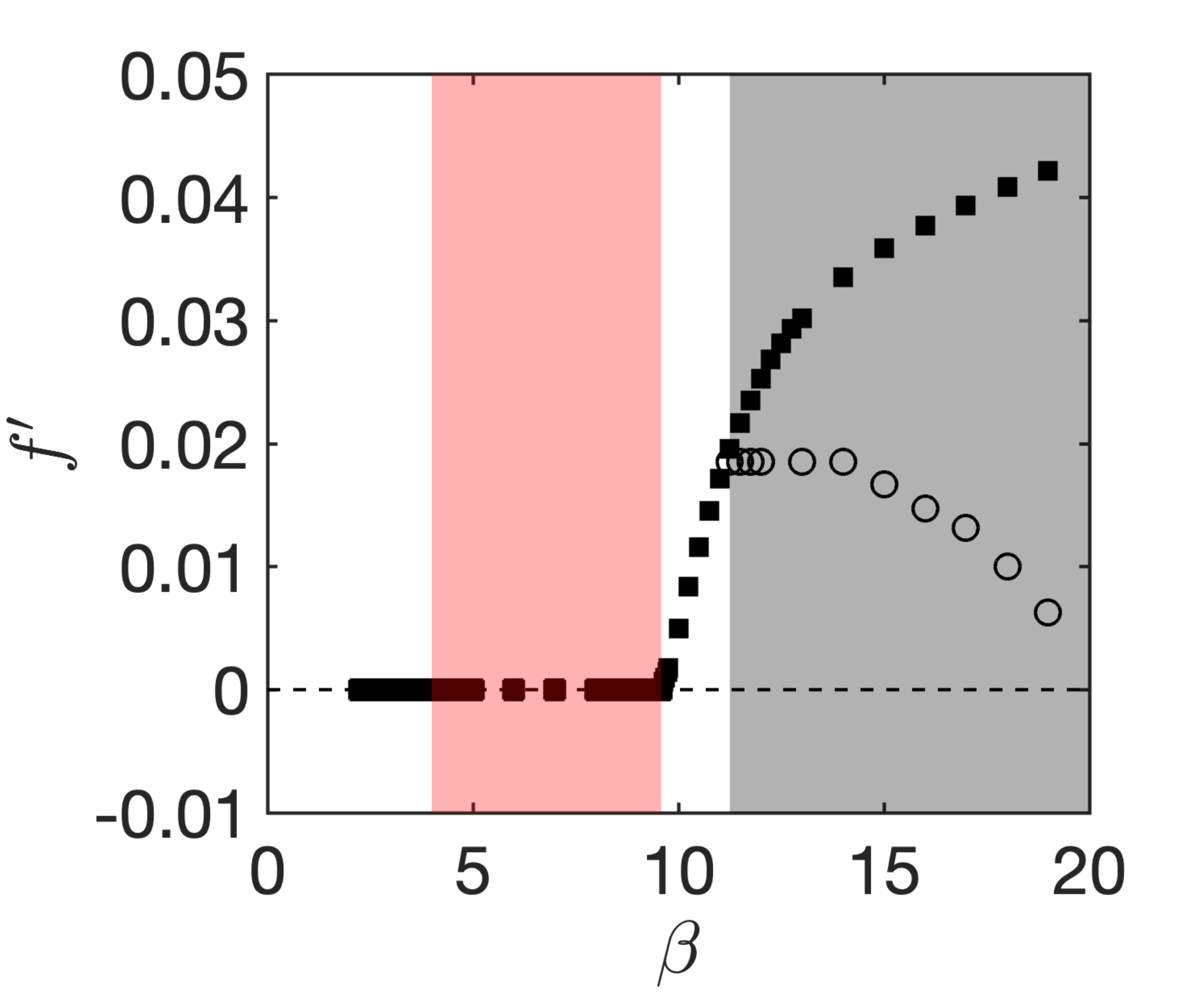}
    \end{tabular}
    \end{center}
    \caption{Fluid-solid Floquet analysis for $\rho=100$. (a) Absolute value and (b) frequency of the dominant Floquet multiplier (black circles) as a function of the Stokes number $\beta$. The latin numbers indicate regimes of non-linear solution identified in \S \ref{sec:nonlinear}, the red region corresponding to unidirectional self-propelled solutions, while the grey regions correspond to back \& forth self-propelled solutions. In (b), the open circles indicate the low-frequencies characteristic of the Back \& Forth solutions.}
    \label{Fluid-Solid-Floquet}
\end{figure}

Results of the fluid-solid Floquet analysis performed for the  density ratio $\rho=100$ are first depicted in figure \ref{Fluid-Solid-Floquet}, with the modulus and frequency of the leading Floquet multipliers as a function of $\beta$ in figures \ref{Fluid-Solid-Floquet}(a) and \ref{Fluid-Solid-Floquet}(b), respectively.
We can clearly identify two ranges of Stokes number where the leading Floquet modes get unstable ($|\mu_0|>1$) and they compare very well with regions II and IV, where unidirectional and back \& forth self-propelled solutions were obtained in the nonlinear unsteady simulations.\\

In the range $ 4.00 \le \beta \le 9.53$, the unstable Floquet modes are synchronous ($f'=0$). Thus, the perturbation does not break the periodicity of the base solution, in agreement with the unidirectional propulsive solution observed in region II. The frequency $f=2$ characterizing the horizontal speed of this propulsive solution (see figure \ref{2-NonLinear-MeanU}-b) is rather related to the spatio-temporal symmetry of the Floquet mode, as will be seen in \S \ref{subsubsec:synchronousFloquet}. A quantitative comparison of the thresholds is provided in table \ref{tab:comparisonthreshold}. The destabilisation of the synchronous mode at $\beta=4$ is in perfect agreement with the emergence of the unidirectional propulsive solution, i.e. the transition between region I and II. On the other hand, the threshold value $\beta=9.53$ corresponding to the stabilization of this mode is slightly different from the threshold $\beta=9.58$ above which the unidirectional propulsive solution disappears (transition from II to III). This is due to the sub-critical nature of the bifurcation at this threshold, clearly seen in figure \ref{2-NonLinear-MeanU}(c). But the threshold value $\beta=9.53$ found with the linear stability corresponds perfectly to the disappearance of the symmetric solution when decreasing the Stokes number (transition from III to II). \\

When further increasing the Stokes number $ \beta \ge 11.25$, an asynchronous Floquet mode gets unstable, with a very low frequency ($f'\sim 0.01$) compared to the flapping frequency ($f=1$). The destabilization of this asynchronous mode occurs at the same value of the Stokes number for which the back \& forth solution appears. The frequency of the asynchronous mode is compared to the frequency of this solution in figure \ref{Fluid-Solid-Floquet}(b). They compare very well at the threshold $\beta=11.25$, but when the Stokes number is increased, the agreement gets worse. Opposite trends are observed, with an increase of the Floquet mode frequency and a decrease of the back \& forth frequency. The flow nonlinearities, i.e. which can be mean-flow distortion or higher-harmonics generation/interaction, clearly play a role in the frequency selection of this back \& forth solutions. Note that such discrepancy between the linear and nonlinear frequency has been observed for the unsteady wake of a fixed cylinder flow, and is predominantly due to mean-flow distortion in that case \citep{Barkley2006,Sipp2007}. \\

\begin{figure}
   \begin{center}
    \begin{tabular}{ll}
    (a) & (b) \\
    \includegraphics[width=0.45\linewidth]{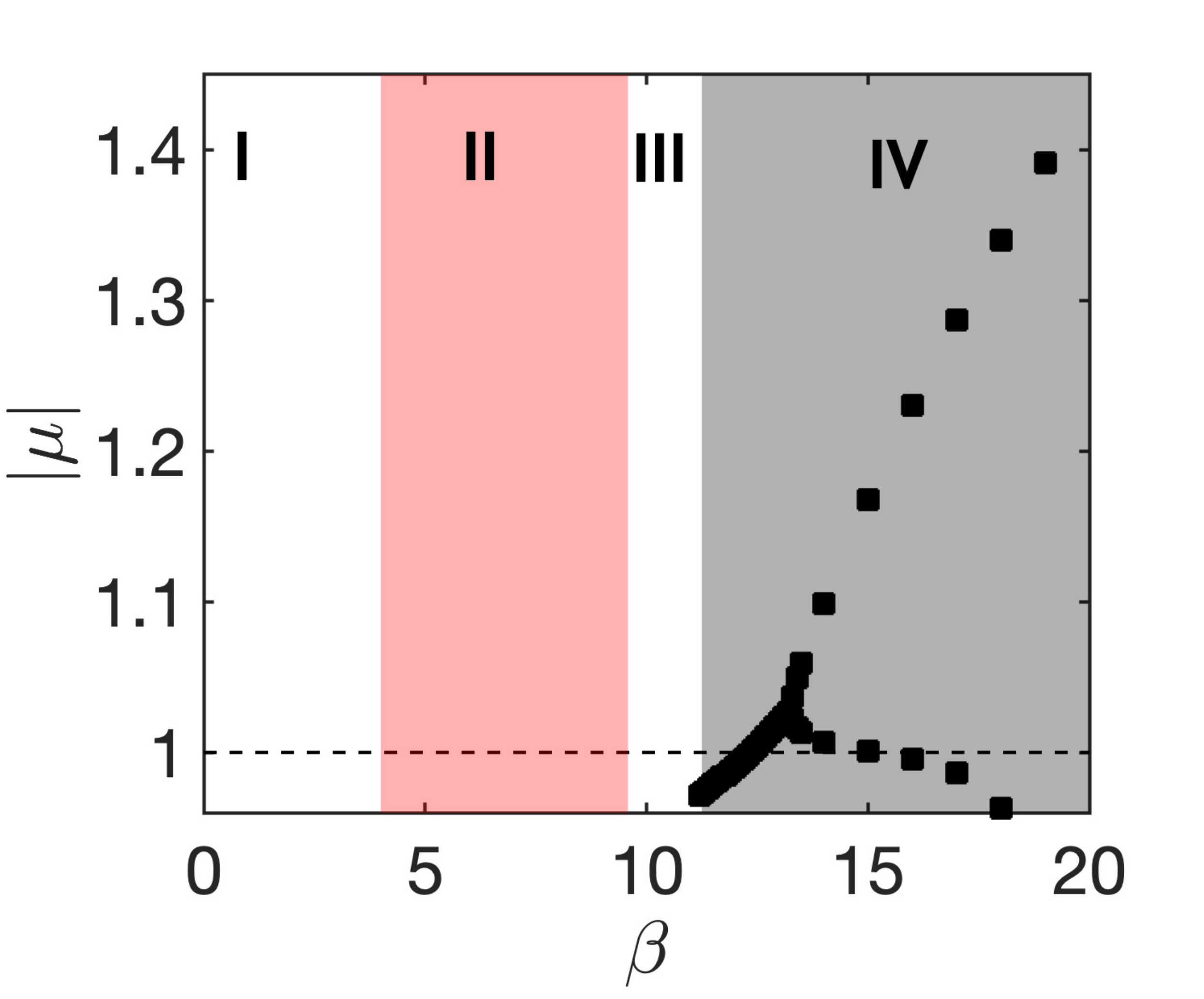}
    & \includegraphics[width=0.45\linewidth]{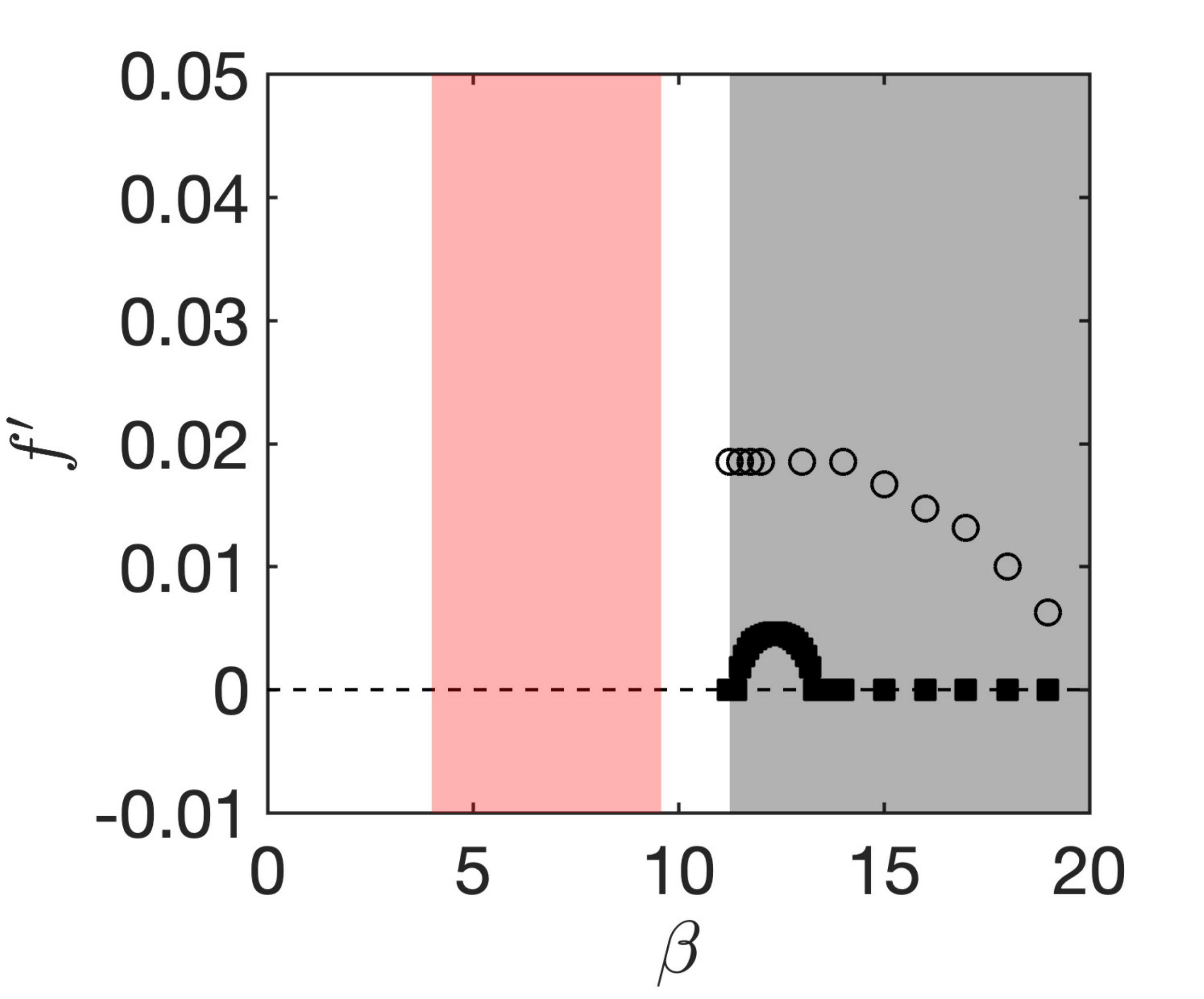}
    \end{tabular}
    \end{center}
\caption{Hydrodynamic Floquet analysis.  Same legend as in previous figure.}
    \label{fig:hydrofloquet}
\end{figure}

Before examining the unstable synchronous and asynchronous Floquet modes, we describe results of the purely-hydrodynamic Floquet analysis, performed in the same range of Stokes number and displayed in figure   \ref{fig:hydrofloquet} with black squares. First, we note that no unstable mode is found in the range of Stokes number corresponding to the region II. So, the purely-hydrodynamic Floquet analysis cannot explain the emergence of the unidirectional propulsive solution. One unstable Floquet mode is found only for larger Stokes number $\beta \ge 12.20$. Just above the threshold, this mode is asynchronous, but when the Stokes number is increased to $\beta=13.3$, the pair of complex asynchronous modes collapses on the real axis becoming two real synchronous modes. One of these modes is further destabilized when increasing $\beta$, while the other one is stabilized for $\beta > 15$. The spatial structure of the unstable asynchronous modes found with the hydrodynamic stability analysis are very similar to the asynchronous mode obtained with the fluid-solid analysis, that are described in \S \ref{sec:asynchronous}, and thus will not be further described. Their frequency is much smaller, as displayed in figure \ref{fig:hydrofloquet}(b), failing to predict the frequency of the back \& forth solution even at the threshold. As indicated in table \ref{tab:comparisonthreshold}, this threshold is under-predicted by the purely-hydrodynamic analysis. Unlike the fluid-solid Floquet analysis, the purely-hydrodynamic Floquet analysis in one hand cannot explain the emergence of the unidirectional propulsive solutions, and on the other hand does not accurately predict the occurrence of the back \& forth solutions. These two observations offer a possible explanation to observations made by \citet{Deng2016a} when comparing the results of unsteady nonlinear simulations and a purely-hydrodynamic stability analysis. In their study, the purely-hydrodynamic analysis did not estimate the enhancement (earlier transition) of the quasi-periodic nonlinear solutions for ellipses of aspect-ratio $h=0.5$ and did not predict the unidirectional propulsion threshold for ellipses of aspect-ratio $h=0.1$. Visibly, the onset of self-propulsion cannot be explained by the flow symmetry breaking instability alone and the self-propelled fluid-solid coupling is vital for its prediction.
\\

\begin{table}
    \centering
    \begin{tabular}{c c c c}
        Transition between regimes & I $-$ II &  II $-$ III & III $-$ IV \\
    \hline
        Fluid-solid stability analysis ($\rho=100$) & $4.00$ & $9.53$ & $11.25$ \\
        Purely-hydrodynamic stability analysis & $\times$ & $\times$ & $12.20$  \\
        \hline 
        Nonlinear simulations & $4.00$ & $9.53$ and $9.58$ & $11.25$ 
       \end{tabular}
    \caption{Critical thresholds obtained with the unsteady nonlinear simulations and the stability analyses. $\times$ indicates that no unstable modes were obtained.}
    \label{tab:comparisonthreshold}
\end{table}{}

\subsubsection{Synchronous Floquet modes: emergence of unidirectional propulsion}
\label{subsubsec:synchronousFloquet}

\begin{figure}
   \centering
            \begin{subfigure}{0.24\textwidth}
    \includegraphics[width=\linewidth]{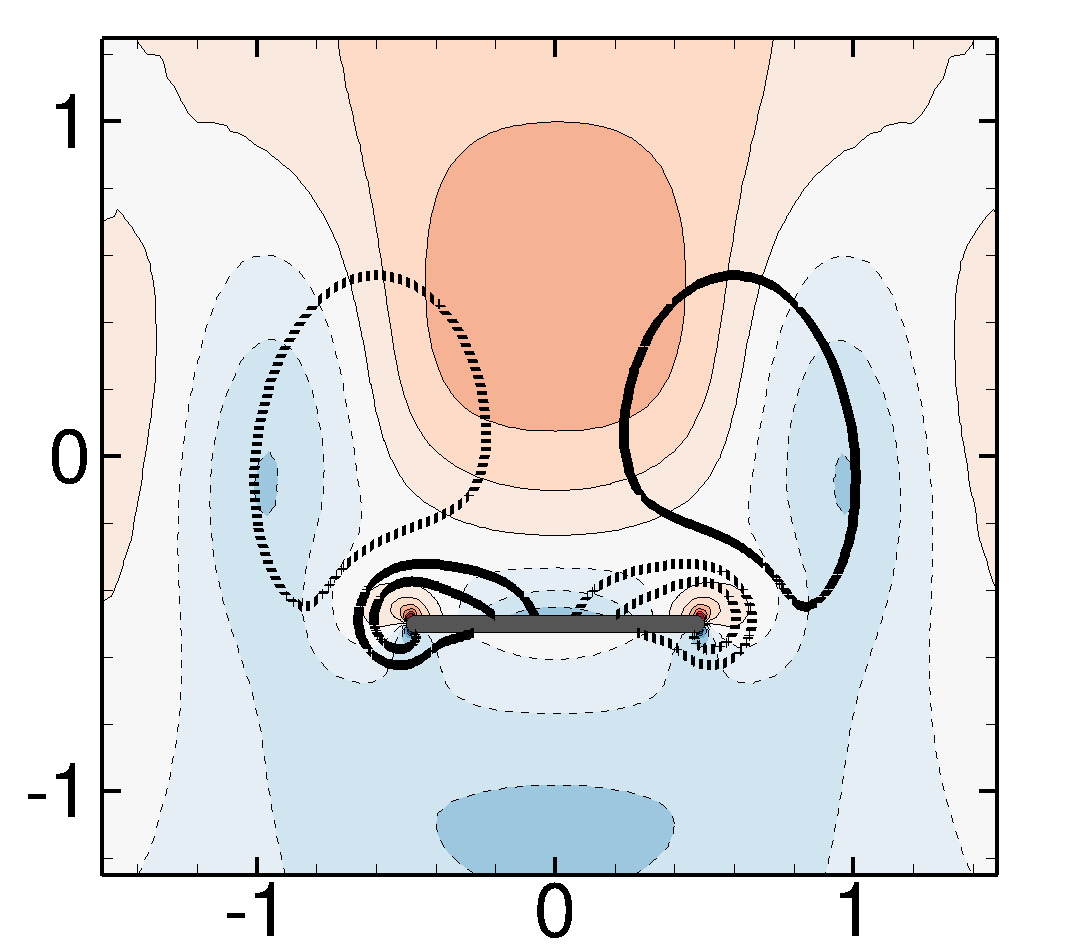}
    \caption{$t=0$}
    \end{subfigure}
       \begin{subfigure}{0.24\textwidth}
    \includegraphics[width=\linewidth]{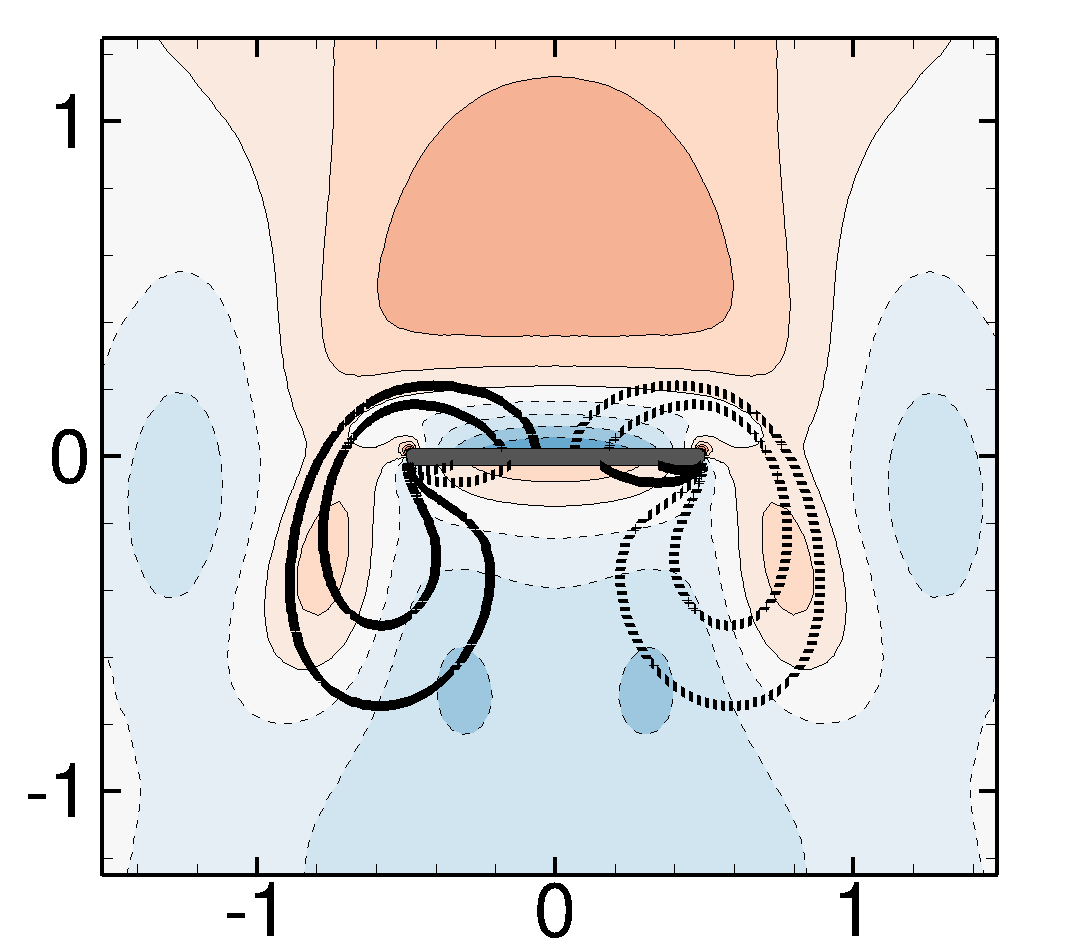}
    \caption{$t=1/4$}
    \end{subfigure}
       \begin{subfigure}{0.24\textwidth}
    \includegraphics[width=\linewidth]{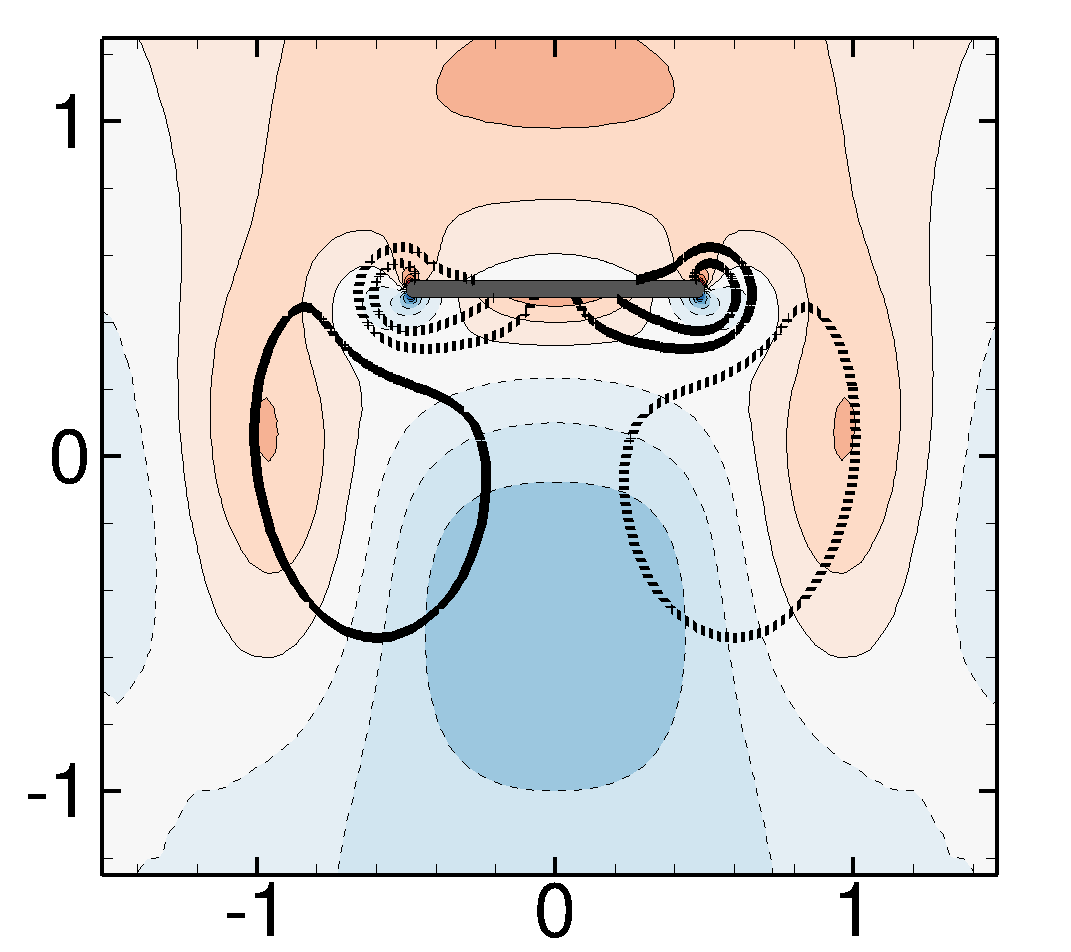}
    \caption{$t=1/2$}
    \end{subfigure}
    \begin{subfigure}{0.24\textwidth}
    \includegraphics[width=\linewidth]{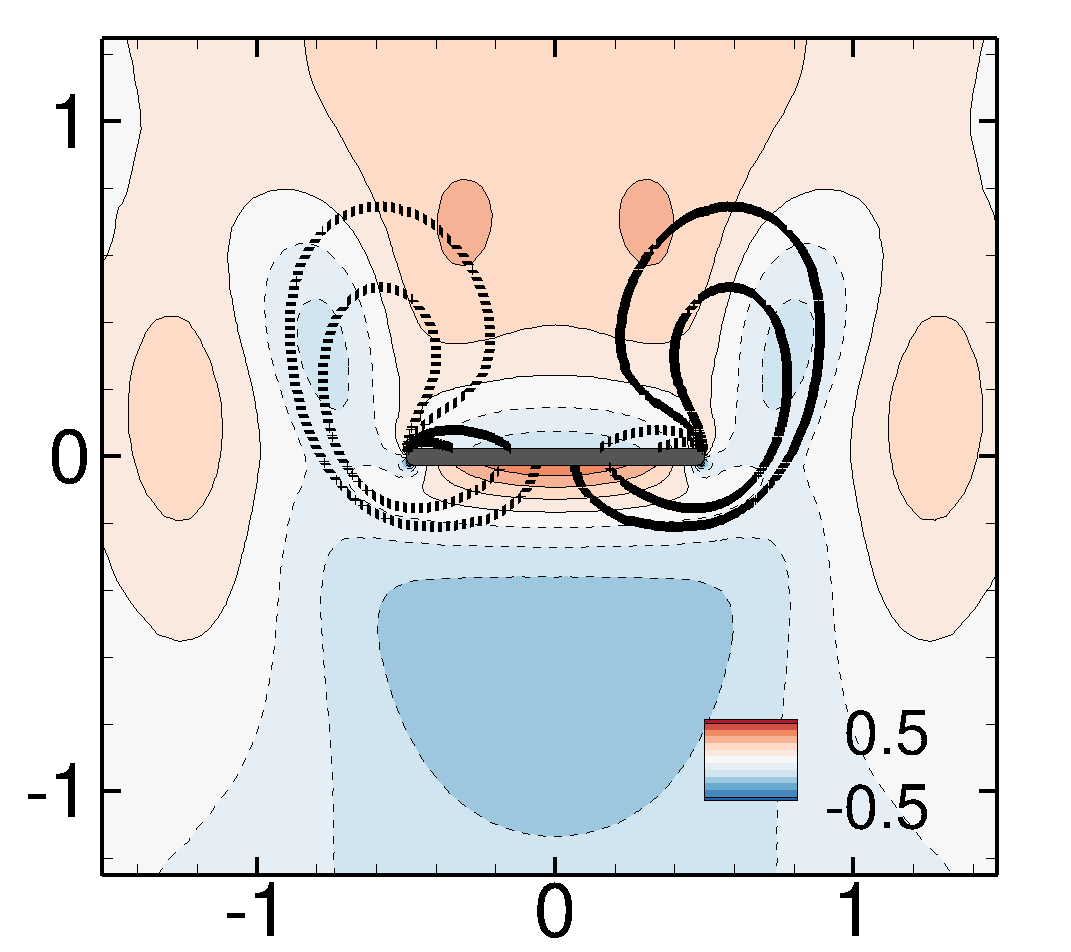}
    \caption{$t=3/4$}
    \end{subfigure}
    \begin{subfigure}{0.4\textwidth}
    \includegraphics[width=\linewidth]{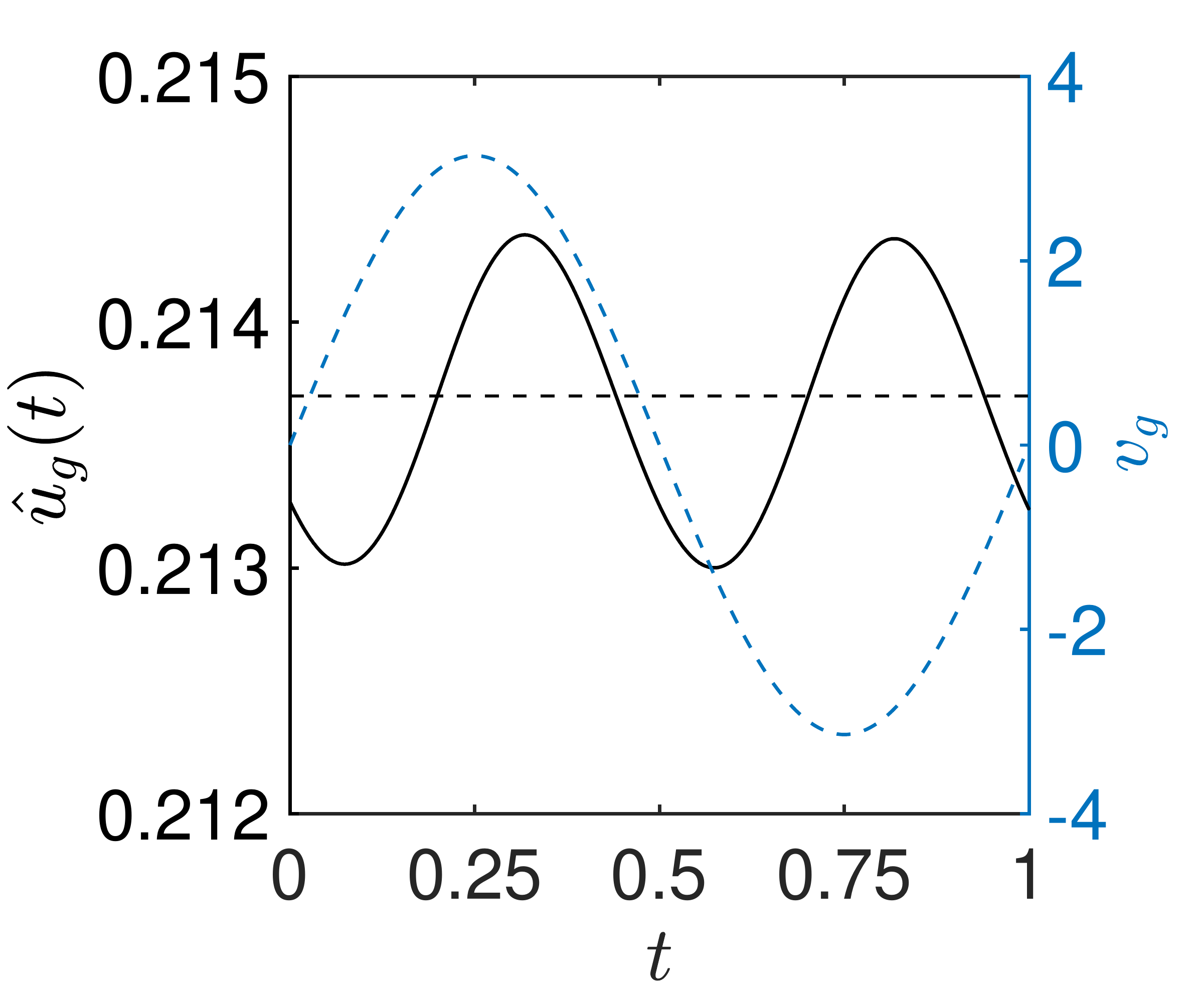}
    \caption{}
    \end{subfigure}
    \begin{subfigure}{0.4\textwidth}
    \includegraphics[width=\linewidth]{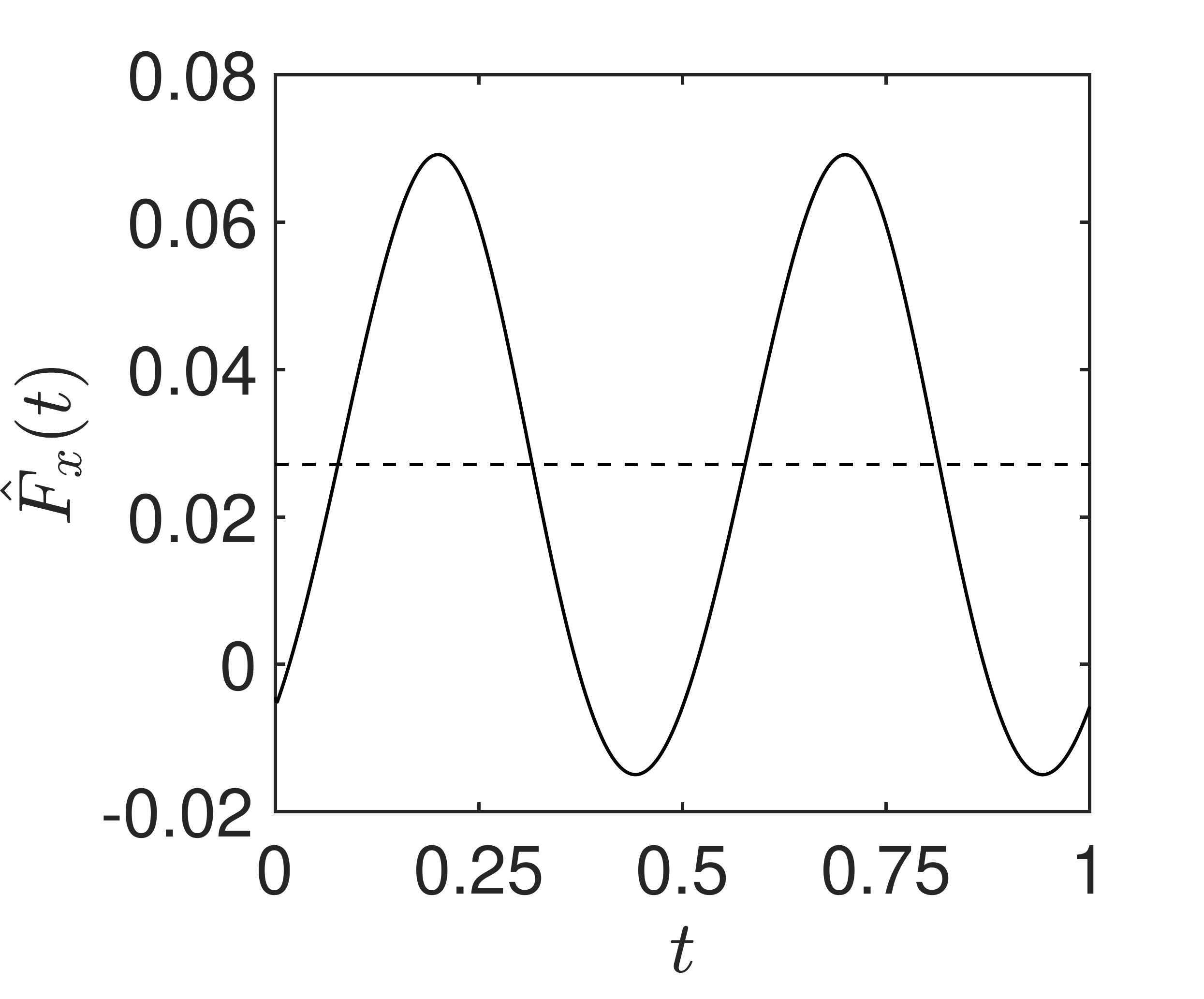}
    \caption{}
    \end{subfigure}
    \caption{Unstable synchronous Floquet mode for $\beta=6$: (a-d) Vorticity contours for a flapping period (base flow positive (resp. negative) values represented by solid (resp. dashed) lines). (e) Foil horizontal speed and (f) force. Time-averaged value along a flapping period of both values is represented by a dashed line in (e,f). In (e) the vertical speed $v_g$ of the base-flow (light blue dashed line) is represented in the right axis.}
            \label{3-Linear-FloquetMode-Beta6}
\end{figure}

Turning back to results of the fluid-solid Floquet analysis, the synchronous Floquet mode is depicted in figure \ref{3-Linear-FloquetMode-Beta6} for $\beta=6$. The vorticity field of this periodic mode is displayed with colored map at four instants of the flapping period in figure \ref{3-Linear-FloquetMode-Beta6}(a-d), and the vorticity of the periodic base flow is superimposed using black (dashed) isolines for positive (negative) values. First we note that the synchronous Floquet mode breaks the left-right symmetry \eqref{spatialsymmetry} of the base flow, since the perturbative vorticity is an even function of $x$ while the base vorticity is an odd function of $x$. But, as the base flow, it still satisfies the spatio-temporal symmetry \eqref{spatiotemporalsymmetry}, so that we can restrict our description of the mode to the up-stroke phase $0\le t \le 1/2$. During the acceleration phase of this up-stroke motion $(t<1/4)$, a patch of positive vorticity exists above the foil, in a region where the vorticity of the base flow is weak, since the latter is rather generated under the foil during the upstroke. This patch of vorticity corresponds to a shear region in the flow perturbation, that induces an increase in the horizontal forces exerted on the foil, as seen in figure \ref{3-Linear-FloquetMode-Beta6}(f).  During the second-half of the up-stroke ($1/4 < t < 1/2$) where the vertical velocity of the foil decreases, the patch of positive vorticity also decreases in size and amplitude. Meanwhile, a patch of negative vorticity appears under the foil, leading to a decrease of the horizontal force. This oscillation of the horizontal force results in an out-of-phase oscillation of the horizontal velocity, shown in figure \ref{3-Linear-FloquetMode-Beta6}(e).  
Interestingly, the average of the horizontal force and velocity over a flapping period (indicated with dashed line in figures) are non-zero and positive here. Therefore, this synchronous Floquet mode is clearly at the origin of the propulsion of the foil in the horizontal direction. Note that the direction of propulsion is not determined by the Floquet mode, since its amplitude is arbitrarily positive (here) or negative, leading to a right (here) or left displacement of the foil.

\begin{figure}
   \centering
   \begin{tabular}{lll}
(a) & (b) & (c) \\
\includegraphics[width=0.32\linewidth]{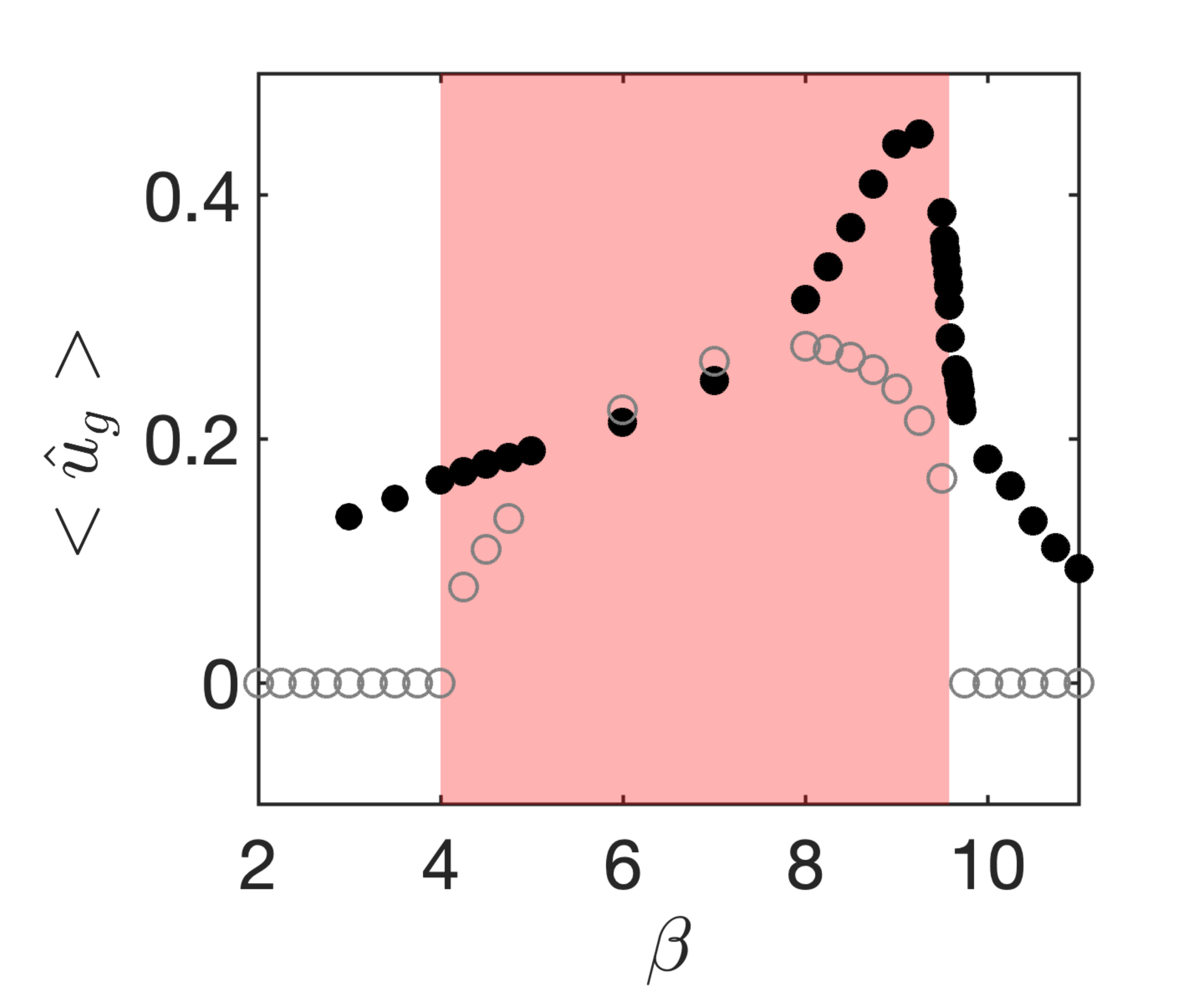}
&    \includegraphics[width=0.32\linewidth]{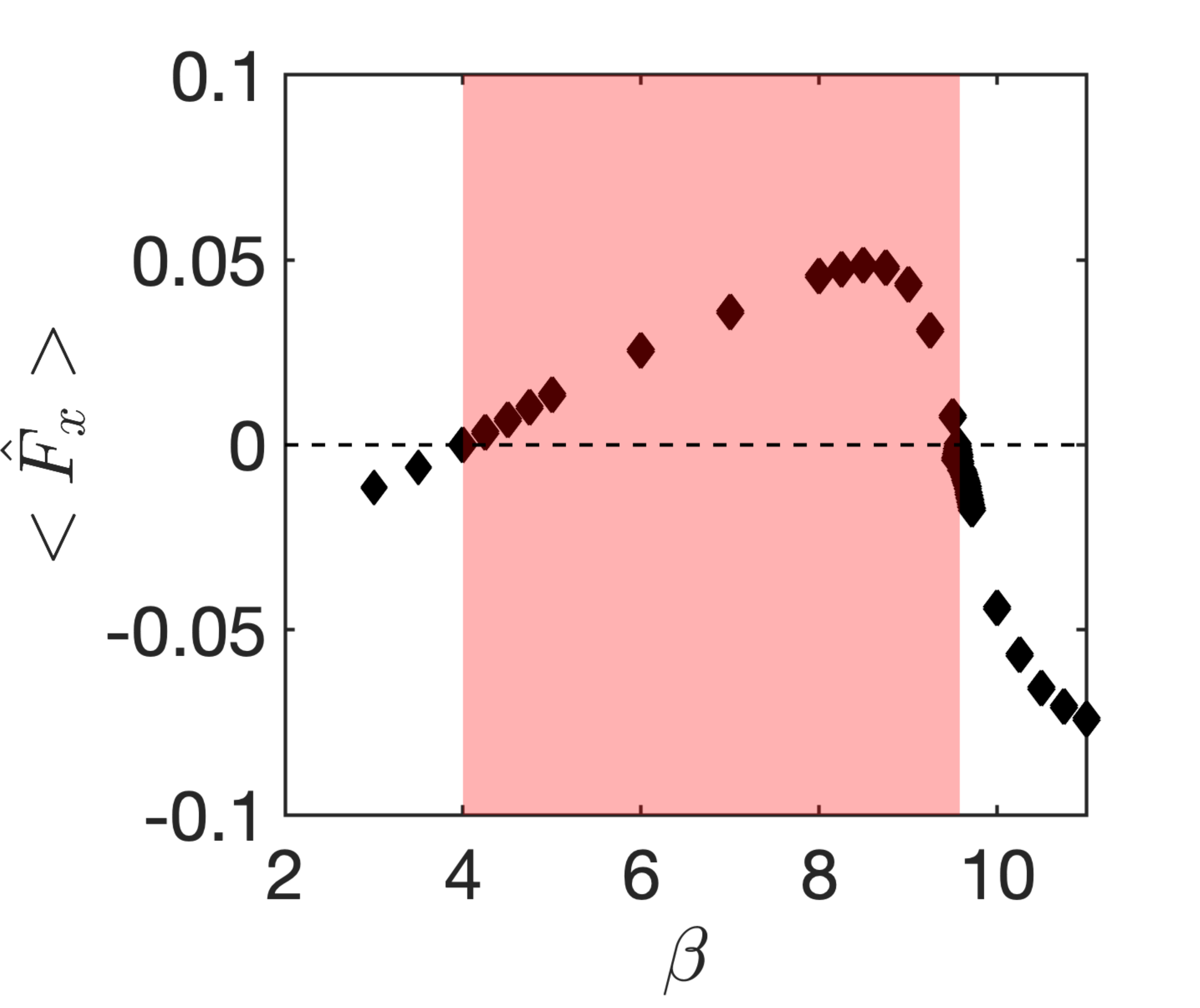}
&   \includegraphics[width=0.32\linewidth]{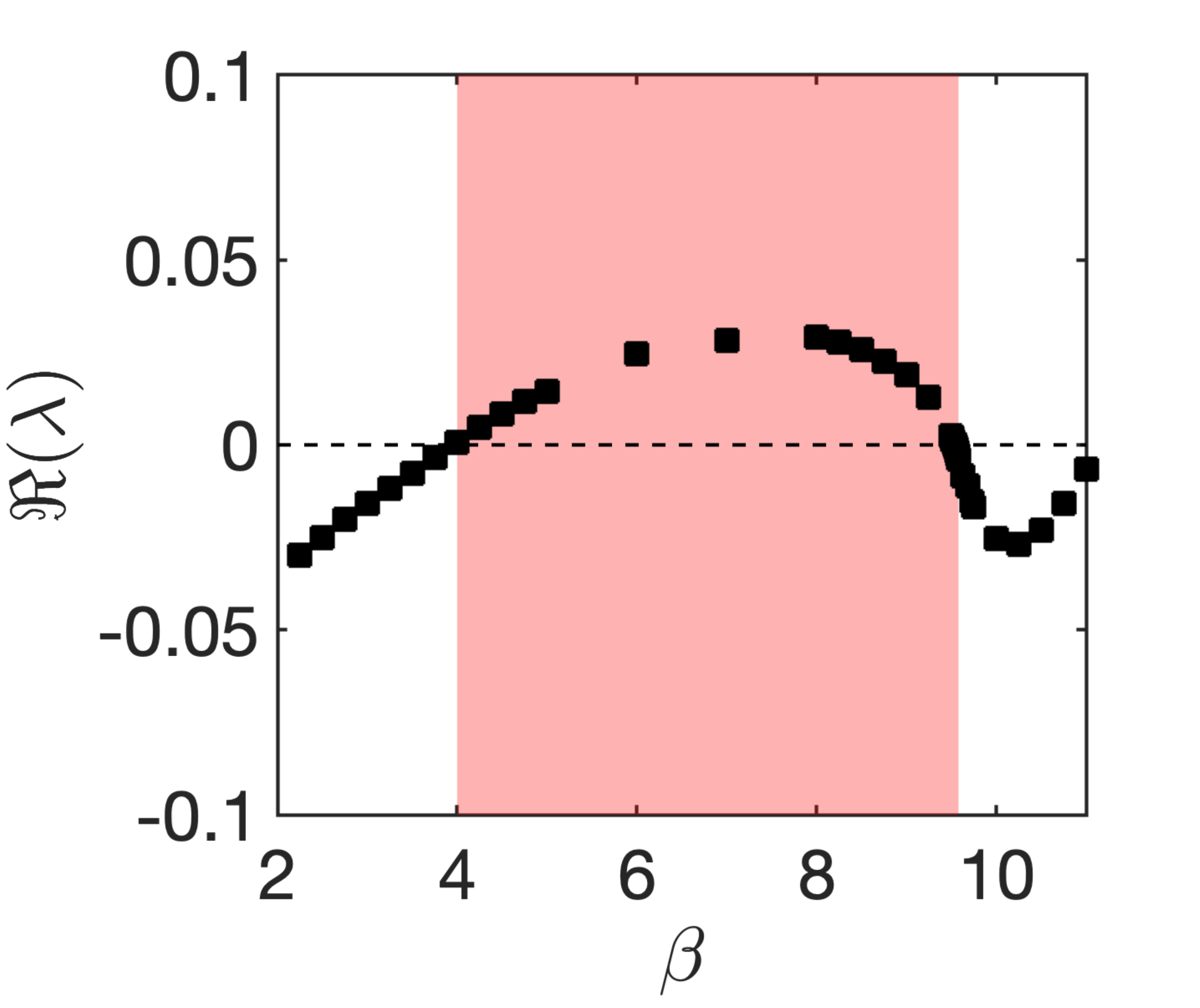}
   \end{tabular}
    \caption{Time-averaged horizontal (a) velocity and (b) force for the synchronous Floquet mode as a function of $\beta$. (c) Real part of the Floquet exponent. In (a) the mean horizontal speed obtained with nonlinear simulations is represented with empty grey circles for comparison.}
            \label{Fig-Mean-Force-Speed-Modes-Syn}
\end{figure}

To stress again the role of the fluid-solid coupling in the destabilization of the mode, we consider the time-averaged analysis of the Floquet mode exposed  in \S \ref{subsec:timeaveragedanalysis}. For synchronous modes, it was shown that  their growth rate is given by
\begin{equation*}
    \lambda_{r} = \frac{1}{\rho S} \frac{\left< \hat{F}_x\right>}{\left<\hat{u}_g\right>} \;, 
\end{equation*}
i.e. the ratio between the time-averaged horizontal force and velocity, weighted by the foil mass. These quantities are plotted  
in figure \ref{Fig-Mean-Force-Speed-Modes-Syn} as a function of $\beta$.
The time-averaged horizontal velocity, shown with black circles in (a), is positive for all values of the Stokes number. We note that its evolution is different from the time-averaged velocity of the foil computed with temporal simulation (open circles), indicating the importance of flow nonlinearities in the terminal foil velocity. Examining now the horizontal force in figure \ref{Fig-Mean-Force-Speed-Modes-Syn}(b), its changes of sign correspond clearly to the destabilization and stabilization of the Floquet mode (c). When this force is positive (resp. negative), the growth rate is also positive (resp. negative), in agreement with the above relation (recalling that the horizontal velocity is always positive). Finally, we conclude that this synchronous Floquet mode is responsible for the emergence of the unidirectional propulsion solution obtained in the region II, delimited in red in the figure. Retaining the fluid-solid coupling at the perturbation level is fundamental to explain the destabilization of this mode.

\subsubsection{Asynchronous Floquet modes: emergence of back \& forth solution}\label{sec:asynchronous}

\begin{figure}
   \centering
         \begin{subfigure}{0.24\textwidth}
    \includegraphics[width=\linewidth]{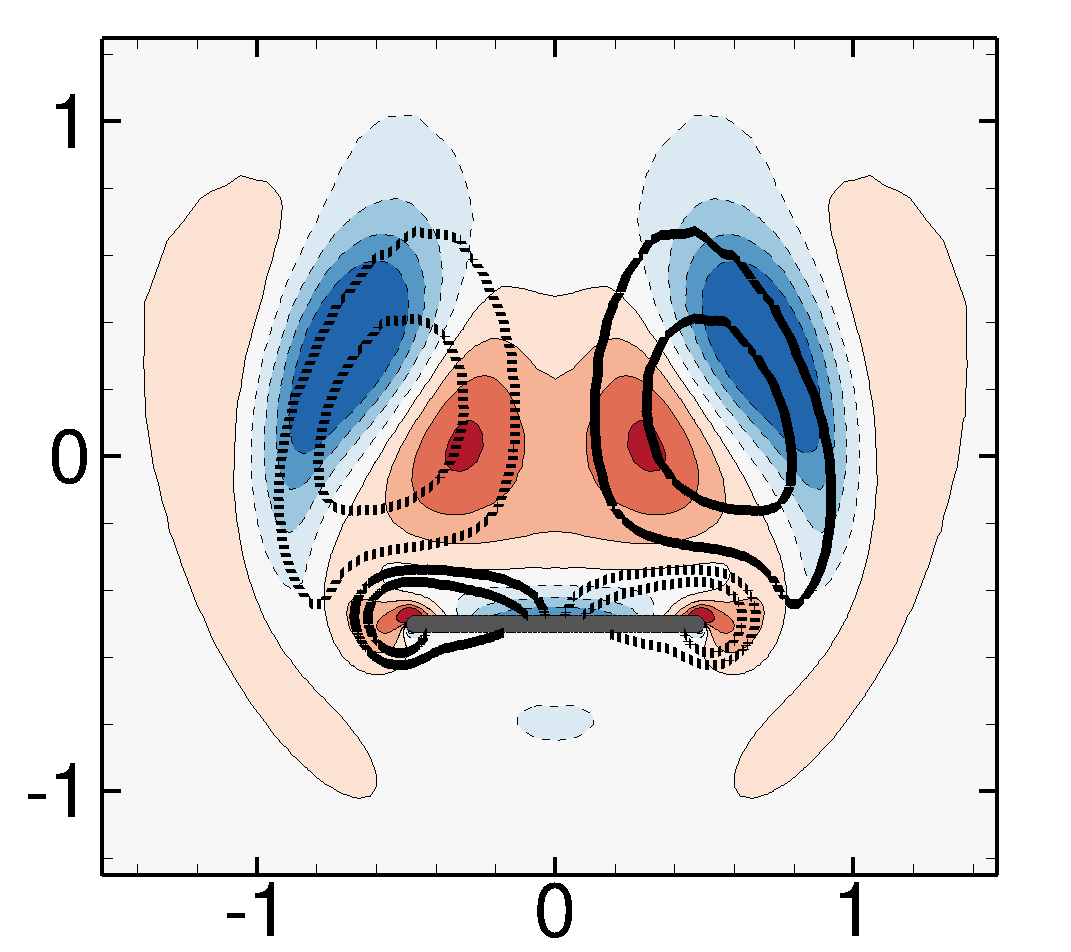}
    \caption{$t=0$}
    \end{subfigure}
       \begin{subfigure}{0.24\textwidth}
    \includegraphics[width=\linewidth]{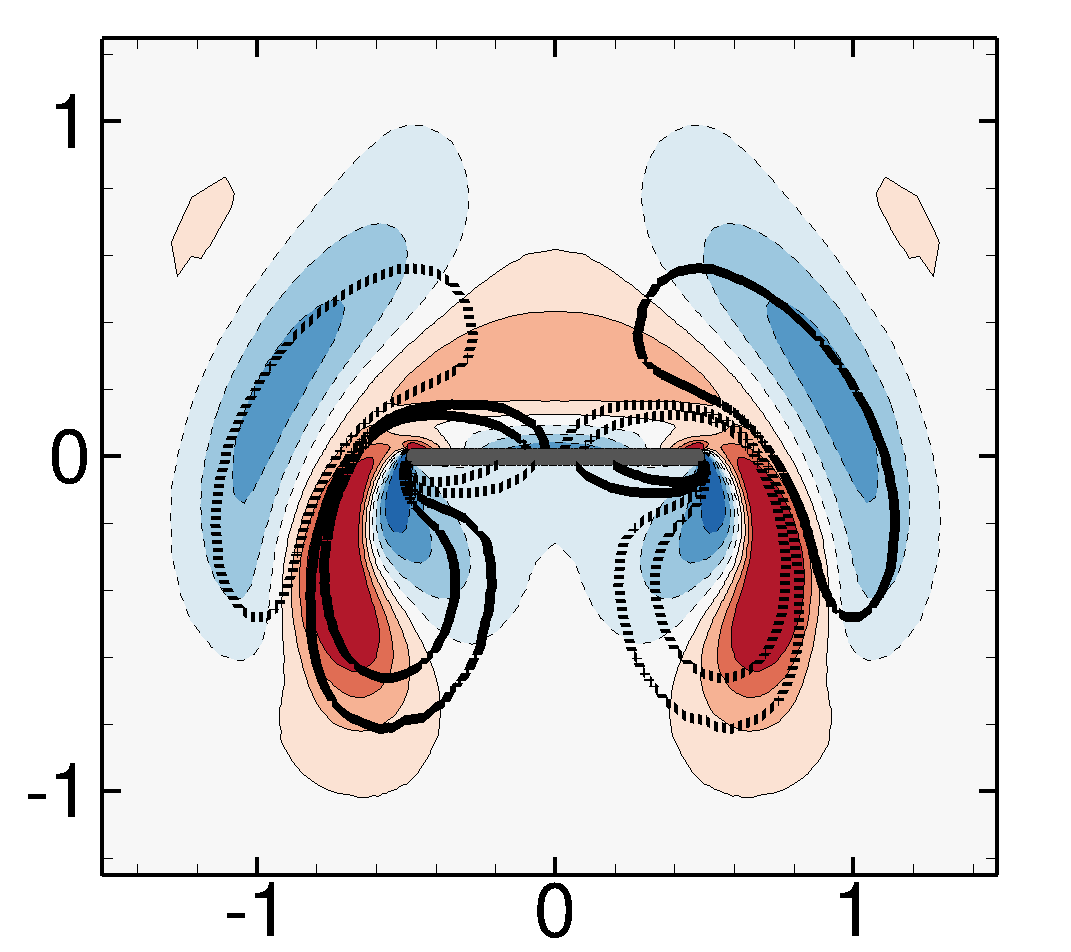}
    \caption{$t=1/4$}
    \end{subfigure}
       \begin{subfigure}{0.24\textwidth}
    \includegraphics[width=\linewidth]{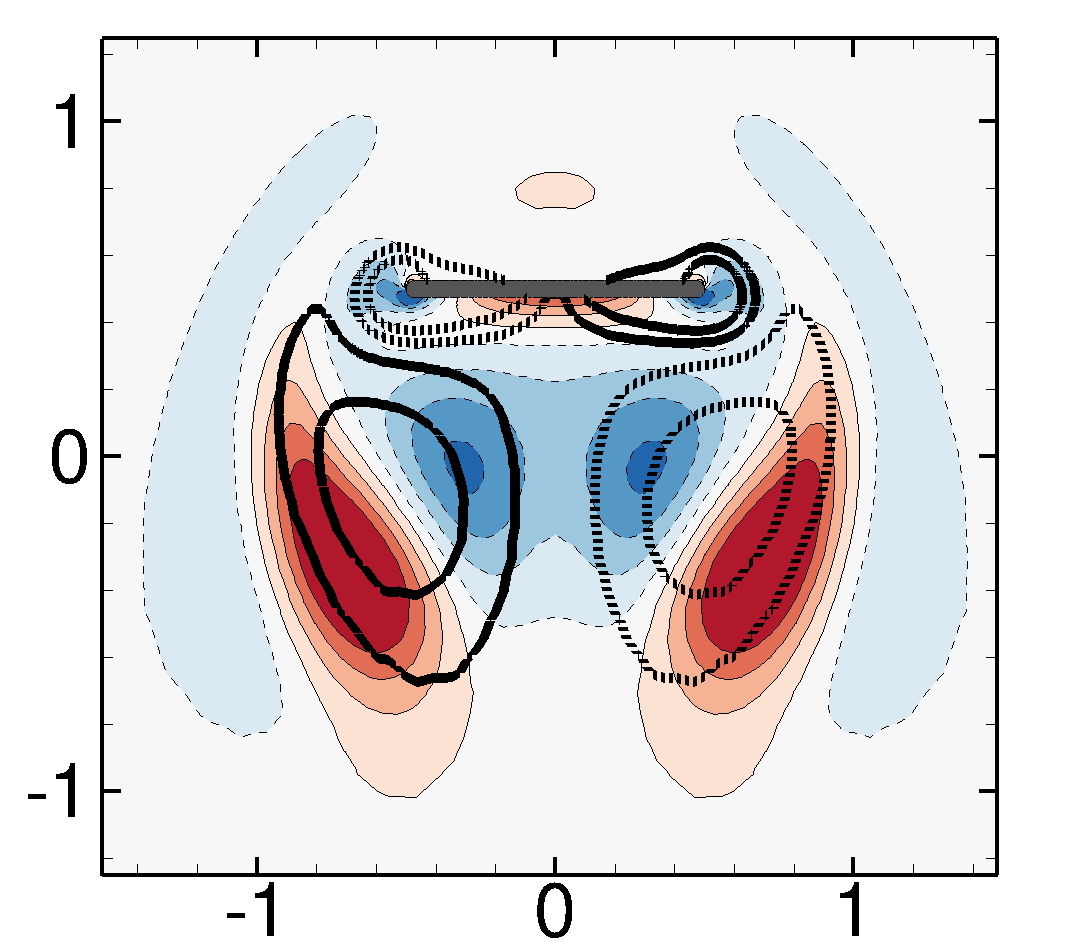}
    \caption{$t=1/2$}
    \end{subfigure}
    \begin{subfigure}{0.24\textwidth}
    \includegraphics[width=\linewidth]{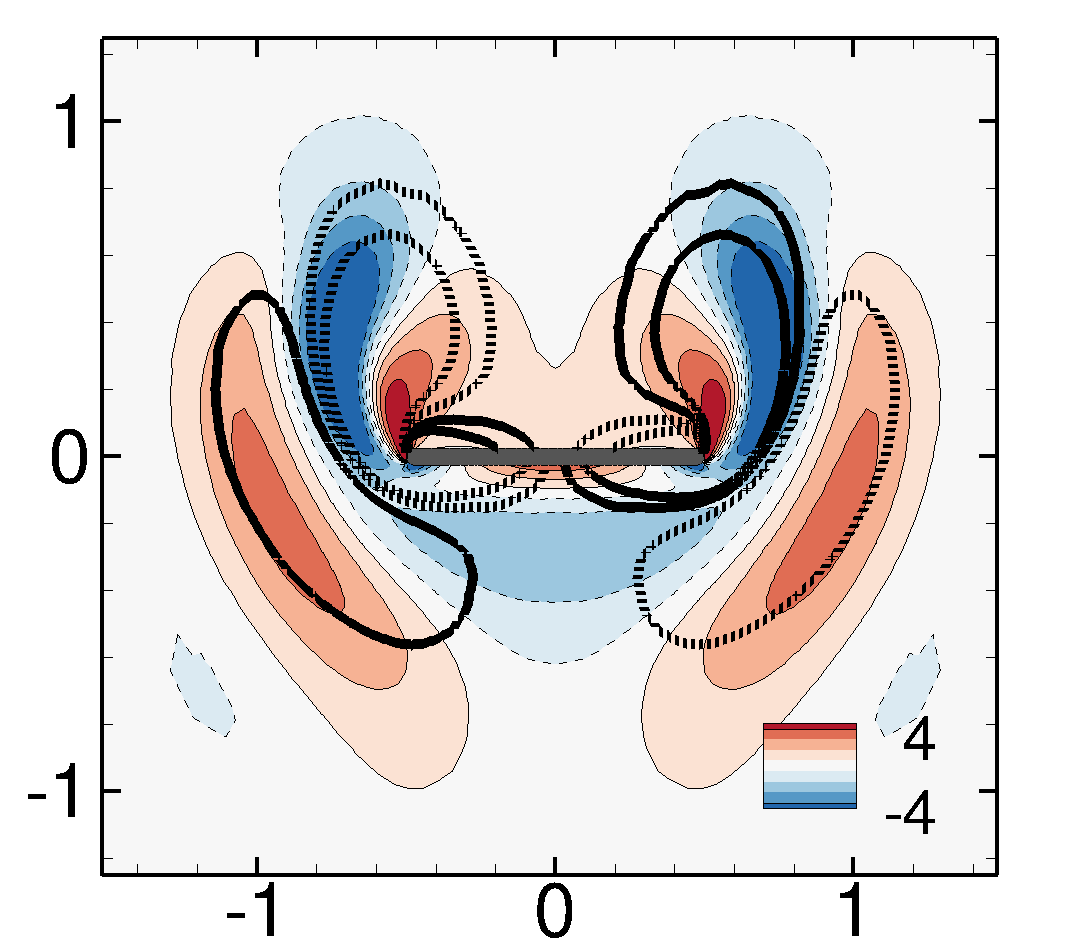}
    \caption{$t=3/4$}
    \end{subfigure}
      \begin{subfigure}{0.24\textwidth}
    \includegraphics[width=\linewidth]{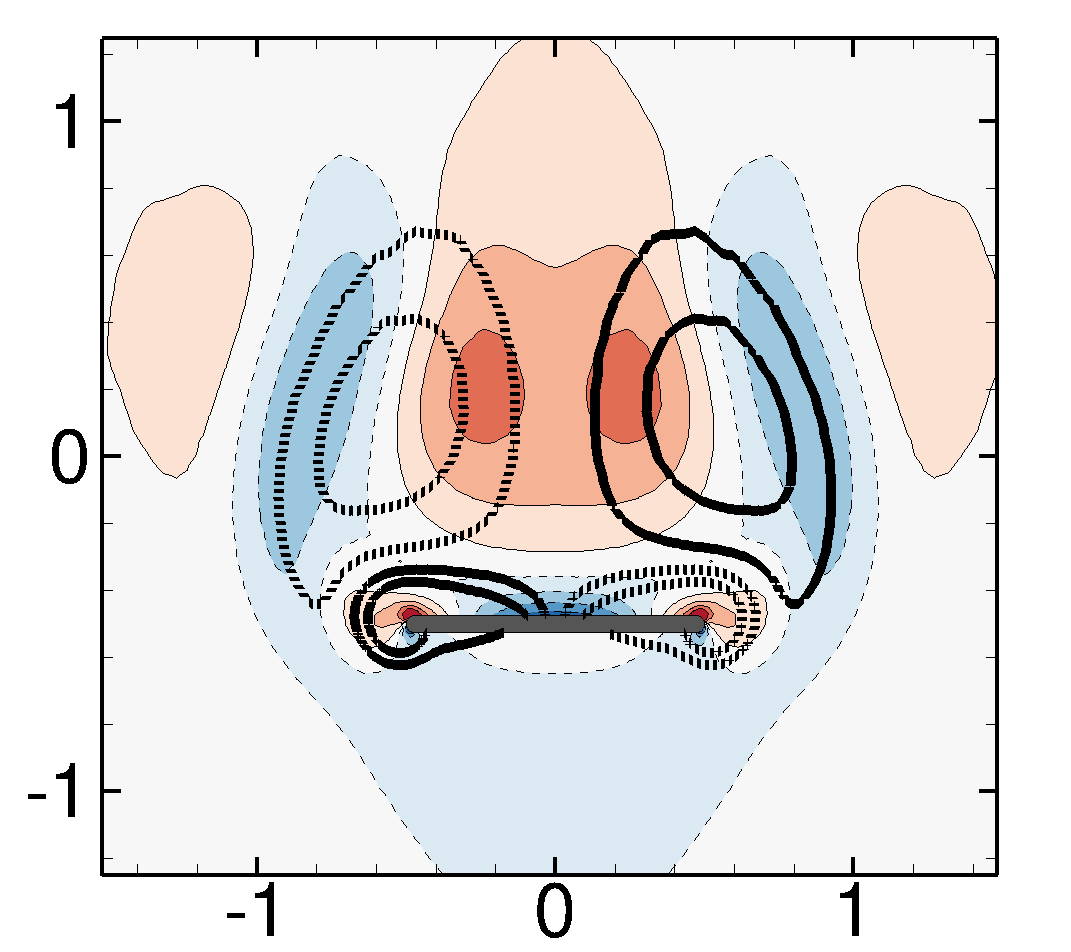}
    \caption{$t=0$}
    \end{subfigure}
      \begin{subfigure}{0.24\textwidth}
    \includegraphics[width=\linewidth]{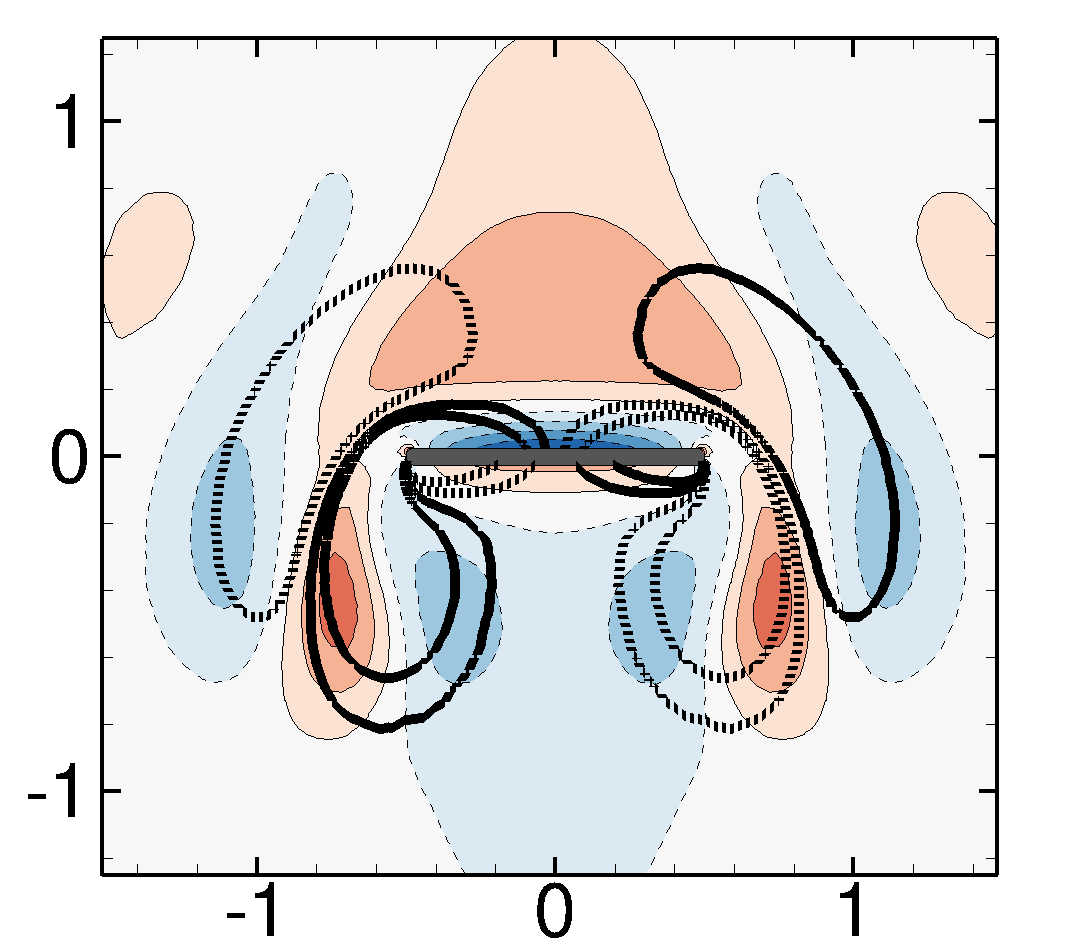}
    \caption{$t=1/4$}
    \end{subfigure}
      \begin{subfigure}{0.24\textwidth}
    \includegraphics[width=\linewidth]{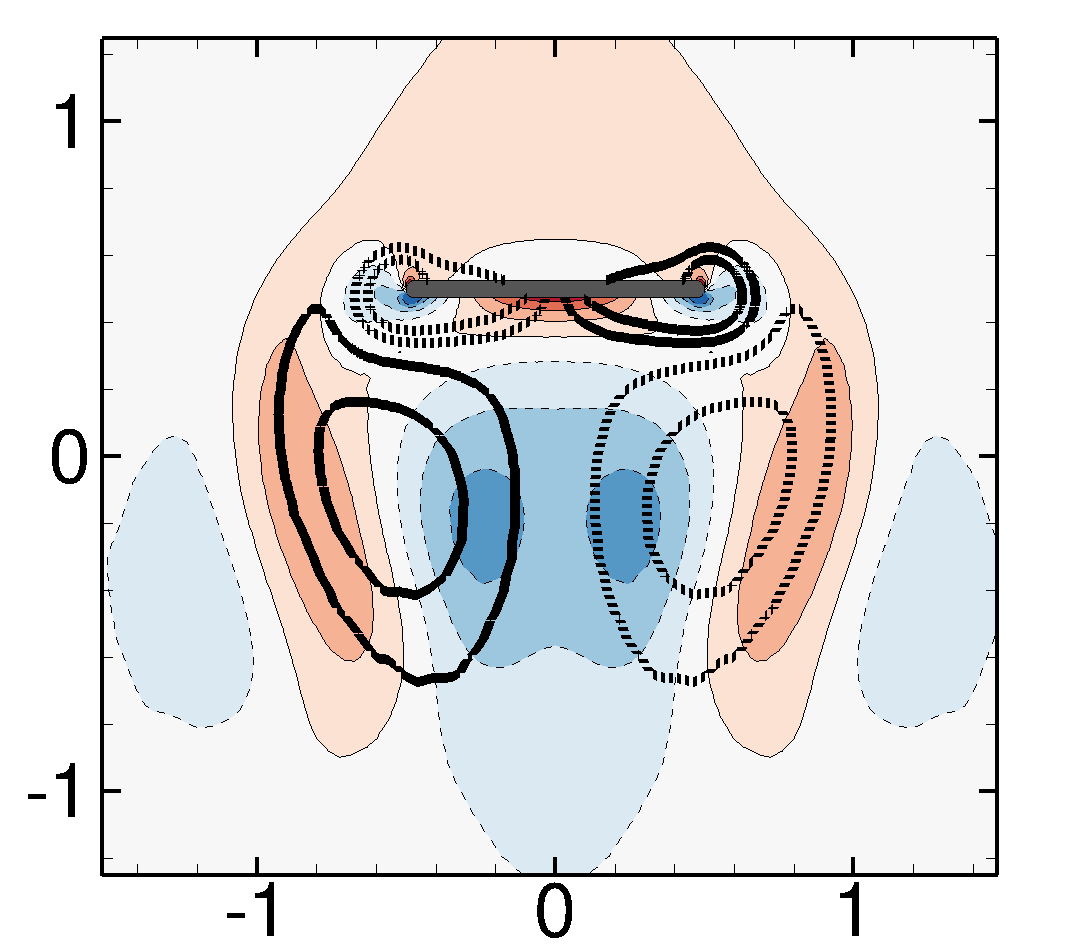}
    \caption{$t=1/2$}
    \end{subfigure}
    \begin{subfigure}{0.24\textwidth}
    \includegraphics[width=\linewidth]{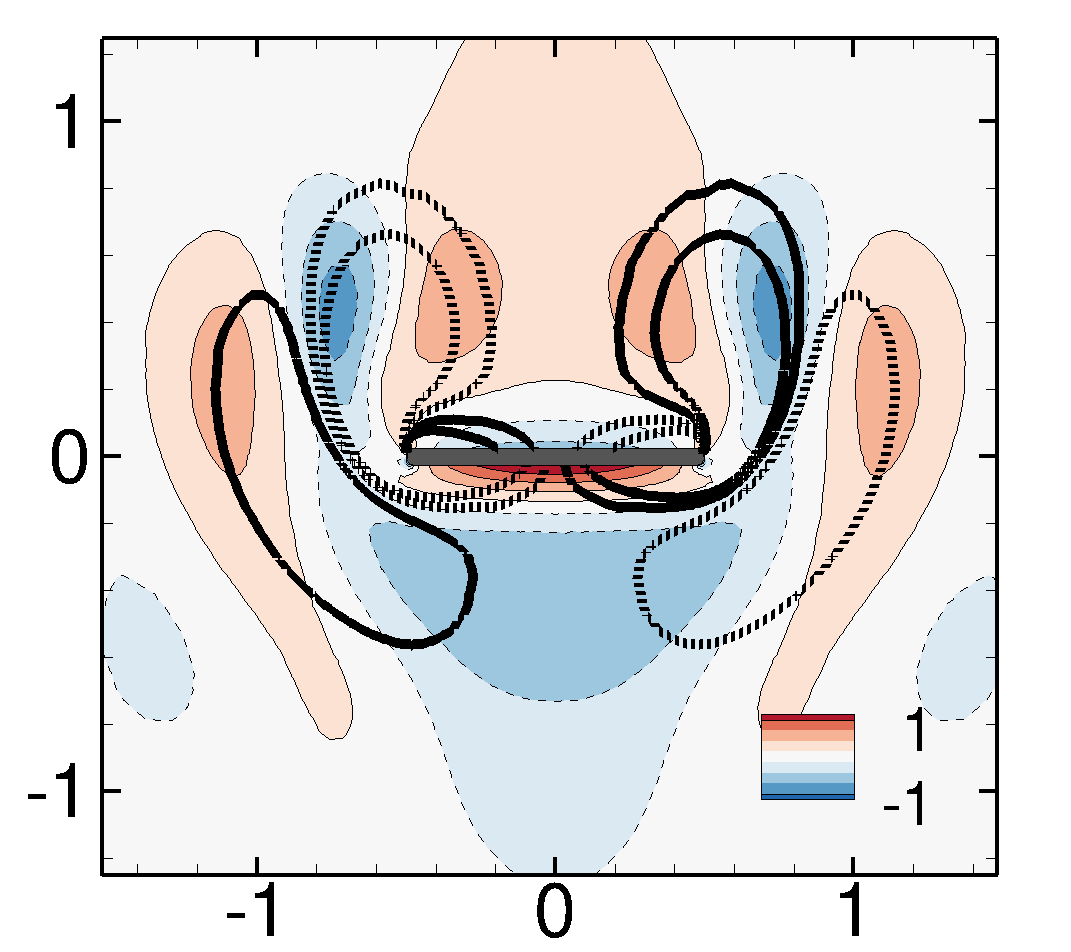}
    \caption{$t=3/4$}
    \end{subfigure}
    \begin{subfigure}{0.24\textwidth}
    \includegraphics[width=\linewidth]{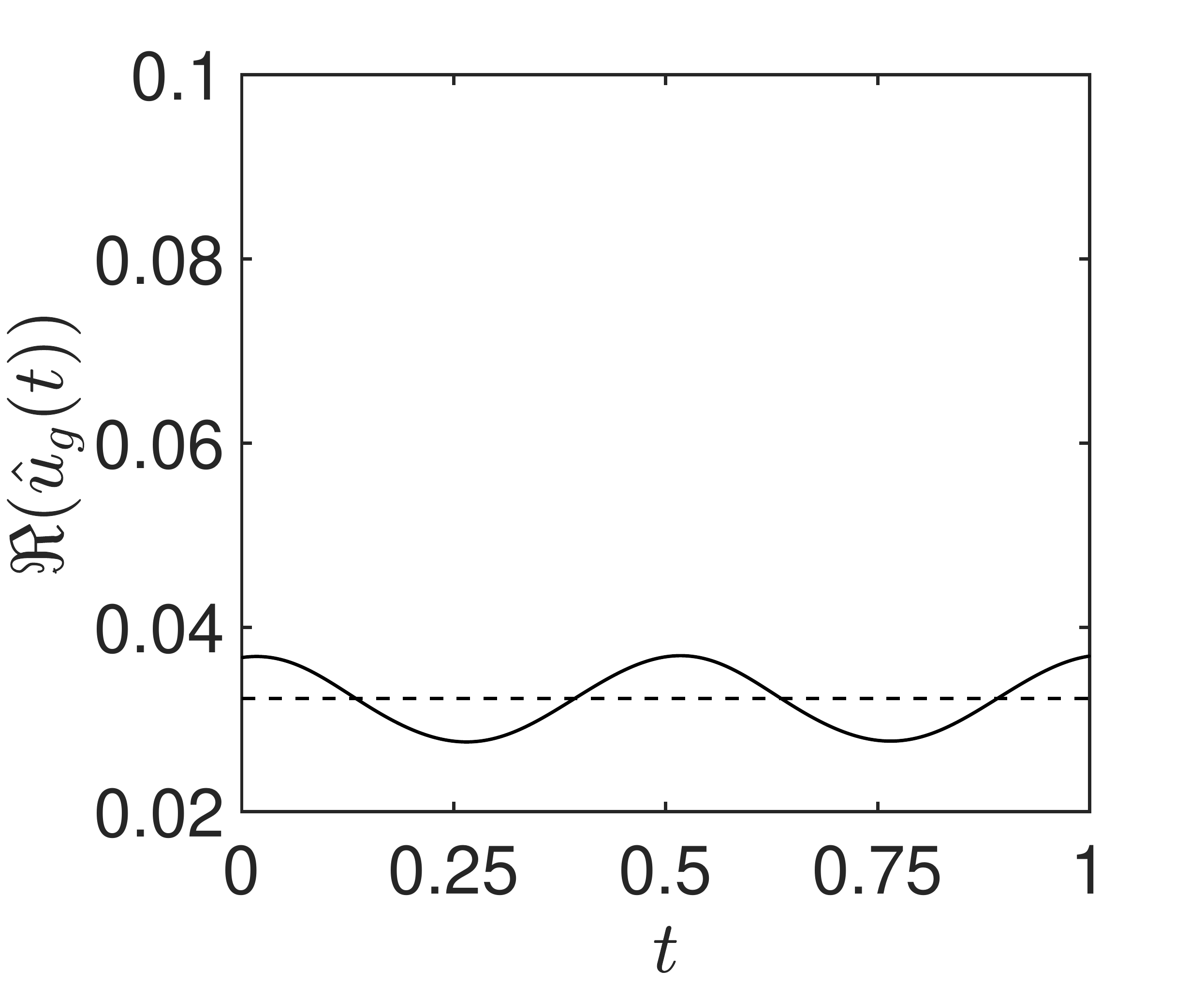}
    \caption{}
    \end{subfigure}
    \begin{subfigure}{0.24\textwidth}
    \includegraphics[width=\linewidth]{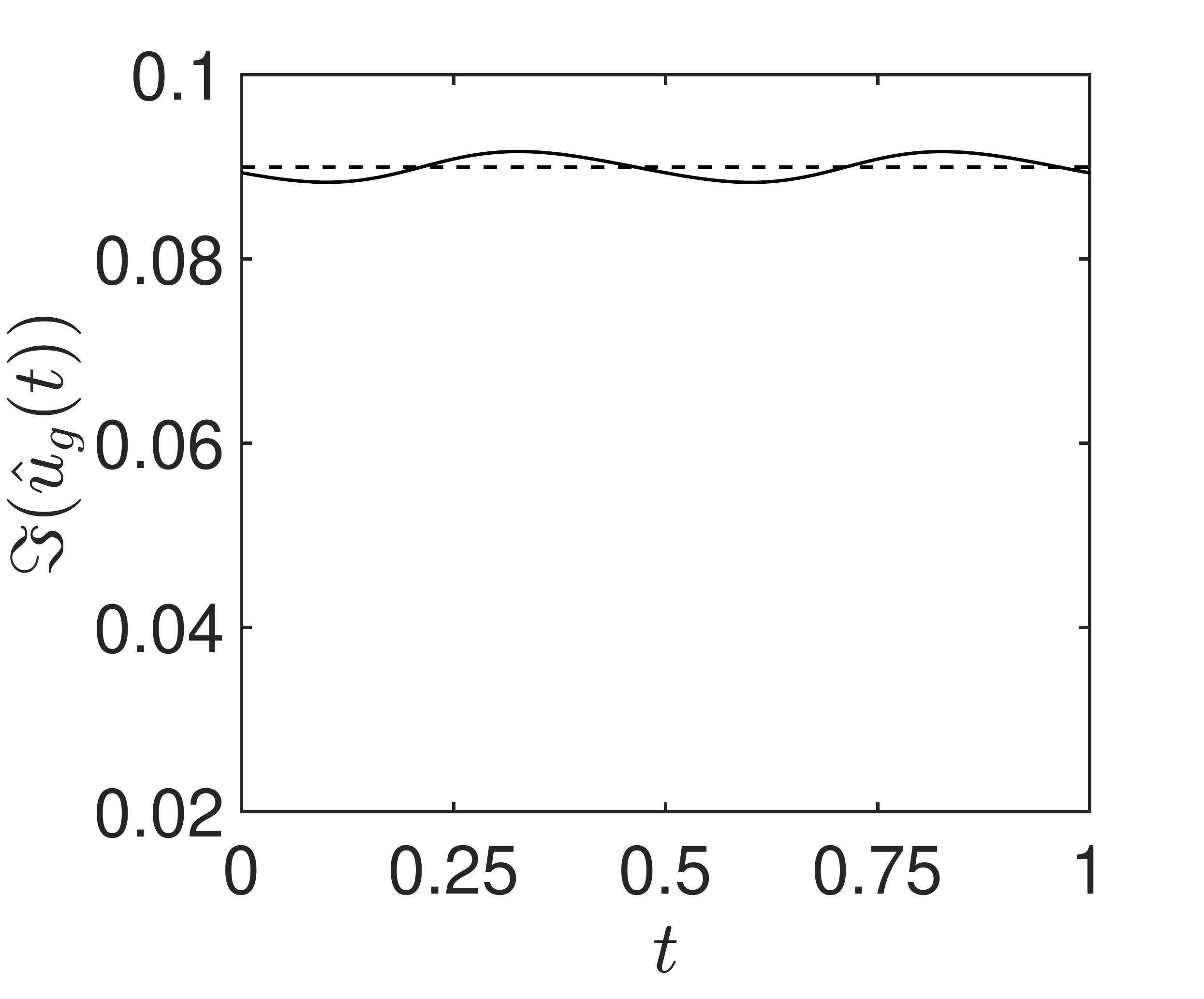}
    \caption{}
    \end{subfigure}
    \begin{subfigure}{0.24\textwidth}
    \includegraphics[width=\linewidth]{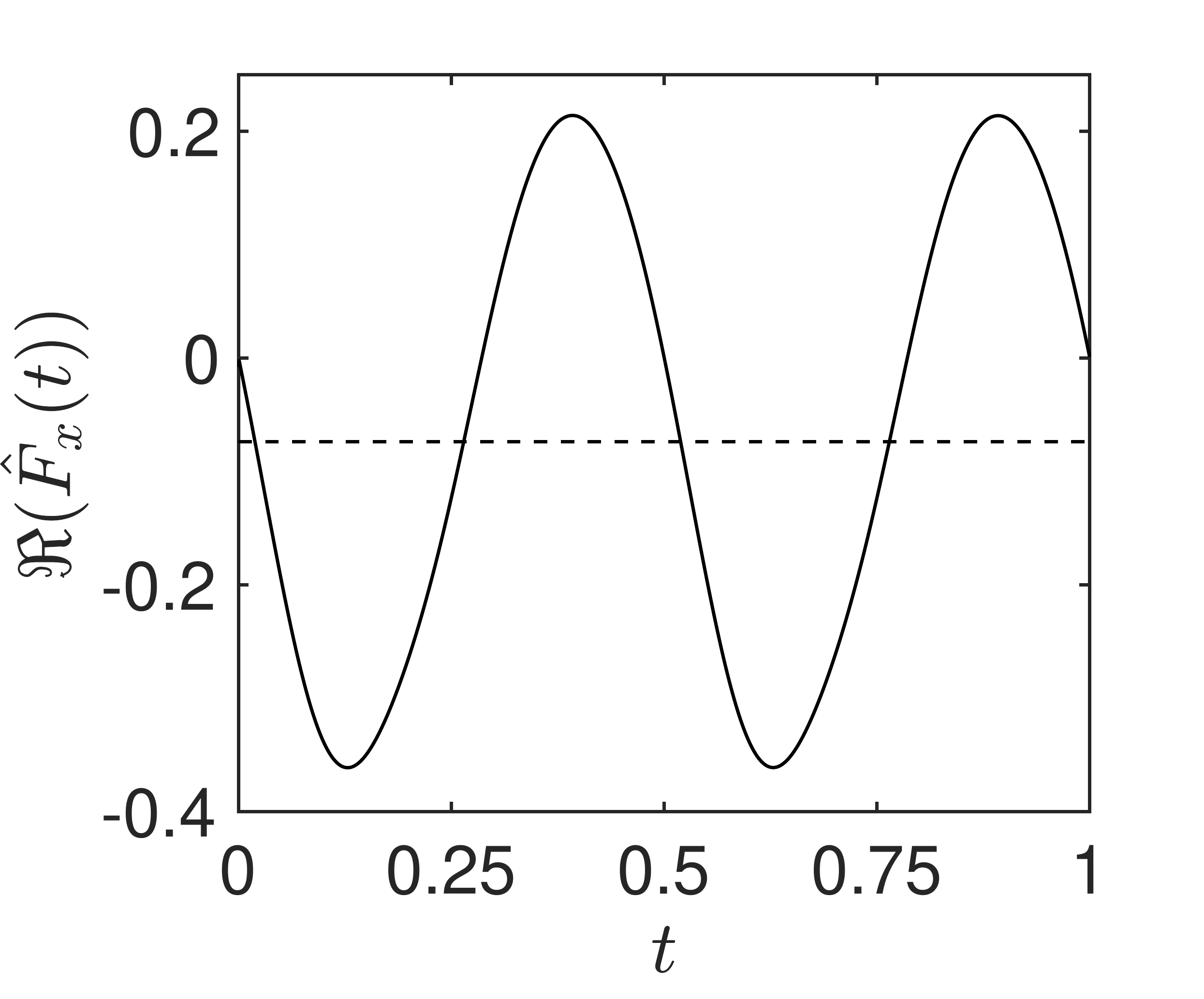}
    \caption{}
    \end{subfigure}
    \begin{subfigure}{0.24\textwidth}
    \includegraphics[width=\linewidth]{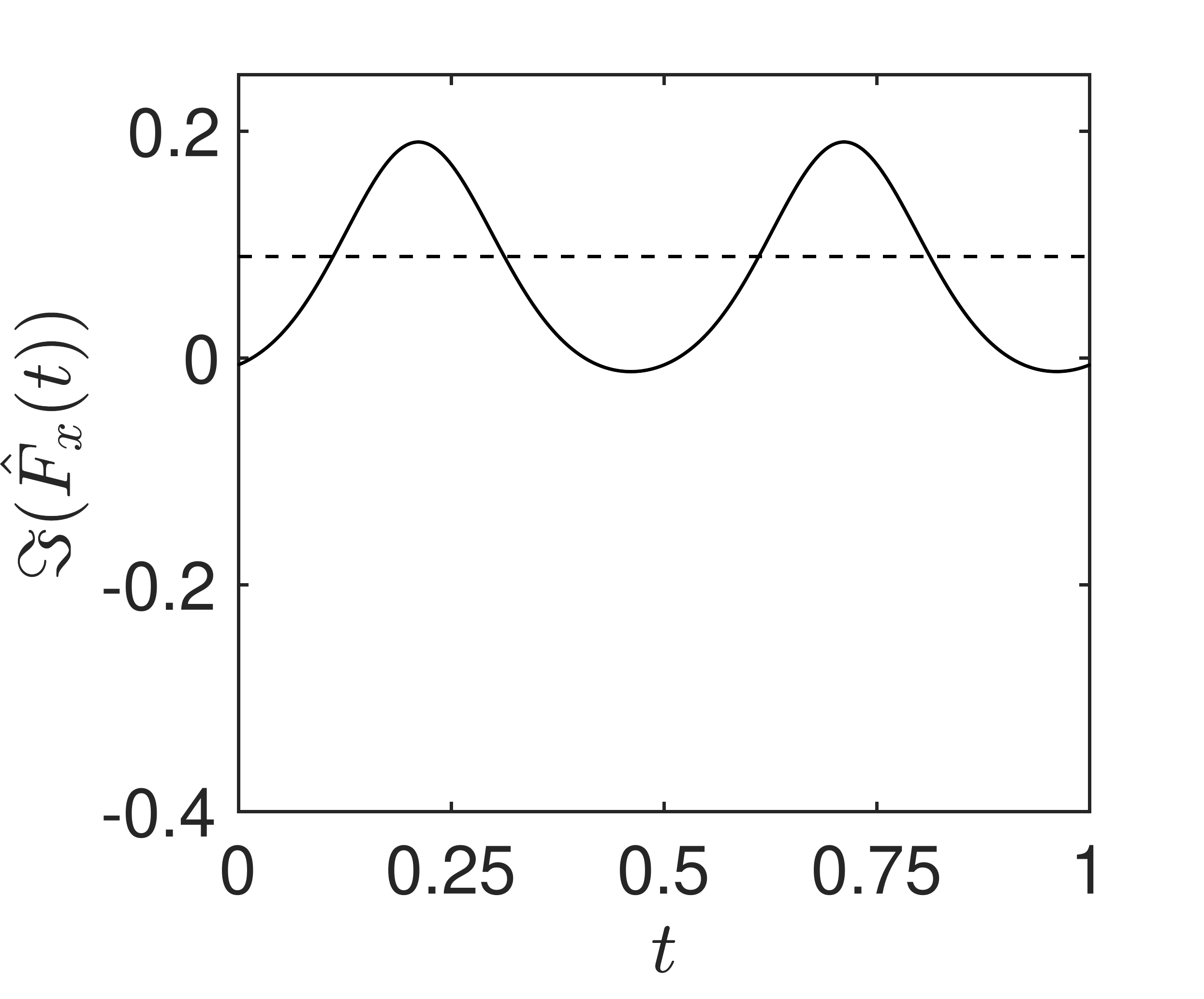}
    \caption{}
    \end{subfigure}

    \caption{Unstable asynchronous Floquet mode for $\beta=13$. Vorticity contours of the (a-d) real and (e-h) imaginary parts of the mode at four instants of the flapping period. Positive and negative values of the base-flow vorticity are depicted with solid and dashed contours. (i-j) Real and imaginary parts of the horizontal velocity $\hat{u}_{g}$. (k-l) Real and imaginary part of the horizontal force $\hat{F}_{x}$.  In (i-l), dashed lines represents the time-averaged value of the plotted quantity.}
            \label{3-Linear-FloquetMode-Beta13}
\end{figure}

We now examine the asynchronous Floquet modes that gets unstable for larger Stokes number. The complex mode, obtained at $\beta=13$ and displayed in figure \ref{3-Linear-FloquetMode-Beta13}, also breaks the left-right symmetry and satisfies the spatio-temporal symmetry \eqref{spatiotemporalsymmetry}. The instantaneous real (resp. imaginary) part of the vorticity is shown in figures (a-d) (resp. e-h) at four instants of the flapping period. 
The amplitude of the real part is noticeably larger than that of the imaginary part, and their spatial structures are quite different. The real part of the mode bears similarities with the  synchronous Floquet mode found by \cite{Jallas2017} to explain the deviation of propulsive wakes in flapping wings and the  displacement modes of vortices \citep{Fabre2006,Brion2014}. Let us consider the right solid dark line representing the base flow vortex of positive vorticity in figure \ref{3-Linear-FloquetMode-Beta13}(a). The perturbation has positive vorticity on the lower left and negative vorticity on the upper right part of the monopole. This superposition strengthens the lower left part of the monopole while weakening the upper right one, resulting in a net displacement of its center to the lower left. The displacement of the dipolar vortex structure results in an horizontal force exerted on the foil whose temporal evolution is shown in figure \ref{3-Linear-FloquetMode-Beta13} (k). Due to the spatio-temporal flow symmetry, the frequency is twice the flapping frequency.  Interestingly, the horizontal force strongly oscillates around a negative time-averaged value.
Compared to the oscillation of the foil horizontal velocity displayed in figure \ref{3-Linear-FloquetMode-Beta13}(i), we first note that they are not in phase. As explained later, this phase difference is crucial to understand the destabilization of the asynchronous mode. Then, we also remark that the fluctuation of the horizontal velocity is much smaller and around a time-averaged value that is positive. Therefore, the real part of the asynchronous mode produces a mean resistive force over the flapping period, that decreases the foil horizontal velocity. \\

Turning now to the imaginary part of the asynchronous mode depicted in figure \ref{3-Linear-FloquetMode-Beta13}(e-h), its spatial structure is of much smaller amplitude than for the real part. It looks like a combination between the synchronous mode (see figure   \ref{3-Linear-FloquetMode-Beta6}), responsible for the unidirectional self-propulsion of the foil, and the real part of the asynchronous mode, which creates a mean resistive force during a flapping period. The temporal evolution of the foil velocity and horizontal force are displayed in figure \ref{3-Linear-FloquetMode-Beta13}(j) and (l), respectively. The fluctuation of the force (l) is now much smaller than for the real part (k). The real and imaginary horizontal forces are out-of-phase by $1/4$. During the upstroke of the foil ($t<0.5$), the minimal and maximal values of the imaginary horizontal force are obtained at $t=1/2$ and $t=1/4$, respectively, while they are obtained at  $t=1/8$ and $t=3/8$ for the real part.
Interestingly, the time-averaged value of the imaginary part is now positive, as for the horizontal velocity (j). Therefore, the imaginary part of this asynchronous mode produces a mean propulsive force that increases the foil velocity.   \\

\begin{figure}
   \centering
      \begin{subfigure}{0.24\textwidth}
    \includegraphics[width=\linewidth]{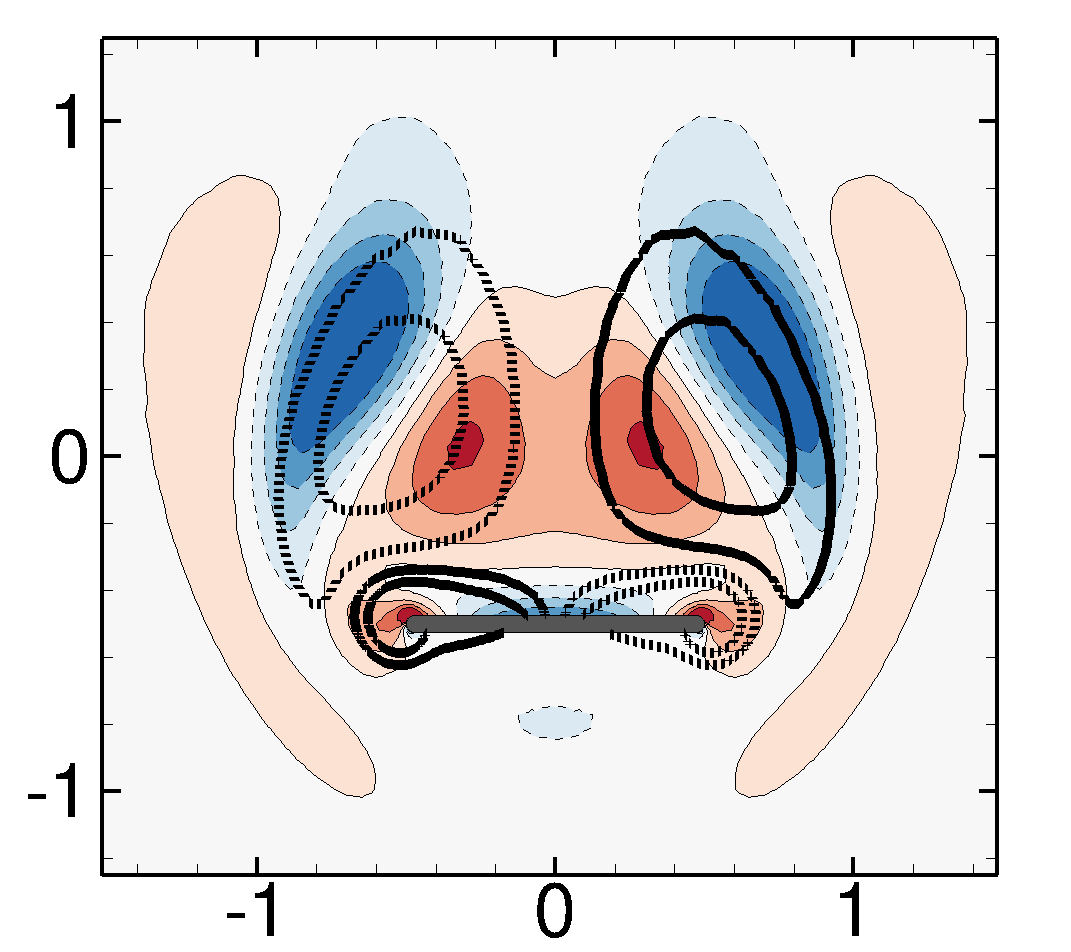}
    \caption{$t_{\epsilon}=0$}
    \end{subfigure}
       \begin{subfigure}{0.24\textwidth}
    \includegraphics[width=\linewidth]{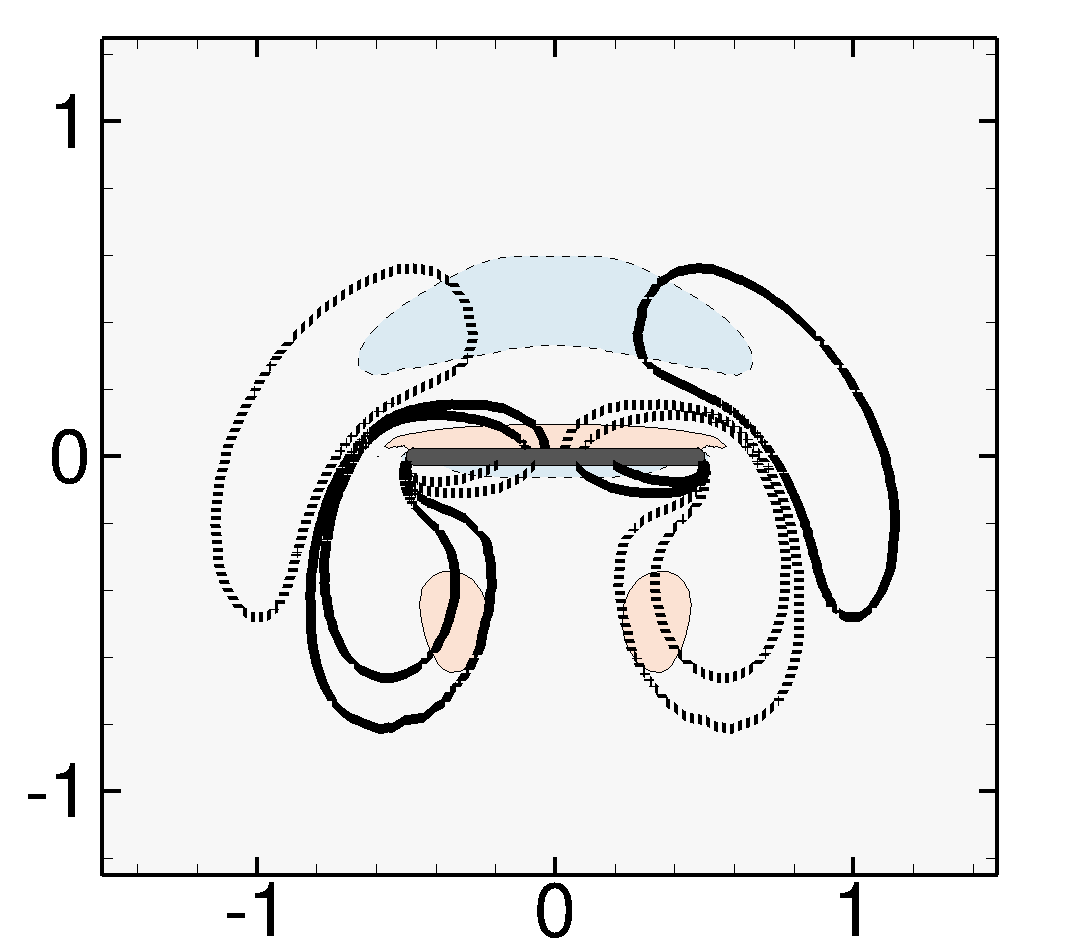}
    \caption{$t_{\epsilon}=0.25$}
    \end{subfigure}
       \begin{subfigure}{0.24\textwidth}
    \includegraphics[width=\linewidth]{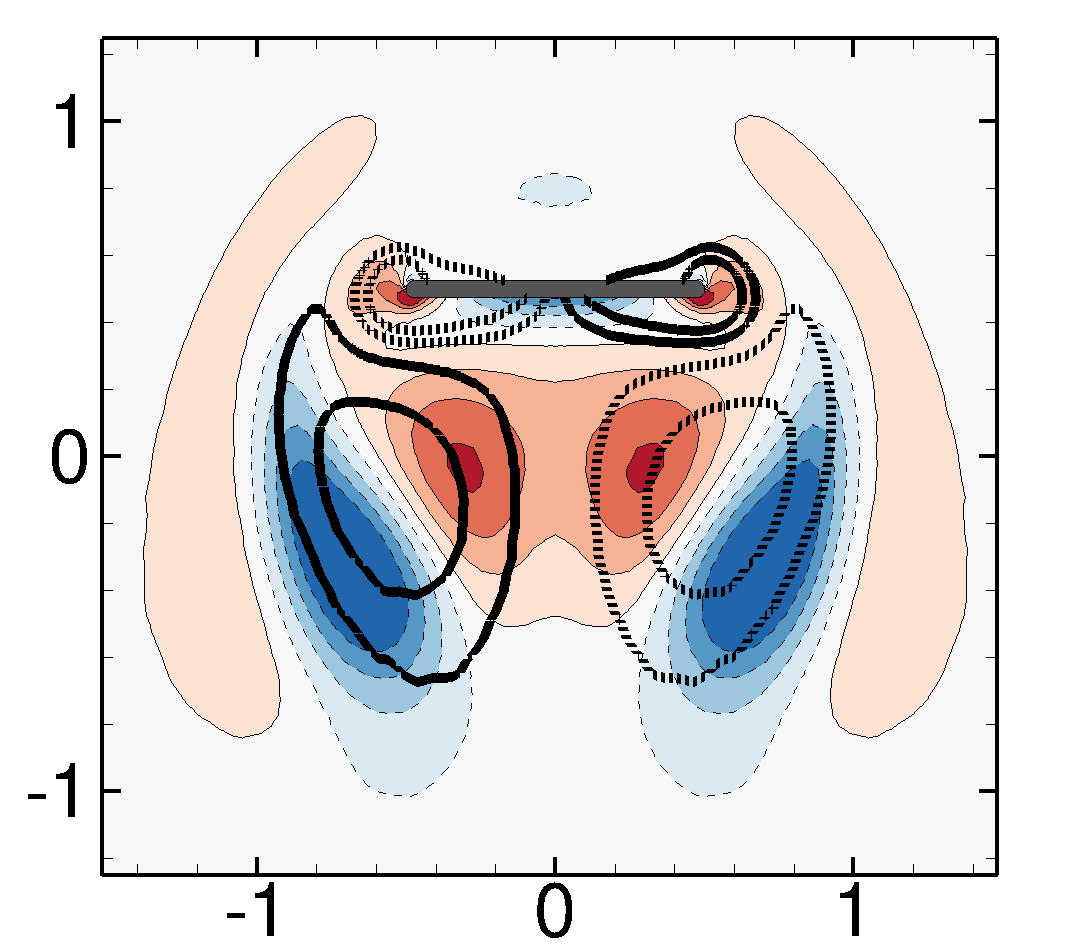}
    \caption{$t_{\epsilon}=0.5$}
    \end{subfigure}
    \begin{subfigure}{0.24\textwidth}
    \includegraphics[width=\linewidth]{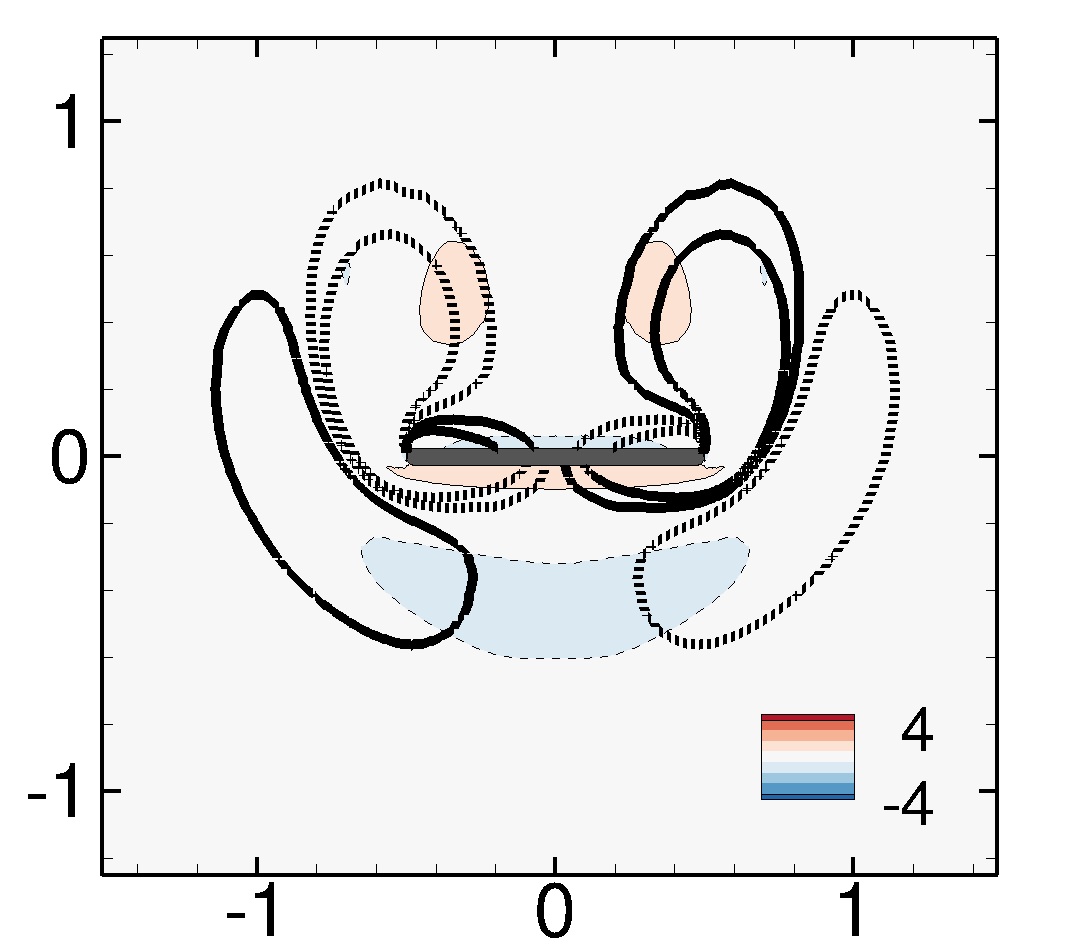}
    \caption{$t_{\epsilon}=0.75$}
    \end{subfigure}
   
     \begin{subfigure}{0.32\textwidth}
       \includegraphics[width=\linewidth]{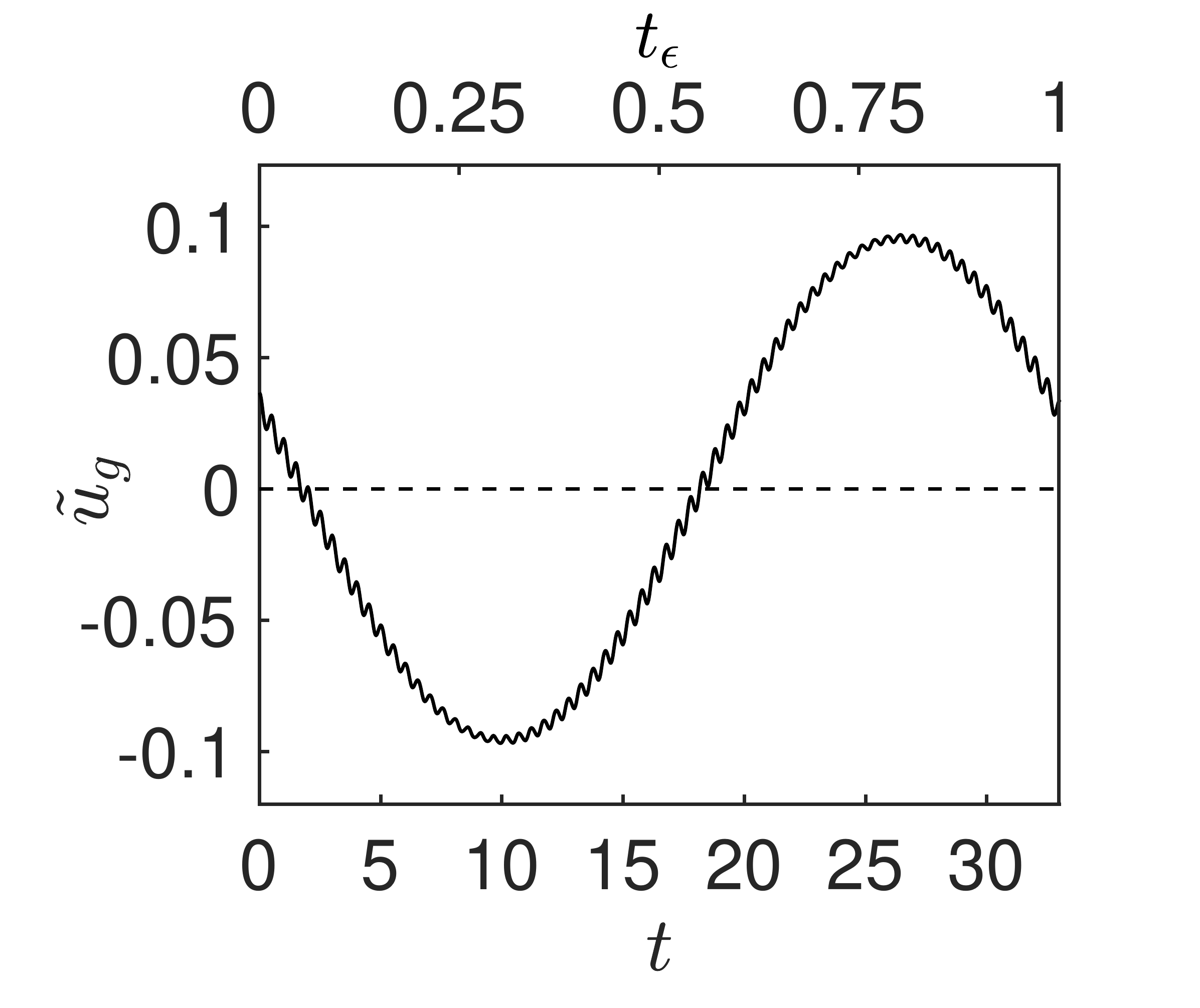}
    \caption{}
    \end{subfigure}
    \begin{subfigure}{0.32\textwidth}
        \includegraphics[width=\linewidth]{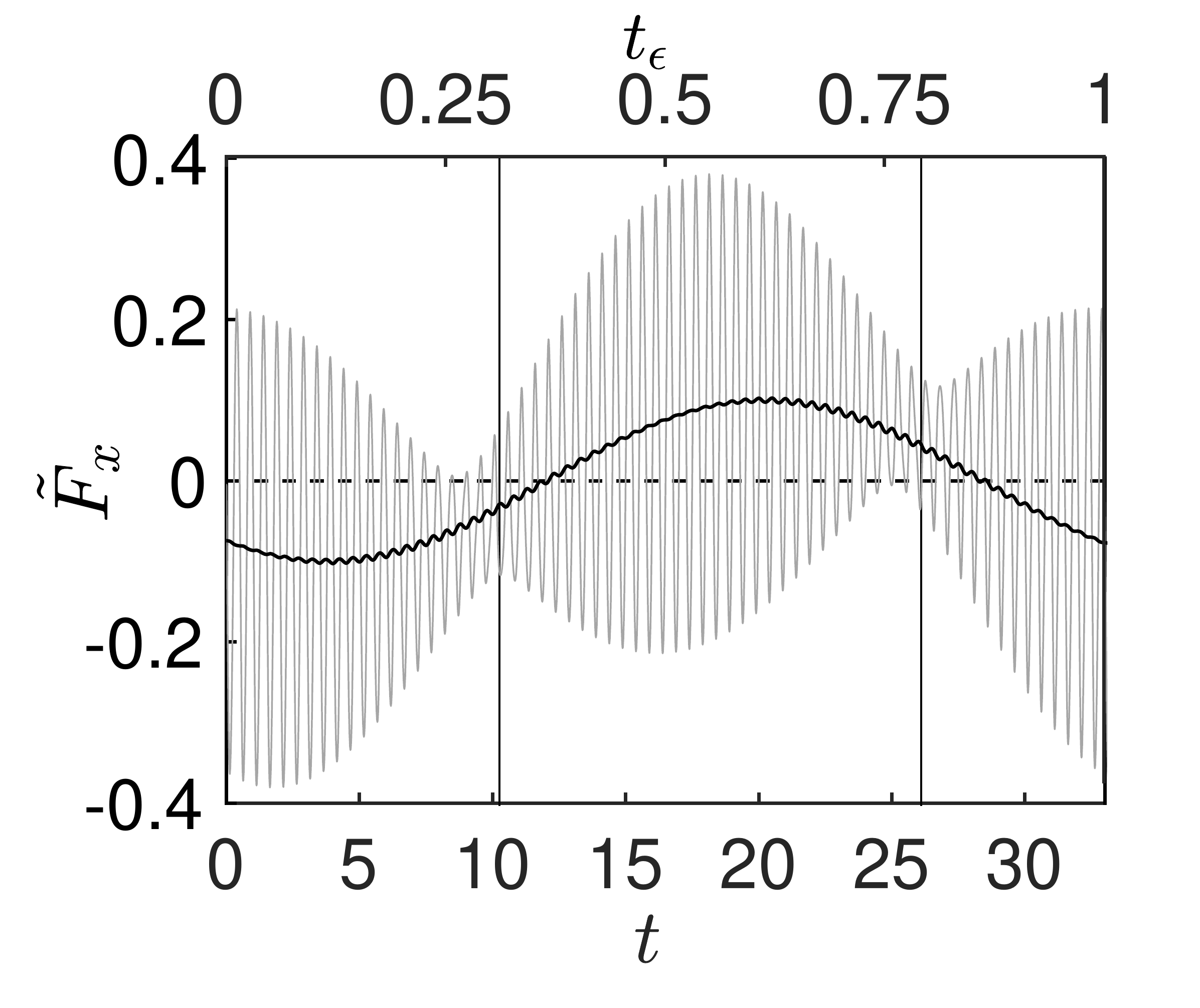}
    \caption{}
    \end{subfigure}
        \begin{subfigure}{0.32\textwidth}
        \includegraphics[width=\linewidth]{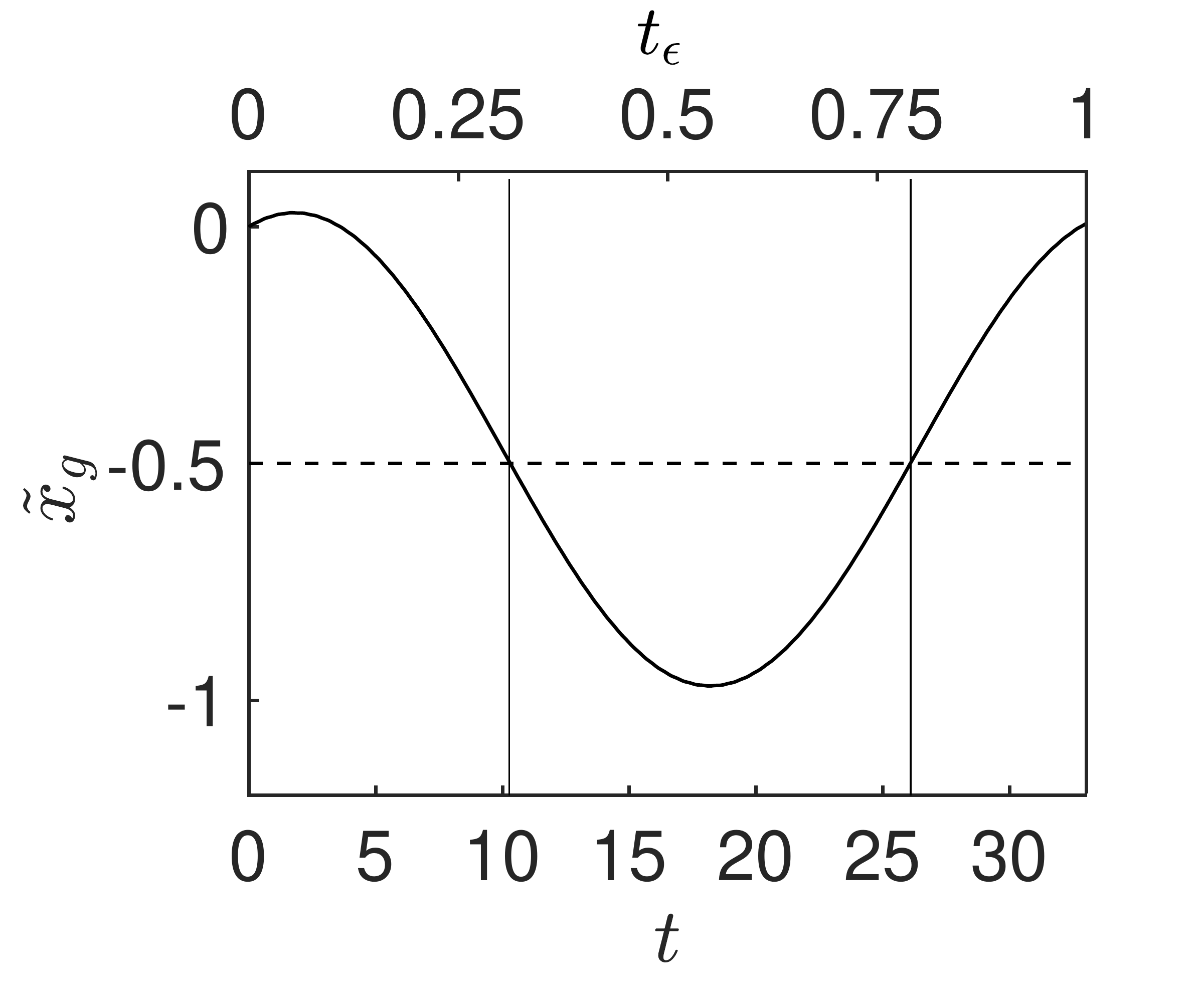}
    \caption{}
    \end{subfigure}
\caption{Temporal evolution of the quasi-periodic perturbation $\mathbf{\tilde{q}}$ for $\beta=13$. (a-d) Vorticity of the perturbation  (colors) and base flow (black lines) at four instants $t_{\epsilon}$ of the slow period $T_{\epsilon}=1/f'=33$. (e-g) Time evolution of the horizontal (e) velocity , (f) force and (g) position of the foil over the slow period, shown as a function of both time $t$ and $t_{\epsilon}$. The horizontal dashed lines are for the time-averaged value of the plotted quantity over the slow time period. In (f), the thick curve depicts the time-averaged value of the horizontal force over the flapping period. The vertical solid lines in (f-g) indicate the instants where the foil pass by its mean horizontal position.}
            \label{3-Linear-FloquetMode-Beta13-TimeEvol}
\end{figure}

To further understand the contrasting actions of the real and imaginary parts of the Floquet mode, we introduce the real \textit{quasi-periodic} perturbation defined as $\mathbf{\tilde{q}} = \mathbf{q}' e^{-\lambda_r t}$. Compared to the real perturbation $\mathbf{q}'$, the exponential growth/decay given by the real part of the Floquet exponent is counteracted. The perturbation $\mathbf{\tilde{q}}$ is quasi-periodic as it retains the low-frequency oscillation given by the imaginary part of the Floquet exponent, in addition to the high-frequency flapping period.
Using the Floquet decomposition \eqref{Floquet}, the temporal evolution of this quasi-periodic perturbation is simply given by   
\begin{equation}\label{eqn:defquasiperio}
    \mathbf{\Tilde{q}}(\mathbf{x},t,t_{\epsilon}) = \Re(\mathbf{\hat{q}})(\mathbf{x},t)\cos(2\pi t_{\epsilon}) - \Im(\mathbf{\hat{q}})(\mathbf{x},t)\sin(2\pi t_{\epsilon}) \, ,
\end{equation}
where  $t_{\epsilon} = f' t$ is  a slow time-scale, as the frequency of the Floquet mode is very small compared to the flapping frequency, i.e. $f' \ll 1$. For $t_{\epsilon} \sim 0$, the above expression shows that the quasi-periodic perturbation is dominated by the real part of the periodic Floquet mode $\Re(\mathbf{\hat{q}})(\mathbf{x},t)$, while for $t_{\epsilon} \sim 1/4$, it is dominated by its imaginary part $\Im(\mathbf{\hat{q}})(\mathbf{x},t)$. The quasi-periodic perturbation thus slowly evolves between the real and imaginary parts of the Floquet mode, on a time scale given by the low frequency of this asynchronous Floquet mode. This slow evolution is depicted in figure \ref{3-Linear-FloquetMode-Beta13-TimeEvol} for the unstable asynchronous mode at $\beta=13$. The contrasting actions of the real and imaginary parts of the Floquet mode are clearly visible in figures \ref{3-Linear-FloquetMode-Beta13-TimeEvol}(a-d), 
that show the vorticity fields at four instants of the slow period. As expected from \eqref{eqn:defquasiperio}, the quasi-periodic perturbation is similar to the real part of the Floquet mode at time $t_{\epsilon}=0$ (compare figures \ref{3-Linear-FloquetMode-Beta13}-a and  \ref{3-Linear-FloquetMode-Beta13-TimeEvol}-a) or to its opposite at time  $t_{\epsilon}=0.5$, while it is of much smaller amplitude and similar to the imaginary part of the Floquet mode at time  $t_{\epsilon}=0.25$ and $t_{\epsilon}=0.75$. Let us now examine the quasi-periodic evolution of the horizontal force exerted on the foil displayed in figure \ref{3-Linear-FloquetMode-Beta13-TimeEvol}(f), as well as the resulting foil speed and displacement shown in  figures \ref{3-Linear-FloquetMode-Beta13-TimeEvol}(e) and and (g), respectively. Around the slow time $t_{\epsilon}=0$, the horizontal force time-averaged over the fast flapping period (thick line in f) is negative and opposite to the positive horizontal velocity (e). So the quasi-perturbation, shown in (a) and dominated by the real part of the Floquet mode, creates a resistive force. When it slowly evolves towards $t_{\epsilon}=0.25$, the horizontal force remains negative but it is then a propulsive force since the foil velocity is negative. For $ t_{\epsilon}>0.32$, the horizontal force gets positive and is a resistive force to its horizontal motion until the sign of the foil velocity changes around $ t_{\epsilon}=0.6$. This slow oscillation of the time-averaged force and velocity creates a back and forth displacement of the foil depicted in \ref{3-Linear-FloquetMode-Beta13-TimeEvol}(g) around a mean position value around $\tilde{x}_{g}=0.5$. Although this displacement is infinitesimally small, it is in agreement with the direction switching observed in the unsteady nonlinear solutions of regime IV. Note also the horizontal forces and speed time-averaged along the slow period $T_{\epsilon}=1/f'$ is zero, so that there is no unidirectional propulsion of the foil at that slow time scale. \\

\begin{figure}
\vspace{0.25cm}
   \centering
       \begin{tabular}{ll}
         (a) & (b) \\
        \includegraphics[width=0.4\linewidth]{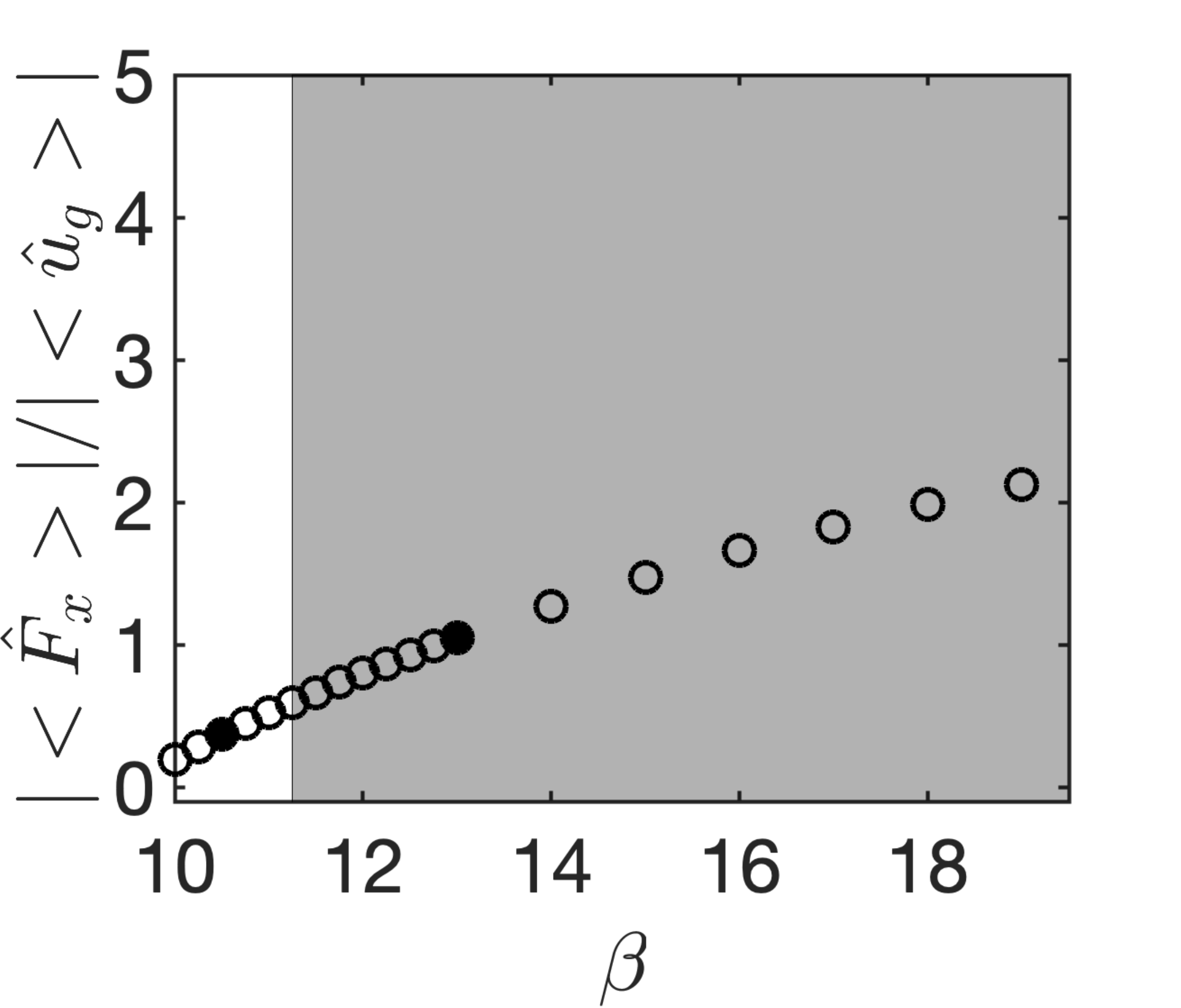}
        &    \includegraphics[width=0.4\linewidth]{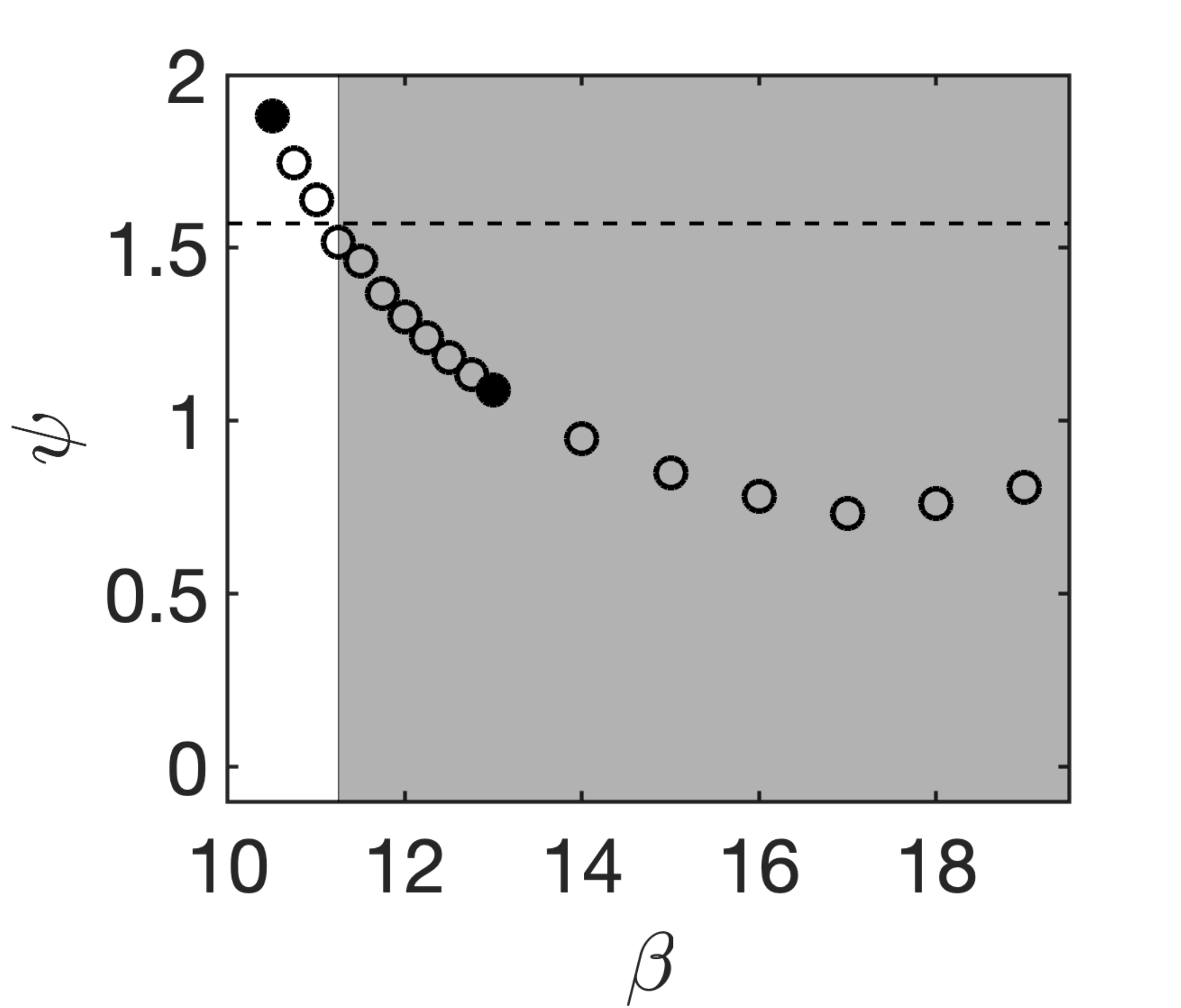} \\
        (c) & (d) \\
        \includegraphics[width=0.48\linewidth]{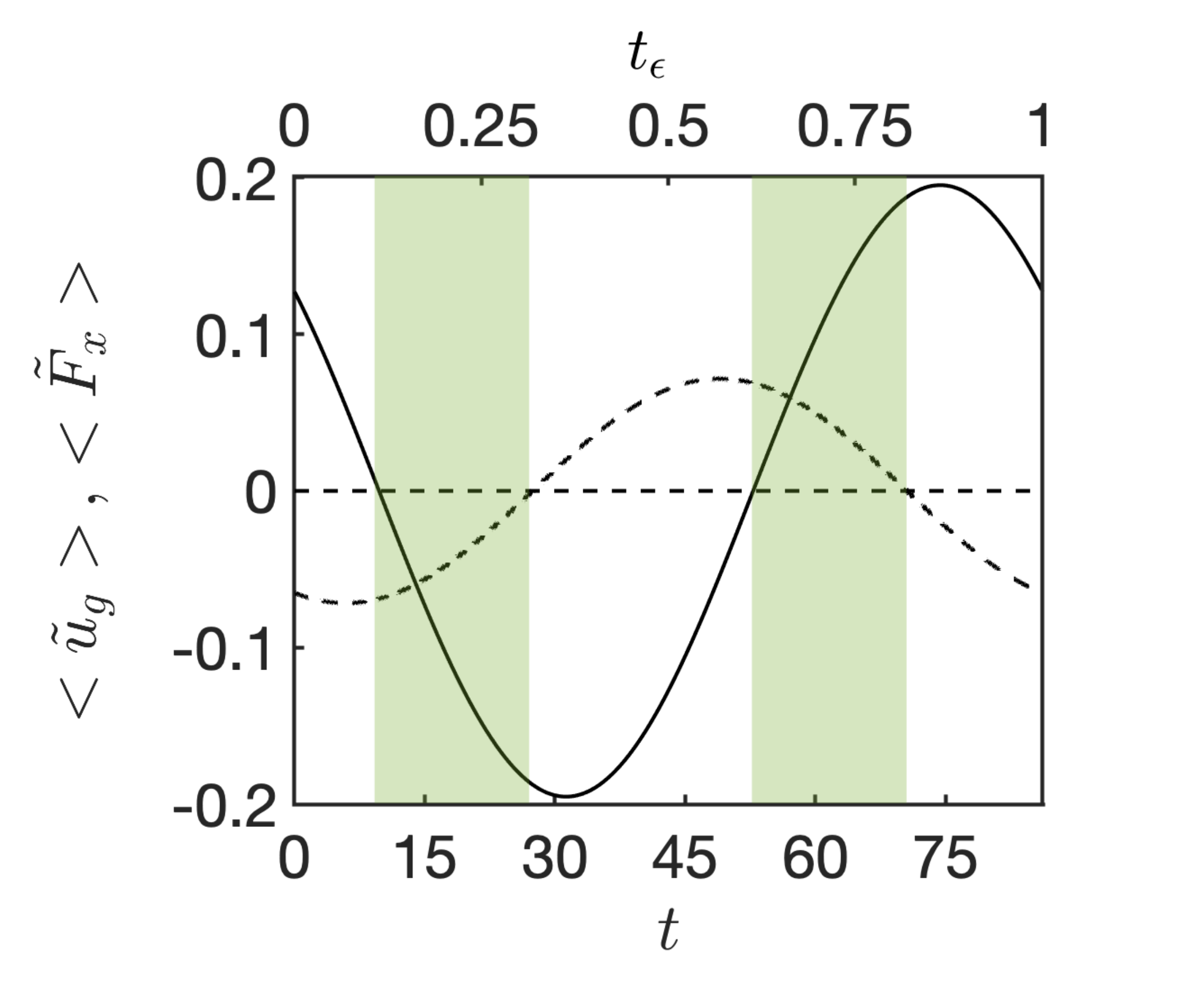} 
        &    \includegraphics[width=0.48\linewidth]{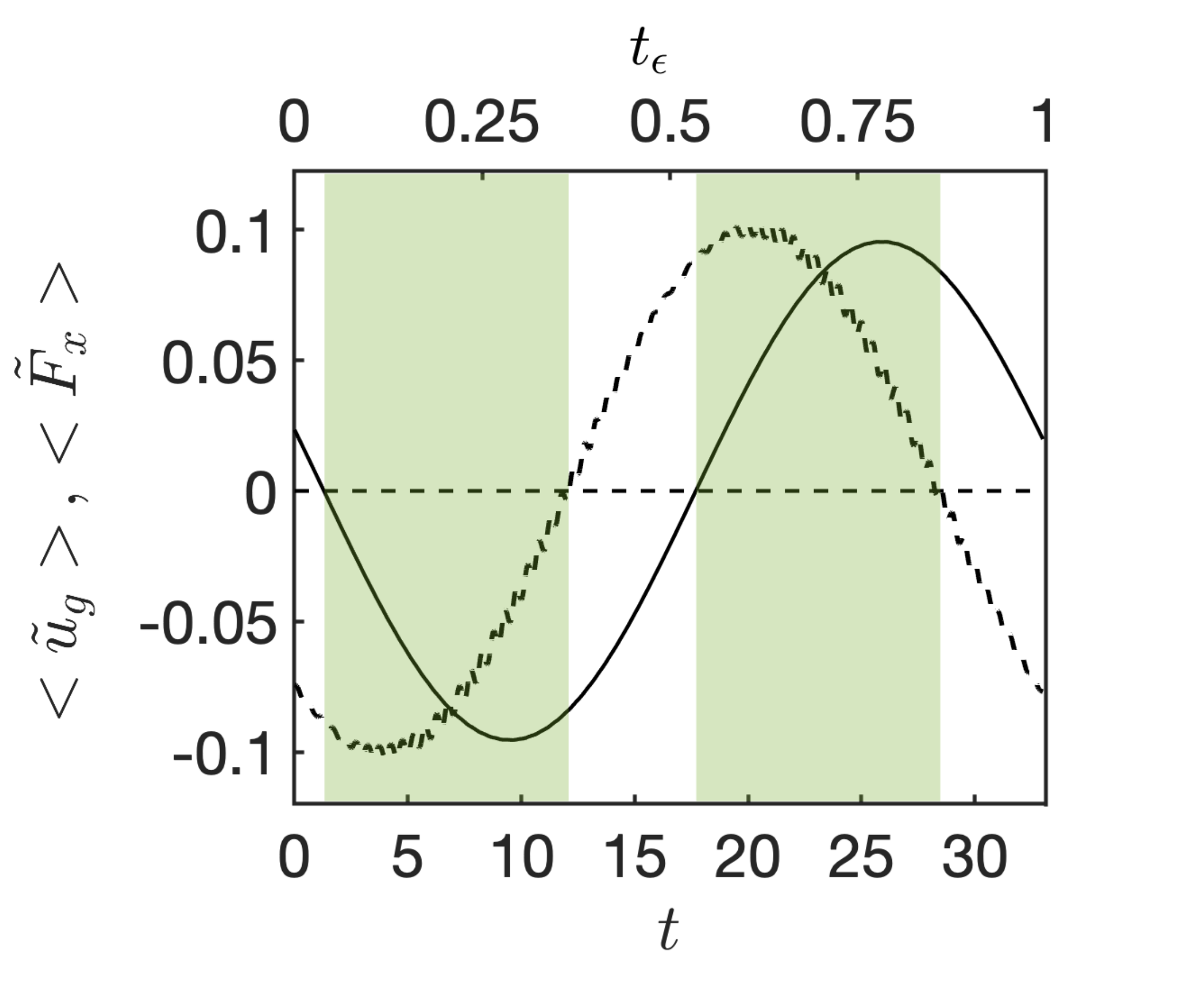}
   \end{tabular}
    \caption{ (a) Amplitude ratio and (b) phase difference of the (time-averaged) horizontal force and foil velocity of the asynchronous modes as a function of the Stokes number. The instability region is marked in grey. The dashed line corresponds to $\psi=\pi /2$.  
    (c,d) Temporal evolution of the quasi-periodic foil velocity $<\tilde{u}_g>$ (solid curve) and force $<\tilde{F}_x>$ (dashed curve) time-averaged along a flapping period for (c) a stable ($\beta=10.5$) and (d) an unstable ($\beta=13$) asynchronous mode, shown in (a,b) with black dots. The green area corresponds to propulsive phases (velocity and force have the same sign) while white area correspond to resistive phases (velocity and force are of opposite sign). }
            \label{Fig-Mean-Force-Speed-Modes}
\end{figure}

Finally, to understand the destabilisation of this asynchronous mode in light of the fluid-solid interaction,  we consider again the time-averaged analysis. The asynchronous mode being complex, the polar decomposition of the time-averaged force and velocity is $\left< \hat{F}_x\right>=|\left< \hat{F}_x\right>| e^{\rm{i} \phi_{F}}$ and $\left< \hat{u}_g\right>=|\left< \hat{u}_g\right>| e^{\rm{i} \phi_{U}}$. As shown in  \S \ref{subsec:timeaveragedanalysis}, the Floquet exponent then satisfies
\begin{equation*}
    \lambda = \frac{1}{\rho S} \frac{|\left< \hat{F}_x\right>| }{|\left<\hat{u}_g\right>| } e^{\rm{i} \psi}
\end{equation*}
where $\psi=\phi_{F}-\phi_{U}$ is the phase difference  between the force and velocity phases. The asynchronous mode is unstable ($\Re(\lambda)>0$) when this phase difference lies in the interval $-\pi/2 < \psi < \pi/2$. 
The crucial role of the phase difference (rather than the force-to-velocity ratio) in the asynchronous mode destabilization is confirmed by examining figures \ref{Fig-Mean-Force-Speed-Modes}(a) and (b) where both quantities are depicted as a function of the Stokes number.
The mode gets unstable (gray area) precisely when the phase difference $\phi<\pi/2$. To better understand the physical meaning of this phase difference, the temporal evolution of the \textit{quasi-periodic} perturbation of the time-averaged velocity $<\tilde{u}_g>$ (solid) and force $<\tilde{F}_x>$ (dashed) are plotted as a function of time in figure \ref{Fig-Mean-Force-Speed-Modes}(c) for a stable and in \ref{Fig-Mean-Force-Speed-Modes}(d) for an unstable mode. Note that the velocity and force are time-averaged over the (short) flapping period and their evolution is depicted over the long period (slow time scale $t_{\epsilon}$). In both figures, the green areas identify phases of motion where the hydrodynamic force is propulsive (since the force and velocity are of same sign), while the white areas correspond to resistive phases of motion (force and velocity of opposite signs). In the case of a stable mode  (see figure \ref{Fig-Mean-Force-Speed-Modes}-c), the phase difference slightly above $\pi/2$ results in a mode with propulsive phases of motion that are shorter than resistive ones. On the other hand, in the case of an unstable mode  (see figure \ref{Fig-Mean-Force-Speed-Modes}-d), the phase difference slightly under $\pi/2$ results in a mode with longer propulsive phases of motion that resistive ones. The phase difference between the horizontal force and velocity is thus related to the cumulative time of propulsive phase over resistive phase. When $-\pi /2 < \psi < \pi /2 $, the propulsive phases last longer than resistive ones, and the asynchronous mode is unstable. A similar criterion was established 
by \citet{Navrose2016} for the instability threshold
of a spring-mounted cylinder flow, based on the global stability analysis of the steady base flow solution.
They showed that the phase difference between the vertical hydrodynamic force and displacement of the cylinder perturbation is related to the transfer of energy from the flow to the cylinder and drives the destabilization of the mode. The present criterion can be viewed as an extension to the instability of periodic fluid-solid interaction problems.

\subsection{Effect of the fluid-solid density ratio }\label{subsec:densityratio}

Before concluding, we investigate the effect of the fluid-solid density ratio $\rho$ on the results of the Floquet analysis. Two limit cases are considered in this section: high density ratios $\rho >> 100$, that lead to a loosely coupled fluid-solid due to the vanishing action of the fluid over the solid problem (as presented in \S \ref{subsec:floquet}), and the range of lower density ratios $\rho < 100$ that tend to the one of swimming organisms.

\begin{figure}
  \centering
\begin{tabular}{ll}
    (a) & (b)  \\
      \includegraphics[width=0.45\linewidth]{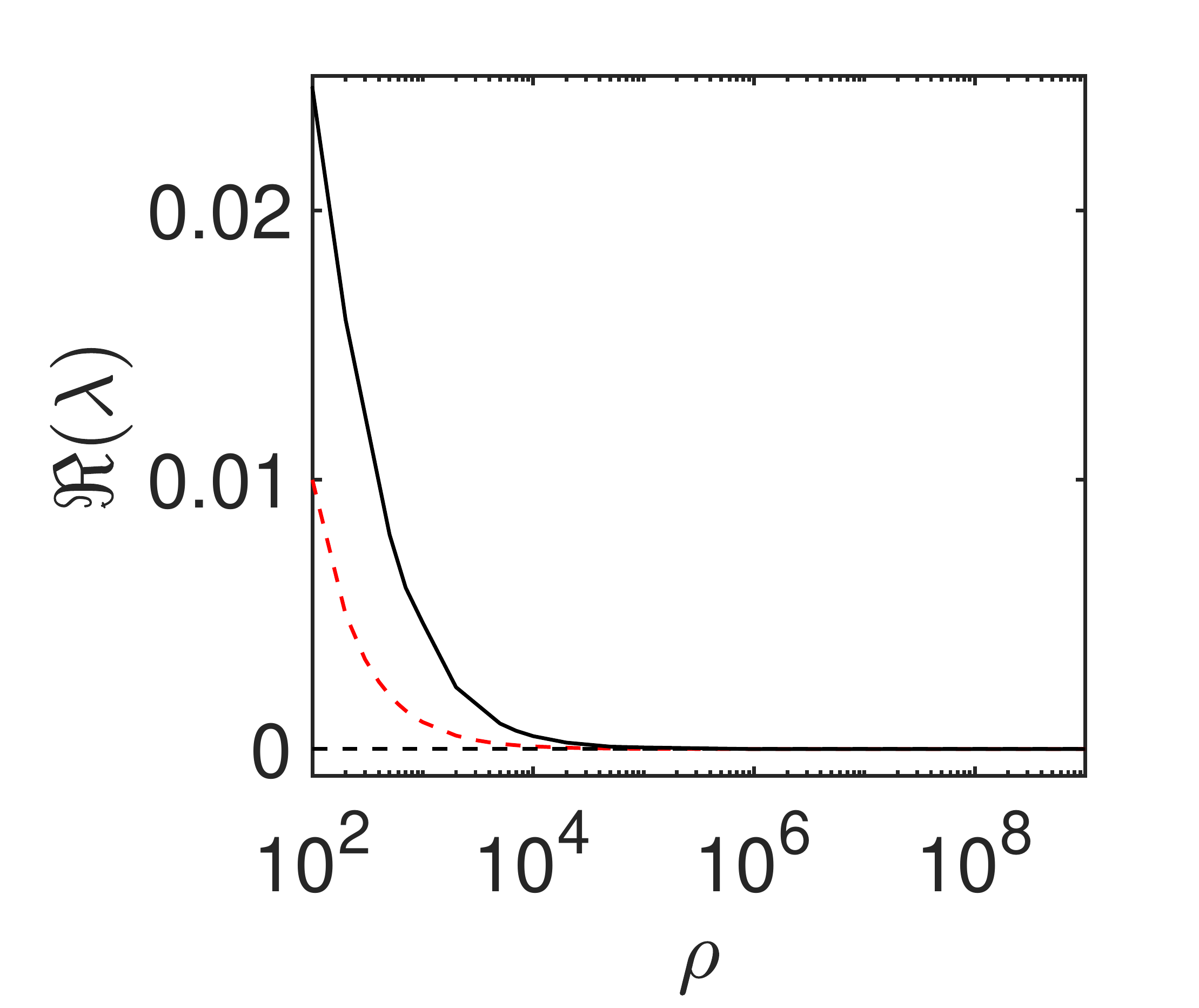}
     &       \includegraphics[width=0.45\linewidth]{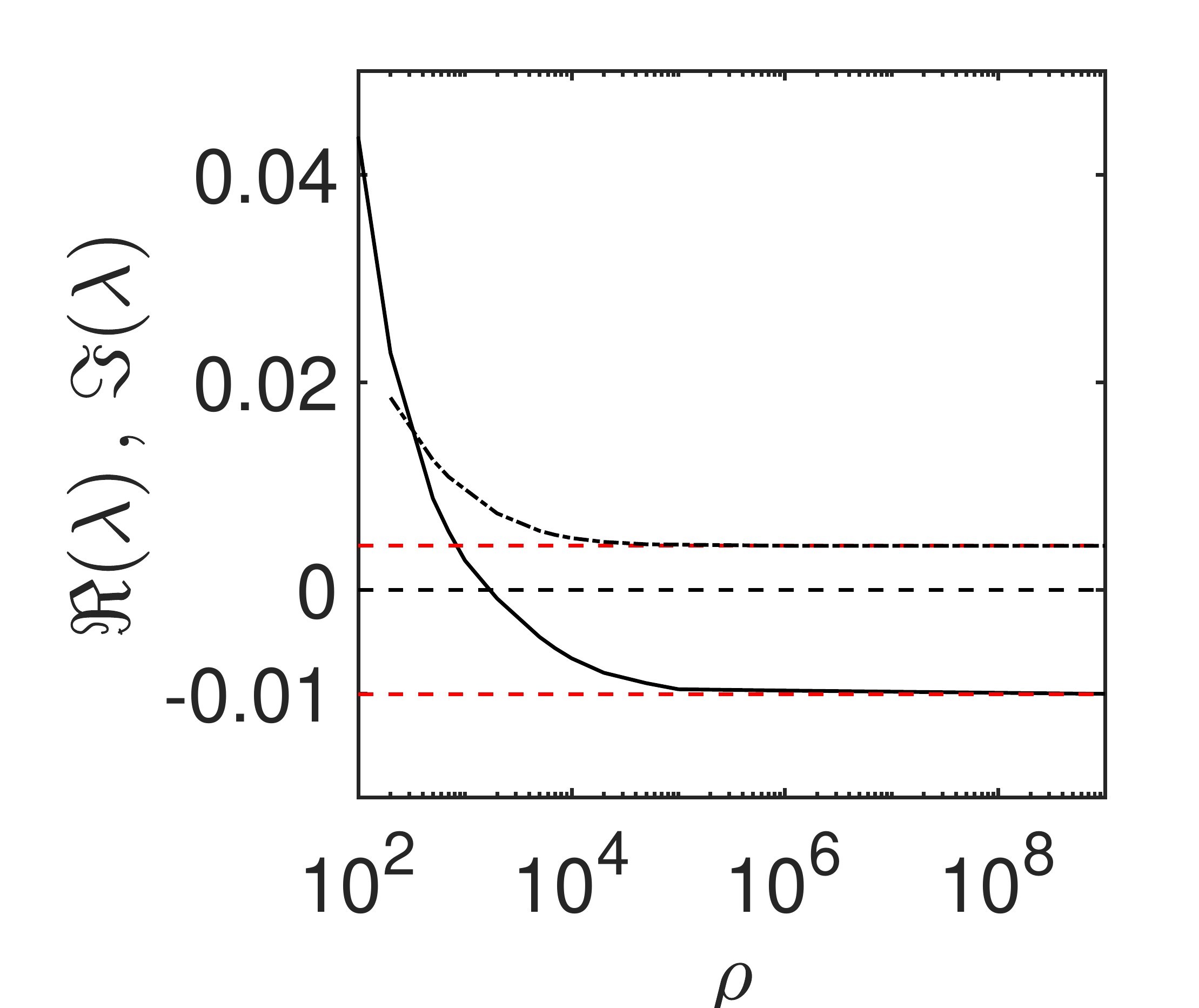} \\
   (c) & (d)  \\
      \includegraphics[width=0.45\linewidth]{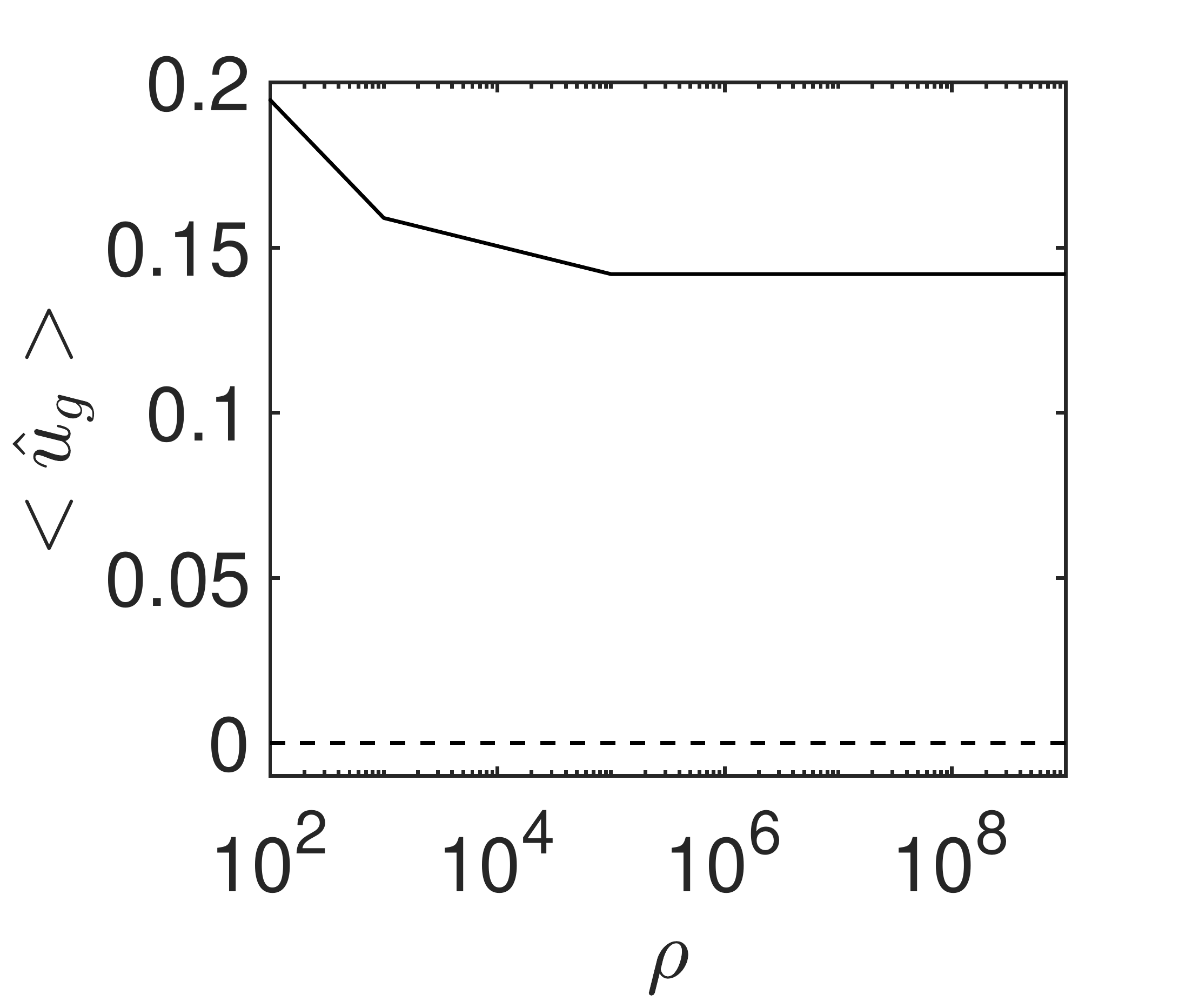}
 &
      \includegraphics[width=0.45\linewidth]{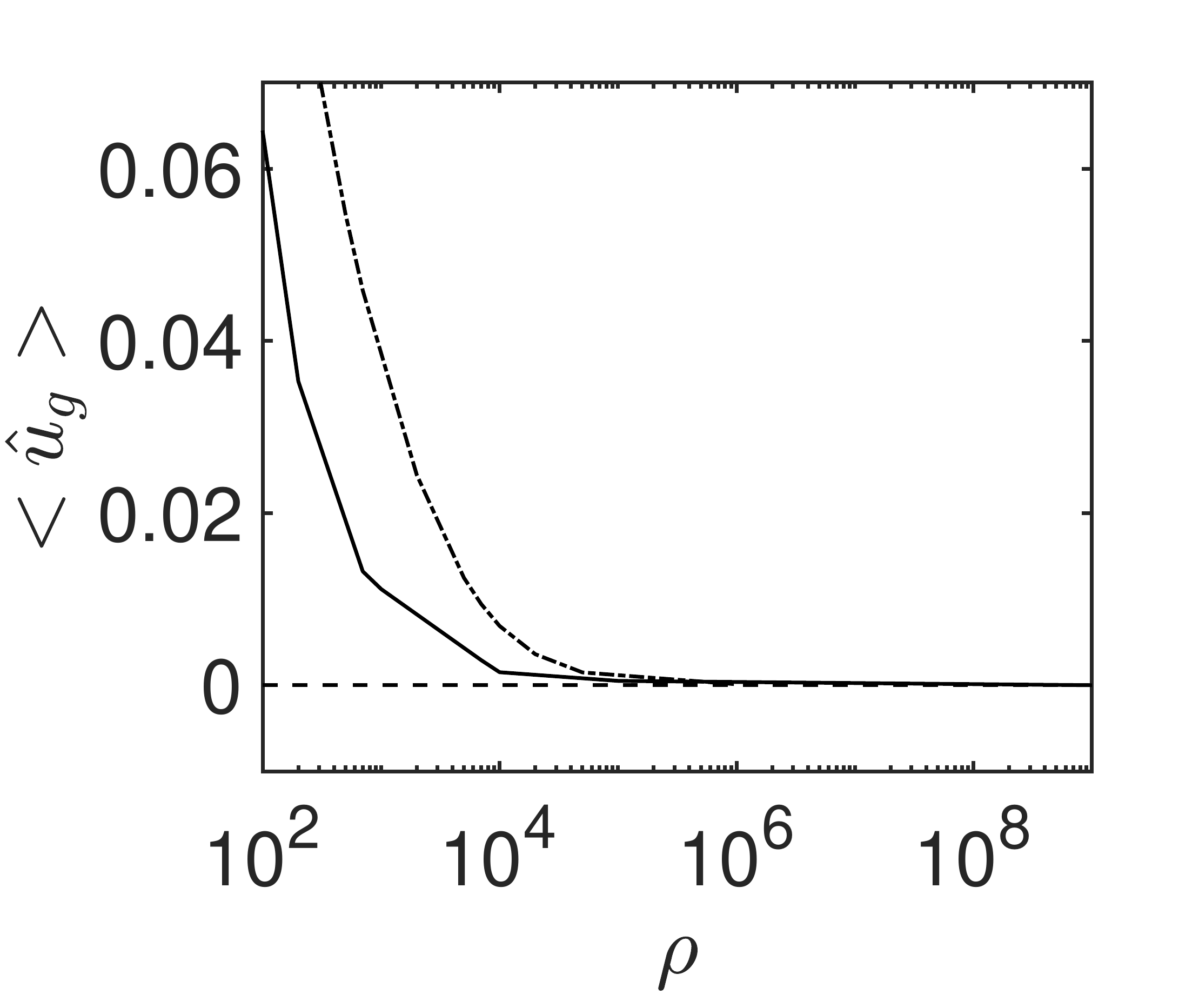}
\end{tabular}
    \caption{Effect of the fluid-solid density ratio $\rho$ on (a,b) the Floquet exponent and (c,d) the perturbation of the foil horizontal speed for (a,c) the synchronous mode ($\beta=6$) and (b,d) the asynchronous mode ($\beta=12$). In (b) and (d) the real and imaginary part of respectively the Floquet exponent and the horizontal speed are represented by solid and dash-and-dot lines. The solid curves correspond to results of the fluid-solid analysis. The red dashed curve correspond to the asymptotic limits of the exponent: (a) $1/\rho$ curve and (b) the values of the purely-hydrodynamic analysis $\Re(\lambda)=-0.01$ and  $\Im(\lambda)=0.0043$. The dashed horizontal lines delimit in (a,b) $\Re(\lambda)=0$ and in (b) $<u_g>=0$.}
    \label{3-Linear-Floquet-Mu-Rho}
\end{figure}

Let us first consider the high density ratio limit. The evolution of the Floquet exponent is shown in figure \ref{3-Linear-Floquet-Mu-Rho}(a) and (b) for the synchronous and asynchronous modes, respectively. Starting from the value $\rho=100$ considered until now, and increasing the density ratio, the absolute value of the Floquet multipliers (black solid curve) decreases for both modes. However,
their asymptotic behaviour in the limit $\rho \rightarrow \infty$, displayed with the dashed red curves in the two figures, is different. The synchronous mode evolves as $1 / \rho$ and thus only gets marginally stable. The asynchronous mode is stabilized for $\rho>10^3$ and its growth rate tends towards that of the purely-hydrodynamic analysis.

To further understand why the Floquet exponents of the synchronous and asynchronous mode exhibit different behaviour in that limit case, we propose to reconsider the time-averaged analysis. In the infinite density ratio limit, the relation  \eqref{eqn:acc-force}, that links the growth rate to the mean value of the horizontal solid velocity and force, gets:
\begin{equation*}
     \lambda <\hat{u}_g> = 0 \, .
\end{equation*}
Thus, either the Floquet exponent or the mean horizontal velocity is zero. The synchronous mode corresponds clearly to the case $\lambda=0$ (figure \ref{3-Linear-Floquet-Mu-Rho}-a). The foil mean velocity does not necessarily vanish for high density ratios, as observed in figure \ref{3-Linear-Floquet-Mu-Rho}(c) that displays the evolution of the time-averaged horizontal speed with the density ratio. The asynchronous modes corresponds to the second case $<\hat{u}_g> = 0$, as seen in \ref{3-Linear-Floquet-Mu-Rho}(d). As a matter of fact, not only the mean horizontal velocity converges to zero, but so does the entire temporal evolution due to the negligible acceleration generated by high density ratios. In this case, the Floquet exponent does not tend to zero  (see figure \ref{3-Linear-Floquet-Mu-Rho}-b), but rather to the value predicted by the purely-hydrodynamic stability analysis (red dashed line). Indeed, in the limit $\rho \rightarrow \infty$, the fluid-solid linearized operator is block triangular  as seen in \eqref{eqn:fluid-solid-infinitedensity} and the purely-hydrodynamic Floquet multipliers are included in the fluid-solid Floquet spectrum. 
A similar asymptotic behaviour was observed by \citet{Fabre2011} when investigating the dynamics of free falling bodies in fluids using a fluid-solid stability analysis of the steady base solution (not periodic as in the present case).  
In the limit case, results of the purely-hydrodynamic and fluid-solid stability analyses converged to the well know Von-Kármán vortex wake instability without effect on the path of the falling body. The present result is somehow an extension to the fluid-solid stability analysis of periodic solution.\\

\begin{figure}
  \centering
  \begin{tabular}{ll}
      (a) & (b) \\
      \includegraphics[width=0.48\linewidth]{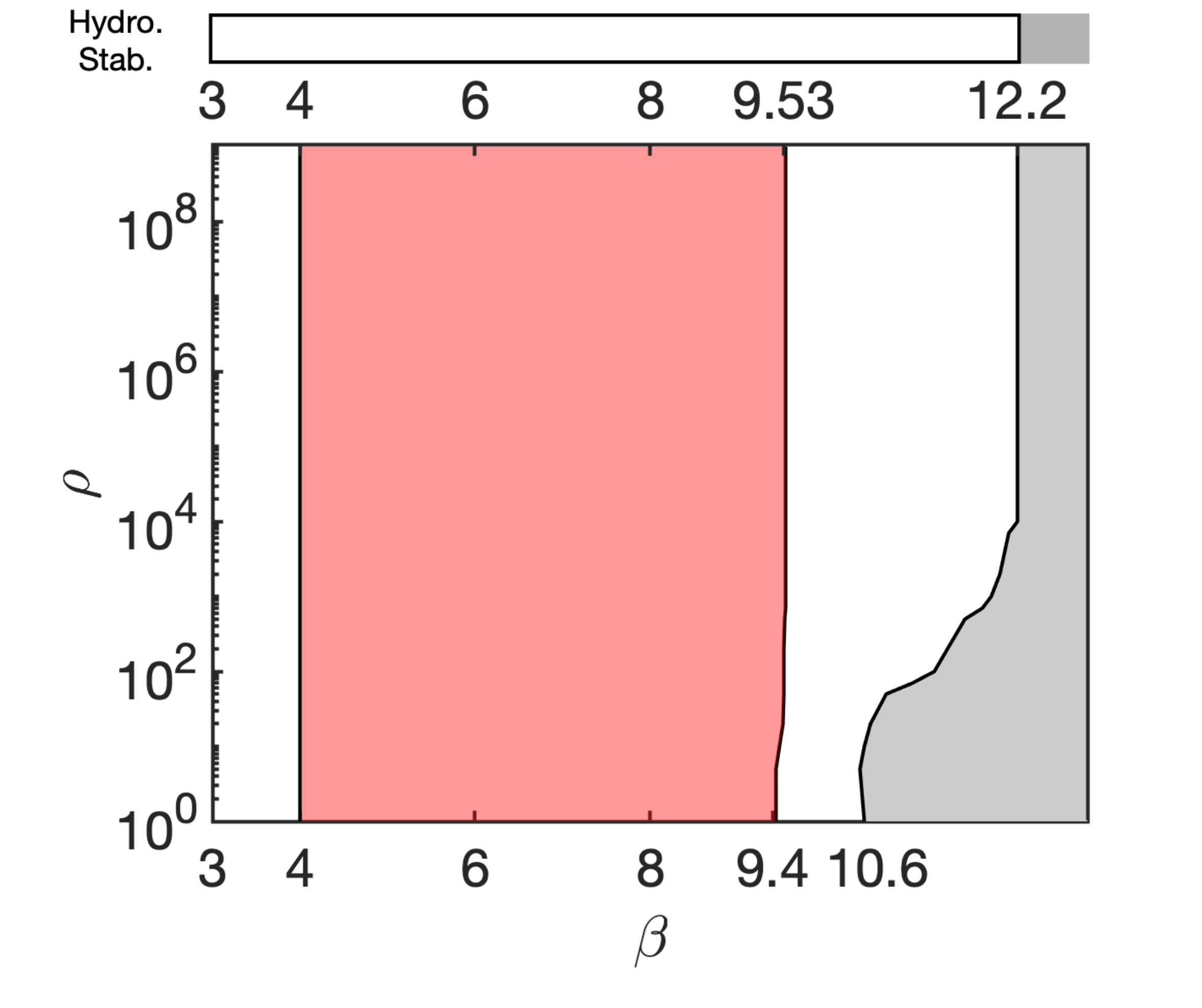}
      & \includegraphics[width=0.48\linewidth]{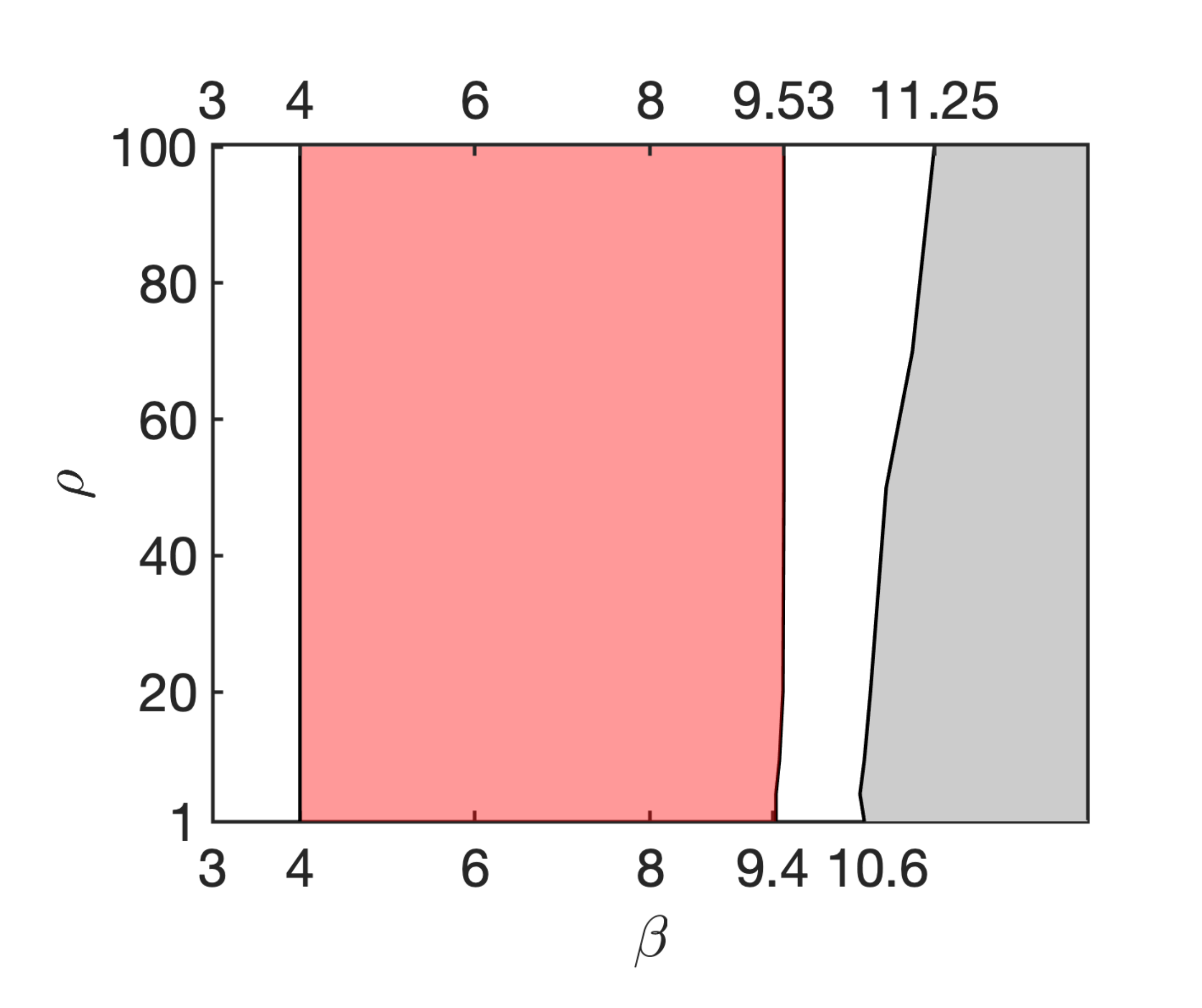}
  \end{tabular}

    \caption{(a) Asymptotic limit ($1 \leq \rho \leq 10^9$) and (b) close-up on low density ratios ($1 \leq \rho \leq 100$) of the instability regimes of the symmetric non-propulsive periodic solution identified with the fluid-solid Floquet analysis in the parameters plane ($\beta$,$\rho$). White regions correspond to stable solutions, while red and grey regions indicate unstable solutions to synchronous and asynchronous Floquet modes, respectively. Results of the purely-hydrodynamics stability analysis are displayed in the top of (a) for comparison.}
    \label{fig:map-rho-stokes}
\end{figure}

Finally, figure \ref{fig:map-rho-stokes} presents results of the fluid-solid stability analysis in the parameter space $(\beta,\rho)$. The white area correspond to stable regions, while the red and grey areas corresponds to regions of unstable synchronous and asynchronous modes, respectively. We first consider the asymptotic limit of very large density ratios $\rho >> 100$. As one can see in figure \ref{fig:map-rho-stokes}(a), the synchronous unstable modes (red region) thresholds are barely modified for high density ratios, always taking place for $\beta=4$ and $\beta=9.53$ for $\rho>100$. The instability threshold of asynchronous modes (grey region), on the other hand, is strongly modified with the increase of $\rho$ going from $\beta=11.25$ for $\rho=100$ to $\beta=12.2$ for $\rho=10
^4$. For Stokes numbers between these two thresholds, the density ratio presents thus a stabilizing effect. Accordingly to our previous analysis, as $\rho \rightarrow \infty$ the asynchronous modes threshold tends to the purely hydrodynamic analysis one (displayed at the top of the figure). Synchronous modes, on the other hand, become marginally unstable and do not converge to the purely hydrodynamic analysis, their marginal instability explaining the minimal threshold changes for high density ratios.

For small density ratios $1 \leq \rho \leq 100$, figure \ref{fig:map-rho-stokes}(b) the synchronous mode thresholds are again not strongly modified with the $\rho$ decrease. While the transition from stable to unstable synchronous modes remains constant at $\beta=4$, the stabilization threshold presents a significant variation going from $\beta=9.42$ for $\rho=1$ to $\beta=9.53$ for $\rho=100$. The instability threshold of the asynchronous mode is strongly modified, as for high density ratios. Its onset is now encouraged rather than delayed, varying from $\beta=11.25$ for $\rho=100$ to $\beta=10.6$ for $\rho=1$.
These results indicate that swimming organisms (nearby an unity density ratio) are significantly more prone to non-coherent motions than flying organisms.

All instability thresholds and their trends are coherent with the unsteady nonlinear results previously reported. As shown for $\rho=1$ in table \ref{tab:comparisonthresholdrho1}, the destabilisation of the synchronous mode and the onset of unidirectional propulsion take both place for $\beta=4$. The threshold value $\beta=9.42$ of the synchronous mode stabilization is again slightly different from the threshold $\beta=9.50$ of the transition between propulsive and non-propulsive solutions due to the bifurcation sub-critical nature (see figure \ref{2-NonLinear-MeanU-rho}-b). The threshold value $\beta=9.42$ found with the linear stability corresponds, as in the case $\rho=100$, to the disappearance of the symmetric solution when decreasing the Stokes number (transition from III to II). The onset of back \& forth solutions matches again the destabilization of the asynchronous Floquet mode for $\beta=10.6$.

\begin{table}
    \centering
    \begin{tabular}{c c c c}
        Transition between regimes & I $-$ II &  II $-$ III & III $-$ IV \\
    \hline
        Fluid-solid stability analysis & $4.00$ & $9.42$ & $10.60$ \\
        Nonlinear simulations & $4.00$ & $9.42$ and $9.50$ & $10.60$ 
       \end{tabular}
    \caption{Critical thresholds obtained with the unsteady nonlinear simulations and the fluid-solid stability analysis for $\rho=1$.}
    \label{tab:comparisonthresholdrho1}
\end{table}{}

%%%%%%%%%%%%%%%%%%%%%%%%%%%%%%%%%%%%%%%%%%%%%%%%%%%%%%%%%%%%%%%%%%%%%%%%%%%%%%%%%%%%%%%%%%%%%%%%%%%%%%%%%
\section{Conclusions}

The role of linear mechanisms in the transition to horizontal locomotion of a vertically flapping foil has been investigated. First, the occurrence of non-propulsive, unidirectional propulsive and back \& forth solutions was established in the range of Stokes numbers $\beta \in [1,20]$ for a rectangular shaped foil with an aspect ratio $h=1/20$ with a flapping amplitude $A=0.5$ and a solid-fluid density ratio $\rho=100$. Floquet stability analysis of the coupled fluid-solid system was thus performed over symmetry preserved non-propulsive solutions. Our study was finally concluded by analysing the effect of the solid-fluid density ratio on the stability analysis, this last study being compared to predictions of non-propelled foil stability analysis usually employed in the literature.

First, symmetric non-propulsive, unidirectional propulsive and back \& forth solutions were obtained while raising $\beta$, as in \citet{Alben2005,Lu2006}. As expected for a low aspect ratio foil \citep{Deng2016a}, non-propulsive solutions first transition to unidirectional propulsion. The results presented in this paper highlight the existence of a sub-critical transition between propulsive and again non-propulsive solutions (regimes II-III), with back \& forth oscillations finally reached (regime IV) for higher $\beta$. 
The emergence of these nonlinear solutions was then investigated through a self-propelled stability analysis of non-propulsive solutions. This analysis revealed the existence of unstable synchronous and asynchronous Floquet modes in the region of unidirectional and back \& forth solutions, respectively. We therefore studied the characteristics of these unstable modes, investigating their associated fluid flow, horizontal force and speed to finally relate these mechanisms to the locomotion regimes obtained in nonlinear simulations. The evolution of the modes mean horizontal force and speed with $\beta$ allowed us to establish a criterion of instability that link these quantities to the Floquet exponent.

In the case of synchronous modes, spatial symmetry breaking modes with non zero force and speed are obtained, similar to unidirectional propulsion. Hydrodynamic forces that accelerates the horizontal speed lead to unstable Floquet exponents, and the transition to unstable modes is driven by the increase of the hydrodynamic force. The decay of this horizontal force leads subsequently to the mode stabilization and the re-emergence of symmetric non-propulsive solutions.
Concerning asynchronous modes, the direction switching phenomenon observed in nonlinear solutions is explained by the competing action of the complex and real parts of these modes associated to the complex phase of their multiplier. This complex multiplier accurately predicts the low frequency of back \& forth solutions at their onset. The temporal evolution of the quasi-periodic perturbation, resulting from the superposition of the real and imaginary part of the Floquet mode, clearly shows how the horizontal force exerted by the flow perturbation is alternatively propulsive or resistive, i.e of the same or opposite sign to the foil velocity, acting as a restoring spring-like force on the foil position over the slow period. 
The destabilization of the asynchronous modes depends on the phase difference between the time-averaged force and velocity perturbation, as it measures how long this force is propulsive or resistive over the slow period.
A generalization to three-dimensional foils and different flapping movements of this mode might offer an additional path to understand, for example, the \textit{snaking} trajectory presented by \citet{Deng2018a} in the horizontal locomotion of oblate spheroids, first supposed to be connected to nonlinear effects, or the non-coherent motion of living organisms as the planktonic sea butterfly \citep{Murphy2016}.

The influence of the solid-fluid density ratio was finally investigated. For $\rho \in [1,10^9]$ we observed that the $\beta$ range of synchronous modes is not largely affected by the density ratio, whereas the transition to asynchronous modes is greatly impacted. To understand this behaviour, we have studied the coupled stability equations in the limit of high density ratios. We have shown that synchronous modes are asymptotically marginally unstable, whereas asynchronous modes converge to the uncoupled fluid system, since their horizontal speed converge to zero. These results explain the observed sensibility to the density ratio of the asynchronous modes threshold and the small variation of the transition to synchronous modes. Comparing these results to the non-propelled stability analysis, we have highlighted that the later converges to the self-propelled analysis only for asynchronous modes in the limit of very large density ratios, since in this limit the fluid-solid coupling terms disappear.

We conclude thus that the studied fluid-solid Floquet stability analysis, that takes into account the inherently fluid-solid coupling of the studied self-propelled foil interacting with a viscous fluid, is best adapted to predict the onset of unidirectional and back \& forth propulsion. Possible future applications of this coupled stability analysis are the return from back \& forth to coherent unidirectional propulsion observed for higher stokes numbers \citep{Alben2005}, as well as bifurcations of self-propelled heaving foils passively pitching around their leading edge \citep{Spagnolie2010} and of a self-propelled infinite array of flapping wings \citep{Becker2015}.

\vspace{3mm}

\textbf{Declaration of Interests}: The authors report no conflict of interest.

\vspace{3mm}

\textbf{Acknowledgments}
This work was initiated during the PhD thesis of Damien Jallas, co-supervised by David Fabre and the second author, who gratefully acknowledges both of them. This project has received funding from the European Research Council (ERC) under the European Union Horizon 2020 research and innovation program (grant agreement No.638307).

\appendix

%% ----------------------------------------------------------------------------------------------%%
\section{Segregated approach for solving the implicitly coupled fluid-solid problem}
\label{Appendix:TimeMarchingScheme}
We describe here the segregated approach used to solve efficiently
the temporally discretized equations \eqref{eqn:NS-Implicit}. Manipulating the $r$-backward differential formula of the time derivative, we may split  the unknown solid velocity $u_g^{n+1}$ as, 
\begin{equation} \label{eqn:SolidSplitting}
u_g^{n+1} = (\Delta t / \alpha_0) \, a_{g}^{n+1} + \hat{u}_{g} \,,
\end{equation}
where the first term is an unknown component, proportional to the unknown acceleration  $a_{g}^{n+1}=\left( d u_g /dt \right)^{n+1}$, and the second term  is the known velocity component defined as 
\begin{equation*}
\hat{u}_{g}=-\sum_{k=1}^r (\alpha_k / \alpha_0) \, u_g^{n+1-k} \,.   
\end{equation*}
The above decomposition of the horizontal solid velocity is then used to split the fluid  variables as,
\begin{eqnarray} \label{eqn:FluidSplitting}
(\mathbf{u}^{n+1},p^{n+1}) = \left( \frac{\Delta t}{\alpha_0} a_g^{n+1} \right)  (\delta \mathbf{u}, \delta p) + (\hat{\mathbf{u}}^{n+1},\hat{p}^{n+1}) \, ,
\end{eqnarray}
where we have introduced the flow component $(\mathbf{\delta u},\delta p)$, proportional to the solid acceleration, and the flow component $(\mathbf{\hat{ u}}^{n+1},\hat{p}^{n+1})$ that will depend on the solid velocity $\hat{u}_{g}$. Introducing the solid \eqref{eqn:SolidSplitting} and fluid  \eqref{eqn:FluidSplitting} decomposition into the governing flow equations \eqref{eqn:NS-Implicit} , we obtain two independent linear system of equations. The first one
governs the flow component $(\mathbf{\hat{u}}^{n+1},\hat{p}^{n+1})$ as
\begin{eqnarray} \label{eqn:NS-hat}
\frac{\alpha_0}{\Delta t} \mathbf{\hat{u}}^{n+1}
+ \nabla{\hat{p}}^{n+1} -\beta^{-1} \Delta \mathbf{\hat{u}}^{n+1}  &=&  \mathbf{f}^{n+1} \, \nonumber \\
\nabla \cdot \mathbf{\hat{u}}^{n+1} &=& 0 \,  \\
(\hat{u}^{n+1},\hat{v}^{n+1})(\Gamma_w) &=& \left( \hat{u}_{g},2 \pi A sin(2 \pi t^{n+1}) \right) \, \nonumber 
\end{eqnarray}
The boundary conditions at the fluid-solid interface $\Gamma_{w}$ is explicitly known. Therefore, any classical algorithm such as the Uzawa method or the projection splitting method can be used to obtain the solution of this discretized forced unsteady Stokes equations. Following \citet{Jallas2017}, we have used an a preconditioned conjugate gradient algorithm \citep{Cahouet1988} to impose the divergence-free condition.
As it depends on the flow history through the right-hand side forcing terms $\mathbf{f}^{n+1}$ and the boundary conditions, it is solved at each temporal iteration.\\

The second problem governs the flow component $(\mathbf{\delta u},\delta p)$ that is proportional to the solid acceleration in the horizontal direction. It writes
\begin{eqnarray} \label{eqn:NS-delta}
\frac{\alpha_0}{\Delta t} \mathbf{\delta u} 
+ \nabla{\delta p}^{n+1} -\beta^{-1} \Delta \boldsymbol{\delta u}^{n+1}
&=&  0\, \nonumber \\
\nabla \cdot \mathbf{\delta u} &=& 0 \nonumber \\
(\delta u,\delta v)(\Gamma_w) &=& (1,0)
\end{eqnarray}
Again, the boundary conditions at the fluid-solid interface $\Gamma_{w}$ is explicitly known, but it is now independent from the temporal iteration. The solution can thus be obtained prior to the temporal iteration.  It can be viewed
as the short time response of a Stokes flow, initially at rest, to a unitary horizontal velocity.

The solution of the (above) two independent flow problems does not give access to $(\mathbf{u}^{n+1},p^{n+1})$ in \eqref{eqn:FluidSplitting}
since the horizontal acceleration $a_{g}^{n+1}$ is still unknown. The final step of the algorithm is obtained by introducing this decomposition into the last equation of the governing equation \eqref{eqn:NS-Implicit}, yielding
\begin{equation}\label{FluidForcewithDamping}
(\rho S) \, a_{g}^{n+1} = \left( \frac{\Delta t}{\alpha_0} a_{g}^{n+1} \right) F_x(\mathbf{\delta u}, \delta p) + F_x(\hat{\mathbf{u}},\hat{p})\, , 
\end{equation}
The horizontal acceleration is thus given by 
\begin{equation}\label{Acceleration}
a_{g}^{n+1} = \left( \rho S - \frac{\Delta t}{\alpha_0}  F_x(\mathbf{\delta u}, \delta p) \right)^{-1} F_x(\hat{\mathbf{u}},\hat{p})\, , 
\end{equation}
and the velocity and pressure are obtained using \eqref{eqn:FluidSplitting}.

\begin{table}
    \centering
    \begin{tabular}{c c c | c c c}
        Time step & $ \max(F_y)$ & $<u_g>$ & Mesh nb. of triangles &  $ \max(F_y)$ & $<u_g>$   \\
    \hline
        $\Delta t = 0.01 $ & $38.692$ & $0.223$ & 7836 & $38.641$ & $0.226$ \\
        $\Delta t = 0.005 $ & $38.687$ & $0.227$ & 17116 & $38.687$ & $0.227$ \\
        $\Delta t = 0.001 $ & $38.687$ & $0.227$ & 22580 & $38.687$ & $0.227$
    \end{tabular}
    \caption{Convergence of the maximal vertical force $\max(F_y)$ and the time-averaged horizontal velocity $<u_g>$ with time step and mesh number of triangles for an unidirectional propulsive solution ($\beta=6$). For the time-step convergence a fixed mesh of $17116$ triangles was adopted, whereas for the mesh convergence a fixed time step $\Delta t = 0.005$ was used. }
    \label{Appendix:tab:mesh_dt}
\end{table}{}

\section{Validation of the nonlinear and linear fluid-solid solvers}
\label{Appendix:SectionValidation}

The numerical method is primarily validated through the convergence of the time-averaged horizontal velocity and the maximal vertical force acting on the foil with the mesh number of elements and time step. In table \ref{Appendix:tab:mesh_dt} the influence of the time and spatial discretization is evaluated for an unidirectional propulsive solution ($\beta=6$). A time step of $\Delta t = 0.005$ and a mesh of $17116$ triangles are sufficient to guarantee the convergence of the the hydrodynamic force and horizontal velocity up to order $\mathcal{O}(10^{-3})$. This convergence was attained for all other Stokes number explored in this article. The mesh of $17116$ was used in all cases and the time step evolved from $\Delta t=10^{-2}$ for small values of the Stokes number ($\beta=2$) being decreased to $\Delta t=5\cdot10^{-4}$ for larger values ($\beta=19$).\\

\begin{figure}
  \centering
    % \begin{subfigure}{0.45\textwidth}
     \includegraphics[width=0.5\linewidth]{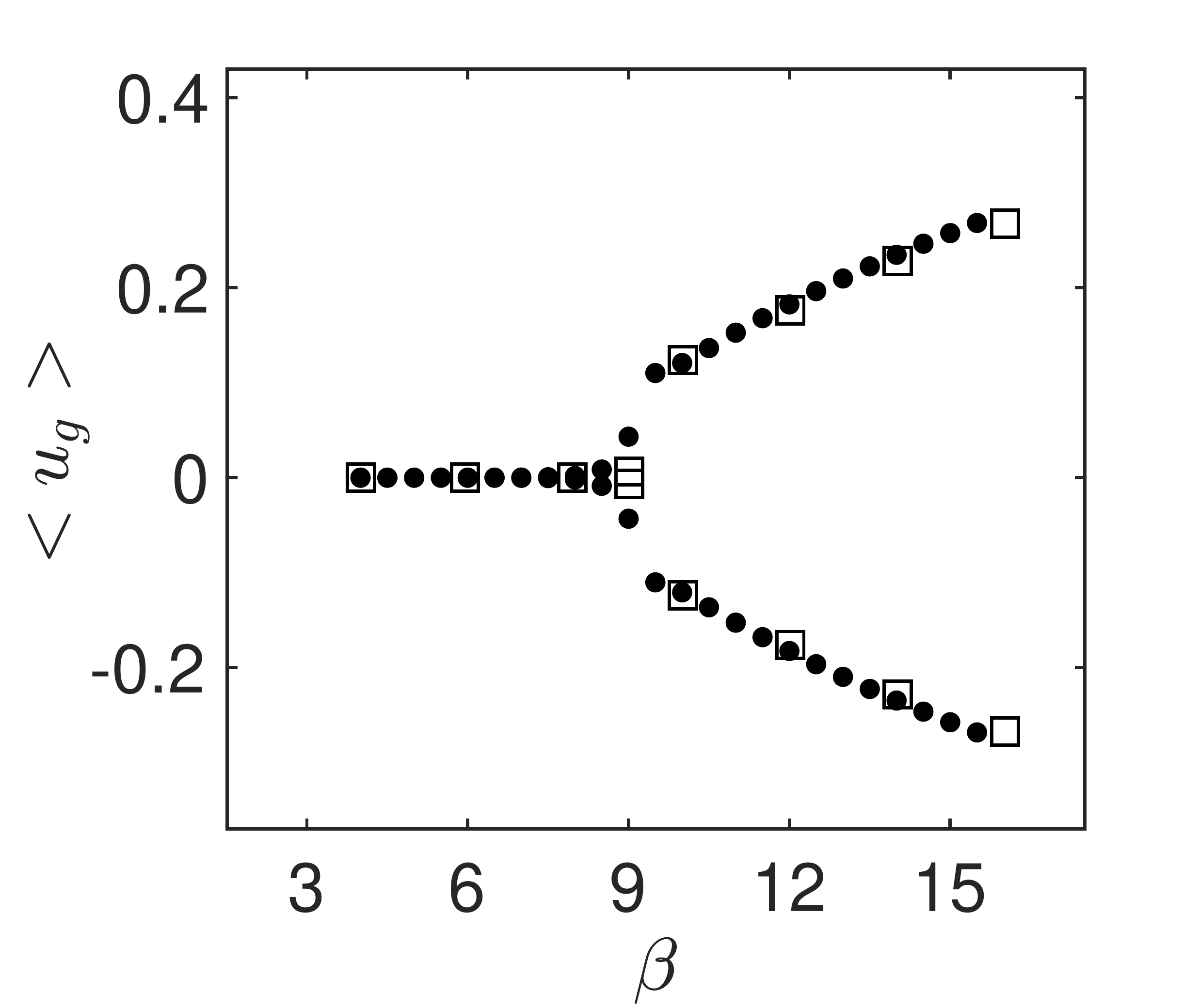}
    % \caption{}
    % \end{subfigure}
    
    \caption{Validation of the nonlinear solver. Time-averaged horizontal velocity of the foil as a function of the Stokes number $\beta$. Results of the  numerical algorithm described in \ref{Appendix:TimeMarchingScheme}
    (filled dots) are compared to results (empty squares)  extracted from figure 13 ( \textit{clamped} case) in \citet{Spagnolie2010}.}
    \label{Appendix-Validation-Spagnolie}
\end{figure}

The nonlinear solver employed in this study has then been validated by simulating the horizontal locomotion of a two-dimensional ellipsoid of aspect ratio $h=0.1$, flapping amplitude $A=0.25$, density ratio $\rho=10$ and Stokes number $\beta \in [2,15]$ as in the numerical study of \citep{Spagnolie2010}. In this work the onset of locomotion is around $\beta>8$, and as seen in figure \ref{Appendix-Validation-Spagnolie} both the emergence of propulsive solutions and their time-averaged horizontal velocities compare very well between the present numerical method and the values found on the reference.\\

\begin{table}
    \centering
    \begin{tabular}{c c c}
         & $\beta=40$, $KC=4.75$ & $\beta=100$, $KC=3.65$  \\
    \hline
        \citet{Elston2006} & 1.1282 & 1.1429 \\
        Present method & 1.1273 & 1.1411
    \end{tabular}
    \caption{Linear solver validation: Comparison of the leading Floquet multiplier $|\mu|$ with the values presented on Table 3 of \citet{Elston2006}}
    \label{Appendix:tab:comparisonElston}
\end{table}{}

The linear solver is validated through the purely hydrodynamic Floquet stability analysis of the flow symmetry breaking around a heaving cylinder. Two distinct flapping amplitudes and Stokes numbers are considered (using the Keulegan-Carpenter number $KC=4\pi A$): $(\beta,KC)=(40,4.75)$ and $(\beta,KC)=(100,3.65)$. We can see in the table \ref{Appendix:tab:comparisonElston} that the absolute value of the leading Floquet multiplier obtained in the two test cases is in a good agreement with values of \citet{Elston2006}.\\

\section{Effect of the foil shape on self-propelled regimes and stability}
\label{Appendix:Alben}

\begin{figure}
  \centering
    \begin{tabular}{ll}
         (a) & (b) \\
         \includegraphics[width=0.45\linewidth]{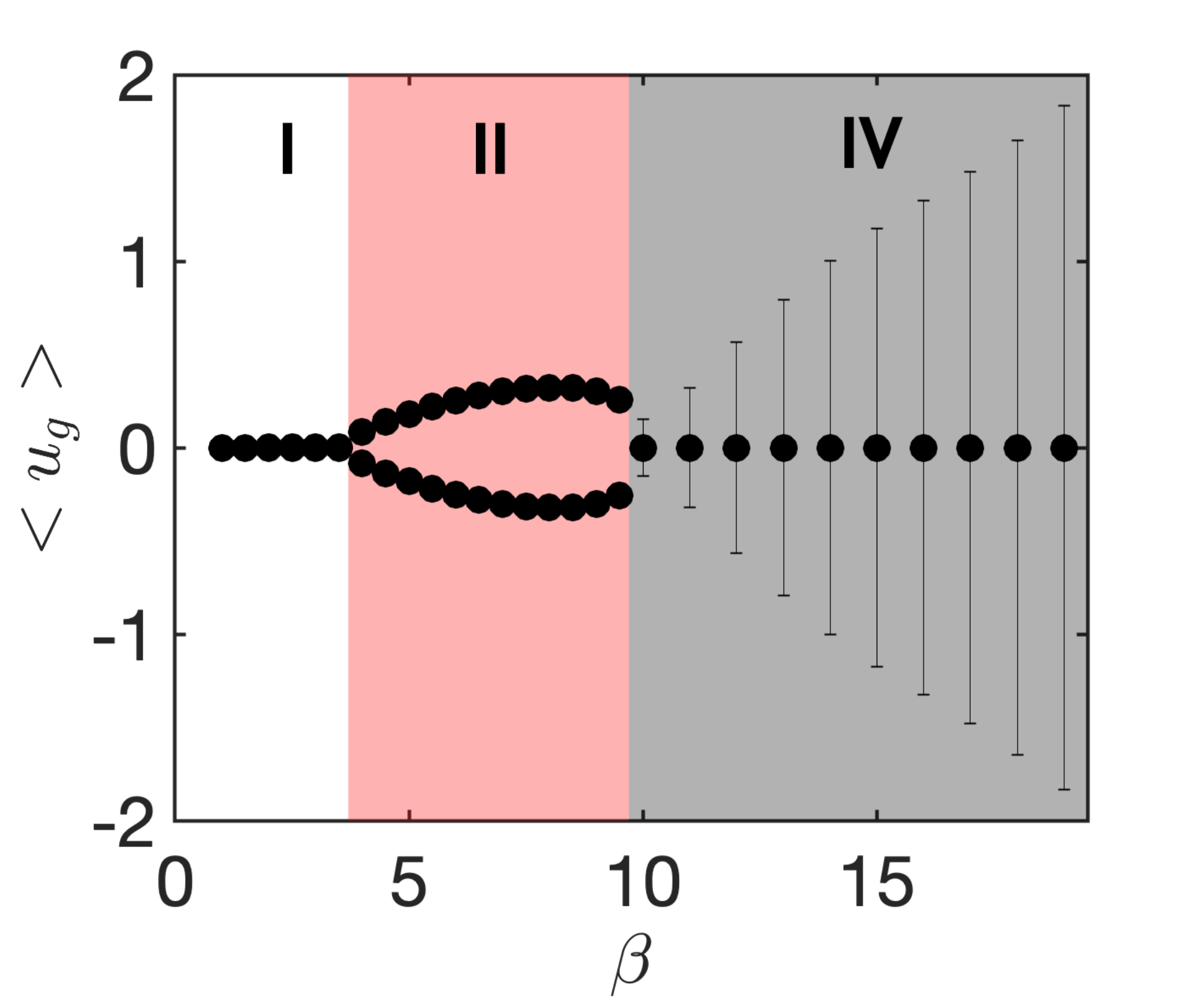}
         & \includegraphics[width=0.45\linewidth]{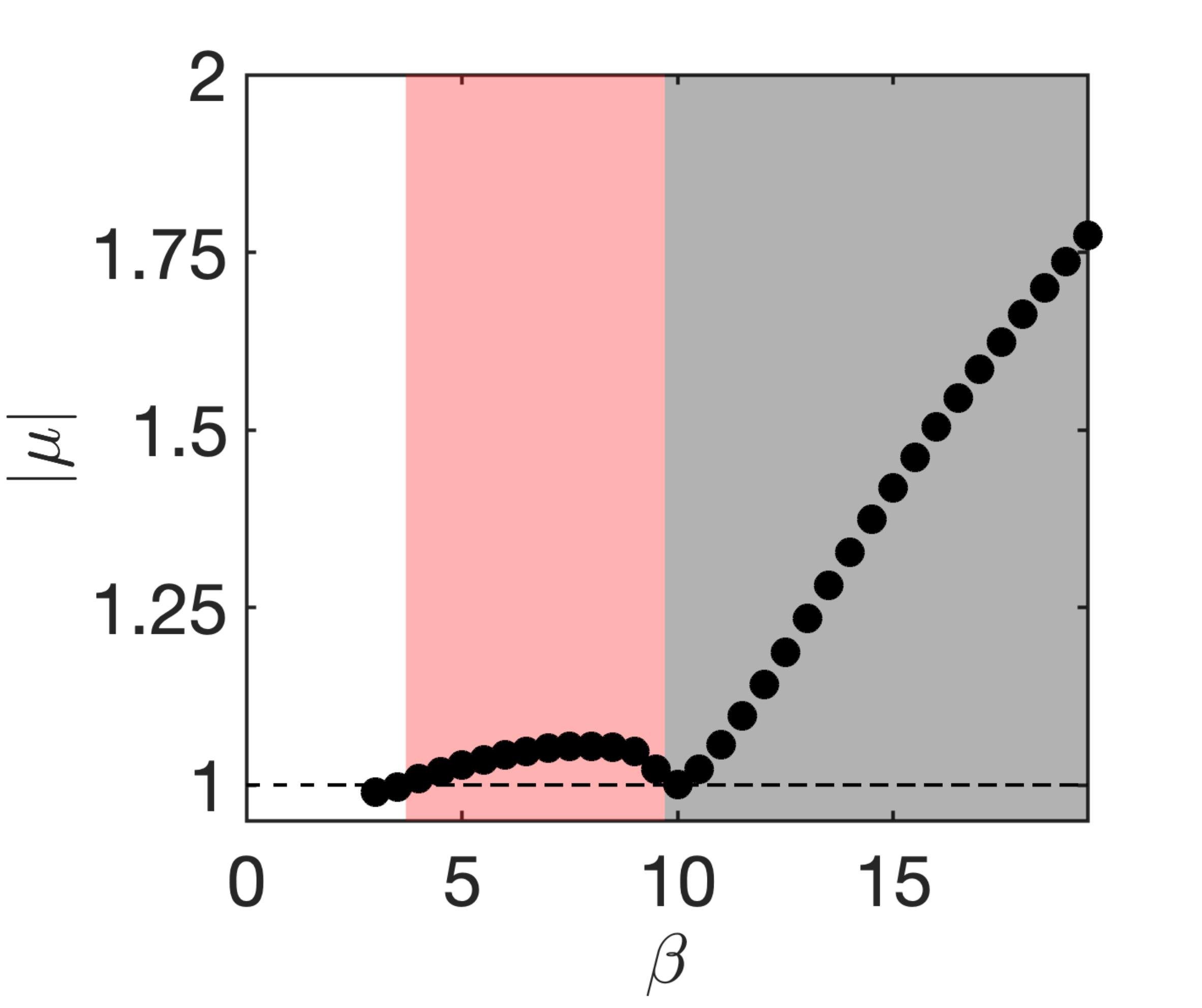}
    \end{tabular}
    \caption{Evolution of the (a) time-averaged horizontal velocity and (b) leading Floquet multiplier absolute value of an elliptical foil of minor/major axis aspect ratio $h=0.1$ with control parameters $A=0.5$ and $\rho=32$. In (a) - resp. (b) - white, red and grey background colors identify symmetric non-propulsive, unidirectional propulsive and back \& forth regimes - resp. stable, synchronous unstable and asynchronous unstable multipliers.}
    \label{Fig:Alben}
\end{figure}

To evaluate the influence of the foil shape and aspect ratio, the evolution with the Stokes number $\beta$ of the unsteady nonlinear dynamics and fluid-solid Floquet stability of an elliptical foil of aspect ratio $h=0.01$ is shown in figure \ref{Fig:Alben}. The flapping amplitude $A=0.5$ (identical to this work) and the density ratio $\rho=32$ have been fixed as to approach one of the configurations explored by \citet{Alben2005}.

The self-propelled regimes and their transition are similar to the rectangular foil with rounded edges. A remarkable difference is, nevertheless, the suppression of the intermediary non-propulsive regime III, between the unidirectional propulsive and the back \& forth regimes. Apparently the increase of the aspect ratio favours, as the decrease of the density ratio (\S \ref{Subsec:Foil_Shape_Rho}), the onset of non-coherent propulsion.
The obtained onset of back \& forth solutions ($\beta>9.5$) closely matches the one of \cite{Alben2005}. The existence of an unidirectional propulsive regime prior to the back \& forth one is, however, to our knowledge newly reported in the literature (apart from being briefly mentioned in the works of \cite{Deng2016a}). We suspect this regime has not yet been characterized due to two factors. In one hand, as illustrated in figure \ref{2-NonLinear-Prop}, this propulsive regime features a very long transitory regime of $\sim 200$ flapping periods. In addition, owing to the low Stokes number and its viscous nature, velocity perturbations added to the quiescent system initially decay. Following this initial decay, more than $100$ flapping periods are commonly needed for the horizontal velocity to grow above an initial perturbation $u_g=0.05$. This threshold is indeed beyond the one employed, for example, by \cite{Alben2005} who differentiated non-propulsive and propulsive solutions by the growth of initial velocity perturbations after $80$ flapping periods.

Unstable synchronous and asynchronous modes, figure \ref{Fig:Alben}(b), are again obtained in the same $\beta$ range as the unidirectional propulsive and back \& forth solutions. Despite the different geometry and the suppression of the intermediary non-propulsive regime, the fluid-solid Floquet stability analysis correctly predicts the onset of unidirectional propulsive and back \& forth solutions.

\bibliographystyle{jfm}
\bibliography{library}

\end{document}